\documentclass[envcount]{llncs}
\usepackage{amsmath}
\usepackage{amssymb}
\usepackage{paralist}
\usepackage{upref}
\usepackage{appendix}
\usepackage{subfigure}
\usepackage[top=1.5in, bottom=1.5in, left=1.2in, right=1.2in]{geometry}

\newtheorem{ob}{Observation}

\usepackage{graphicx}

\title{Complexity of Coloring Graphs without Paths and Cycles}
\author{Pavol Hell and Shenwei Huang}

\institute{School of Computing Science\\
Simon Fraser University, Burnaby B.C., V5A 1S6, Canada\\
\email{pavol@sfu.ca, shenweih@sfu.ca}}
\date{}

\begin{document}

\maketitle

\begin{abstract}
Let $P_t$ and $C_\ell$ denote a path on $t$ vertices and a cycle on $\ell$ vertices, respectively.
In this paper we study the $k$-coloring problem for $(P_t,C_\ell)$-free graphs. Maffray and Morel,
and Bruce, Hoang and Sawada, have proved that $3$-colorability of $P_5$-free graphs has a
finite forbidden induced subgraphs characterization, while Hoang, Moore, Recoskie, Sawada,
and Vatshelle have shown that $k$-colorability of $P_5$-free graphs for $k \geq 4$ does not. These
authors have also shown, aided by a computer search, that $4$-colorability of $(P_5,C_5)$-free
graphs does have a finite forbidden induced subgraph characterization. We prove that for any $k$,
the $k$-colorability of $(P_6,C_4)$-free graphs has a finite forbidden induced subgraph characterization.
We provide the full lists of forbidden induced subgraphs for $k=3$ and $k=4$. As an application, we
obtain certifying polynomial time algorithms for $3$-coloring and $4$-coloring $(P_6,C_4)$-free graphs.
(Polynomial time algorithms have been previously obtained by Golovach, Paulusma, and Song, but those
algorithms are not certifying; in fact they are not efficient in practice, as they depend on multiple
use of Ramsey-type results and resulting tree decompositions of very high widths.) To complement
these results we show that in most other cases the $k$-coloring problem for $(P_t,C_\ell)$-free
graphs is NP-complete. Specifically, for $\ell=5$ we show that $k$-coloring is NP-complete for
$(P_t,C_5)$-free graphs when $k \ge 4$ and $t \ge 7$; for $\ell \ge 6$ we show that $k$-coloring
is NP-complete for $(P_t,C_\ell)$-free graphs when $k \ge 5$, $t \ge 6$; and additionally, for
$\ell=7$, we show that $k$-coloring is also NP-complete for $(P_t,C_7)$-free graphs if $k = 4$
and $t\ge 9$. This is the first systematic study of the complexity of the $k$-coloring problem for
$(P_t,C_\ell)$-free graphs. We almost completely classify the complexity for the cases when
$k \geq 4, \ell \geq 4$, and identify the last three open cases.
\end{abstract}

\section{Introduction}

Since the $k$-coloring problem is known to be NP-complete for any fixed $k\ge 3$, there has
been considerable interest in studying restrictions to various graph classes. For instance the
$k$-coloring problem is polynomially solvable for perfect graphs, since a perfect
graph is $k$-colorable if and only if it has no subgraph isomorphic to $K_{k+1}$. (In fact the
chromatic number of perfect graphs can also be computed in polynomial time \cite{Lovasz}.)
One type of graph class that has been given wide attention in recent years is the class of $H$-free
graphs, for various graphs $H$ \cite{Fomin,Broersma,Short Cycle,Hoang,Schiermeyer,Woe}. For
example, if $H$ contains a cycle, then $k$-coloring is NP-complete for $H$-free graphs. This
follows from the fact, proved by Kami\'{n}ski and Lozin \cite{Lozin} and independently Kr\'al,
Kratochv\' il, Tuza, and Woeginger
\cite{Kral}, that, for any fixed $k \ge 3$ and $g\ge 3$, the $k$-coloring problem is NP-complete
for the class of graphs of girth at least $g$. Similarly, if $H$ is a forest with a vertex of degree at
least 3, then $k$-coloring is NP-complete for $H$-free graphs; this follows from \cite{Holyer}
and \cite{Leven}. Combining these results we conclude that $k$-coloring is NP-complete for
$H$-free graphs, as long as $H$ is not a linear forest, i.e., a union of disjoint paths. This focused
attention on the case when $H$ is a path. Woeginger and Sgall \cite{Woe} have proved that $4$-coloring
is NP-complete for $P_{12}$-free graphs, and that $5$-coloring is NP-complete for $P_8$-free
graphs. Later on, these results were improved by various groups of researchers
\cite{Fomin,Broersma,Short Cycle,Randerath 2007}. The strongest results so far are due to
Huang \cite{Huang MFCS} who has proved that $4$-coloring is NP-complete for $P_{7}$-free
graphs, and that $5$-coloring is NP-complete for $P_6$-free graphs. On the positive side, Ho\`{a}ng,
Kami\'{n}ski, Lozin, Sawada, and Shu \cite{Hoang} have shown that $k$-coloring can be solved in polynomial
time on $P_5$-free graphs for any fixed $k$. These results give a complete classification of the
complexity of $k$-coloring $P_t$-free graphs for any fixed $k \ge 5$, and leave only $4$-coloring
$P_6$-free graphs open for $k=4$. It should be noted that deciding the complexity of $3$-coloring
for $P_t$-free graphs seems difficult. It is not even known that whether or not there exists any $t$
such that $3$-coloring is NP-complete on $P_t$-free graphs. Randerath and  Schiermeyer
\cite{Schiermeyer} have given a polynomial time algorithm for $3$-coloring $P_6$-free graphs.
As far as we know, this result has been extended to $3$-coloring $P_7$-free graphs by Chudnovsky,
Maceli, and Zhong \cite{P_7 1,P_7 2}.

One interesting aspect of the $k$-coloring problem is the number of minimal obstructions, i.e.,
minimal non-$k$-colorable graphs. As noted above, there is a unique minimal non-$k$-colorable
perfect graph, namely $K_{k+1}$. It was shown by Bruce, Hoang and Sawada \cite{Bruce}, that
the set of minimal non-$3$-colorable $P_5$-free graphs is finite, while Hoang, Moore, Recoskie,
Sawada, and Vatshelle \cite{Hoang1} have shown that the set of minimal non-$k$-colorable
$P_5$-free graphs is infinite. These authors have also shown, aided by a computer search, that the set
of minimal non-$4$-colorable $(P_5,C_5)$-free graphs is finite.

In this paper we undertake a systematic examination of  $k$-coloring with inputs restricted to
$(P_t,C_{\ell})$-free graphs. Some results about $k$-coloring these graphs are known. In addition
to the case of $4$-coloring $(P_5,C_5)$-free graphs mentioned just above, it is known that when
$\ell=3$, each $k$-coloring is polynomial for $t \leq 6$, as $(P_6,C_3)$-free graphs have
bounded cliquewidth. On the other hand, for $t \geq 164$, $4$-coloring is NP-complete for
$(P_t,C_3)$-free graphs \cite{Short Cycle}. When $\ell=4$, each $k$-coloring is polynomial
for $(P_t,C_4)$-free graphs \cite{Short Cycle}. When $\ell \geq 5$, $4$-coloring is NP-complete
for $(P_t,C_{\ell})$-free graphs as long as $t$ is large enough with respect to $\ell$ \cite{Short Cycle}.
(For $\ell=5$, the bound on $t$ is $t \geq 21$.)

We first focus on the number of minimal obstructions in a case in which polynomial time algorithms
are known to exist, namely $(P_6,C_4)$-free graphs \cite{Short Cycle}. We prove that, for each $k$,
the set of minimal non-$k$-colorable $(P_6,C_4)$-free graphs is finite. We actually describe all
the minimal non-$k$-colorable $(P_6,C_4)$-free graphs for $k=3$ and $k=4$, and then apply
these results to derive efficient certifying
$k$-coloring algorithms in these cases. We complement these results by showing that in most
cases with $k \ge 4, \ell \ge 4$, the $k$-coloring problem for $(P_t,C_{\ell})$-free graphs is
NP-complete. Specifically, we prove that $k$-coloring is NP-complete for $(P_t,C_5)$-free graphs
when $k \ge 4$ and $t \ge 7$, and that $k$-coloring is NP-complete for $(P_t,C_\ell)$-free graphs
when $\ell \ge 6$ and $k \ge 5, t \ge 6$. We show that $k$-coloring is also NP-complete
for $(P_t,C_7)$-free graphs if $k = 4$ and $t\ge 9$. This almost completely classifies the
complexity of $k$-coloring for $(P_t,C_{\ell})$-free graphs when $\ell \geq 4, k \geq 4$.
The few remaining open problems are listed in the last section.

We say that $G$ is {\em $\mathcal{H}$-free} if it does not contain, as an induced subgraph,
any graph $H\in \mathcal{H}$. If $\mathcal{H}=\{H\}$ or $\mathcal{H}=\{H_1,H_2\}$, we
say that $G$ is $H$-free or $(H_1,H_2)$-free. For two disjoint vertex subsets $X$ and $Y$
we say that $X$ is {\em complete,} respectively {\em anti-complete}, to $Y$ if every vertex
in $X$ is adjacent, respectively non-adjacent, to every vertex in $Y$.
A graph $G$ is called a {\em minimal obstruction} for $k$-coloring if $G$ is not $k$-colorable
but any proper induced subgraph of $G$ is $k$-colorable. We also call $G$ a {\em minimal
non-$k$-colorable graph}. A minimal non-$(k-1)$-colorable graph is also called a {\em k-critical}
graph. A graph is {\em critical} if it is $k$-critical for some $k$. We shall use $n$ and $m$ to
denote the number of vertices and edges of $G$, respectively.

\section{Imperfect $(P_6,C_4)$-Free Graphs}

In this section, we analyze the structure of imperfect $(P_6,C_4)$-free graphs.
Let $G$ be a connected imperfect $(P_6,C_4)$-free graph. By the Strong Perfect
Graph Theorem \cite{SPGT}, $G$ must contain an induced five-cycle, say $C=v_0v_1v_2v_3v_4$.
We call a vertex $v\in V\setminus C$ a {\em $p$-vertex} with respect to $C$
if $v$ has exactly $p$ neighbors on $C$, i.e., $|N_C(v)|=p$.
We denote by $S_p$ the set of $p$-vertices for $0\le p\le 5$.
In the following all indices are modulo 5. Let $S_1(v_i)$ be the subset
of $S_1$ containing all 1-vertices that have $v_i$ as their neighbor on $C$.
Let $S_3(v_i)$ be the subset of $S_3$ containing all 3-vertices
that have $v_{i-1}$, $v_i$ and $v_{i+1}$ as their neighbors on $C$.
Let $S_2(v_i,v_{i+1})$ be the subset of $S_2$ containing all 2-vertices
that have $v_{i}$ and $v_{i+1}$ as their neighbors on $C$. Note that
$S_1=\bigcup_{i=0}^{4}S_1(v_i)$, $S_2=\bigcup_{i=0}^{4}S_2(v_i,v_{i+1})$
and $S_3=\bigcup_{i=0}^{4}S_3(v_i)$.

A subset $S\subseteq V$ is {\em dominating} if every vertex not
in $S$ has a neighbor in $S$. Brandst\"{a}dt and Ho\`{a}ng \cite{Brandstadt}
proved the following fact about induced five-cycles in $(P_6,C_4)$-free graphs.


\begin{lemma}\label{dominating induced C_5}(\cite{Brandstadt})
Let $G$ be a $(P_6,C_4)$-free graph without clique cutset.
Then every induced $C_5$ of $G$ is dominating.
\end{lemma}

In the rest of this section, we collect some information about imperfect
$(P_6,C_4)$-free graphs. Recall that we assume that $G$ is a connected
$(P_6,C_4)$-free graph, and $v_0v_1v_2v_3v_4$ is an induced
five-cycle in $G$. Then the following properties must hold.

\begin{enumerate}

\item[(P0)]\label{p0}
$S_5$ and each $S_3(v_i)$ are cliques and $S_4=\emptyset$.

\vspace{1mm}

\item[(P1)]\label{p1}
$S_1(v_i)$ is complete to $S_1(v_{i+2})$ and anti-complete to $S_1(v_{i+1})$;
moreover, if both sets $S_1(v_i)$ and $S_1(v_{i+2})$ are nonempty, then both
are cliques.

\vspace{1mm}

\item[(P2)]\label{p2}
$S_2(v_i,v_{i+1})$ is complete to $S_2(v_{i+1},v_{i+2})$ and anti-complete to
$S_2(v_{i+2},v_{i+3})$; moreover, if both sets $S_2(v_i,v_{i+1})$ and
$S_2(v_{i+1},v_{i+2})$ are nonempty, then both are cliques.

\vspace{1mm}

\item[(P3)]\label{p3}
$S_3(v_i)$ is anti-complete to $S_3(v_{i+2})$.

\vspace{1mm}

\item[(P4)]\label{p4}
$S_1(v_i)$ is anti-complete to $S_2(v_j,v_{j+1})$ if $j\neq i+2$;
moreover, if $y\in S_2(v_{i+2},v_{i+3})$ is not anti-complete
to $S_1(v_i)$, then $y$ is an universal vertex in $S_2(v_{i+2},v_{i+3})$.

\vspace{1mm}

\item[(P5)]\label{p5}
$S_1(v_i)$ is anti-complete to $S_3(v_{i+2})$.

\vspace{1mm}

\item[(P6)]\label{p6}
$S_2(v_{i+2},v_{i+3})$ is anti-complete to $S_3(v_i)$.

\vspace{1mm}

\item[(P7)]\label{p7}
One of $S_1(v_i)$ and $S_2(v_{i+3},v_{i+4})$ is empty, and
one of $S_1(v_i)$ and $S_2(v_{i+1},v_{i+2})$ is empty.

\vspace{1mm}

\item[(P8)]\label{p8}
One of $S_2(v_{i-1},v_i)$, $S_2(v_i,v_{i+1})$ and $S_2(v_{i+2},v_{i+3})$ is empty.

\vspace{1mm}

\item[(P9)]\label{p9}
If both $S_1(v_{i-1})$ and $S_1(v_{i+1})$ are nonempty, then $S_2=\emptyset$;
if both $S_1(v_{i})$ and $S_1(v_{i+1})$ are nonempty, then $S_2=S_2(v_i,v_{i+1})$.

\vspace{1mm}

\item[(P10)]\label{p10}
Let $x\in S_3(v_i)$. If both $S_2(v_{i+1},v_{i+2})$ and $S_2(v_{i+3},v_{i+4})$ are
nonempty, then $x$ is either complete or anti-complete to
$S_2(v_{i+1},v_{i+2})\cup S_2(v_{i+3},v_{i+4})$. In the former case, both
$S_2(v_{i+1},v_{i+2})$ and $S_2(v_{i+3},v_{i+4})$ are cliques.
Moreover, if $S_2(v_{i+2},v_{i+3})$ is also nonempty, then
$x$ is anti-complete to $S_2(v_{i+1},v_{i+2})\cup S_2(v_{i+3},v_{i+4})$.
\vspace{1mm}

\item[(P11)]\label{p11}
If $S_1(v_i)$ is not anti-complete to $S_2(v_{i+2},v_{i+3})$ then $S_1=S_1(v_i)$.
\vspace{1mm}

\item[(P12)]\label{p12}
If $G$ has no clique cutset, then $S_1(v_i)$ is complete to $S_3(v_i)$.

\end{enumerate}

The proofs of these properties are simple, using the absence
of induced copies of $P_6$ and $C_4$. The proof of property (P12) also
uses Lemma \ref{dominating induced C_5}.

\section{Obstructions to $k$-coloring}


In this section we shall prove our first main result, that for each $k$, there are only finitely many minimal
non-$k$-colorable $(P_6,C_4)$-free graphs. In subsequent sections we then describe all minimal
non-3-colorable and non-4-colorable $(P_6,C_4)$-free graphs, and apply these characterizations
to obtain polynomial time certifying algorithms for the 3-coloring and the 4-coloring problems on
$(P_6,C_4)$-free graphs.

The following lemma is folklore.

\begin{lemma}\label{degree lemma}
A minimal non $k$-colorable graph $G$ has $\delta(G)\ge k$ and no clique cutset.
\end{lemma}

Let $P$ be the graph obtained from the Peterson graph by adding one new vertex that is adjacent
to every vertex of $P$. A graph is called {\em specific} if it results from replacing each vertex of
$P$ by a clique of arbitrary size (including possibly size 0, resulting in deleting the vertex).

\begin{lemma}\label{C_6}(\cite{Brandstadt})
Let $G$ be a $(P_6,C_4)$-free graph without a clique cutset.
Then either $G$ is specific, or every induced $C_6$ of $G$ is dominating.
Moreover, there is a linear time algorithm to decide whether or not $G$ is specific.
\end{lemma}

We are now ready to prove the main result of this section, the finiteness of the number of minimal
obstructions for $k$-coloring $(P_6,C_4)$-free graphs. It should be observed that this result
is best possible in the sense that there are infinitely many minimal non-$k$-colorable $P_6$-free graphs
and infinitely many minimal non-$k$-colorable $C_4$-free graphs. The former fact follows from \cite{Hoang1}
where it is shown that there are infinitely many minimal non-$k$-colorable $P_5$-free graphs, and the latter
fact follows from \cite{erdos} where it is shown that there are non-$k$-colorable graphs of arbitrarily
high girth.

\begin{theorem}\label{finiteness}
For any $k$, there are only finitely many minimal non-$k$-colorable $(P_6,C_4)$-free graphs.
\end{theorem}

\noindent {\bf Proof.}
Let $G$ be a $(P_6,C_4)$-free minimal non-$k$-colorable graph.
By Lemma \ref{degree lemma}, $G$ has $\delta(G)\ge k$ and no clique cutset.
If $G$ contains $K_{k+1}$, then $G=K_{k+1}$. Thus we assume that that $G$ is $K_{k+1}$-free.
If $G$ contains an induced $C=C_6$, then either $G$ is specific or $C$ is dominating by Lemma
\ref{C_6}. In the former case, the size of $G$ is bounded by the definition of specific graph and
the fact that $G$ is $K_{k+1}$-free. In the latter case, we analyze the remaining vertices as to
their connection to $C$, analogously to what we did in the previous section for $C$ being a
five-cycle. We define again, for any $X\subseteq C$ the set $S(X)$ to consist of all vertices not
in $C$ that have $X$ as their neighborhood on $C$. Using the fact that $G$ is $(P_6,C_4)$-free,
we derive easily the fact that $S(X)=\emptyset$ if $X$ has size at most two, and that $S(X)$ is a
clique and thus of size at most $k$, if $|X|\ge 3$. Since there are at most $2^6$ such set $X$,
we conclude that $G$ has at most $64k$ vertices.

Therefore, we assume from now on that $G$ is $K_{k+1}$-free, $C_6$-free, and contains an
induced five-cycle $C=v_0v_1v_2v_3v_4$. Since $G$ is $K_{k+1}$-free, $|S_5|\le k-2$ and
$|S_3(v_i)|\le k-2$ for each $i$.

\begin{lemma}\label{bound lemma}
If $S_1(v_i)$ is anti-complete to $S_2(v_{i+2},v_{i+3})$, then both sets are bounded.
\end{lemma}

\noindent {\bf Proof of Lemma \ref{bound lemma}.} It suffices to prove this for $i=0$.
We bound $S_1(v_0)$ as follows. Let $A$ be a component of $S_1(v_0)$ and $x\in S_3(v_4)$.
If there exist two vertices $y,z\in A$ such that $xy\in E$ and $xz\notin E$, then we may assume
that $yz$ is an edge, by the connectivity of $A$. Thus, $zyxv_4v_3v_2$ induces a $P_6$. This
is a contradiction and therefore $x$ is either complete or anti-complete to $A$.
Moreover, $x$ is complete to $S_3(v_0)$ if $x$ is complete to $A$, as $G$ is $C_4$-free.
The same property holds if $x\in S_3(v_1)$.
Since $G$ has no clique cutset, $A$ must be complete to a pair of vertices
$\{x,y\}$ where $x\in S_3(v_1)$ and $y\in S_3(v_4)$. As $G$ is $C_4$-free, $A$ must be a clique and so of
size at most $k$. Moreover, the number of components of $S_1(v_0)$ is at most $(k-2)^2$. Otherwise
an induced $C_4$ would arise by the pigeonhole principle and the fact there are at most $(k-2)^2$ pairs
of vertices $\{x,y\}$ with $x\in S_3(v_1)$ and $y\in S_3(v_4)$.
Hence, $|S_1(v_0)|\le k(k-2)^2\le k^3$.

Let us now consider $S_2(v_2,v_3)$.
Let $A$ be a component of $S_2(v_2,v_3)$.
Observe first that a vertex $x\in S_3(v_2)\cup S_3(v_3)$,
is either complete or anti-complete to $A$, as $G$ is $P_6$-free.
Let $S'_3(v_3)$ and $S'_3(v_2)$ be the subsets of $S_3(v_3)$ and
$S_3(v_2)$ consisting of all vertices that are complete to $A$, respectively.
Moreover, $S'_3(v_3)$ and $S'_3(v_2)$ are complete to each other. Otherwise $v_0v_1t'ztv_4$
would induce a $C_6$ where $t\in S'_3(v_3)$ and $t'\in S'_3(v_2)$ with $tt'\notin E$, and $z\in A$.
So, if $A$ is anti-complete to $S_3(v_1)\cup S_3(v_4)$,
then $V'=S_5\cup \{v_2,v_3\}\cup S'_3(v_2)\cup S'_3(v_3)$ would be a clique cutset
of $G$.

Therefore, the set $T$ of neighbors of $S_3(v_1)\cup S_3(v_4)$ in $A$ is nonempty.
Let $B$ be  a component of $A\setminus T$. Our goal is to show that $B=\emptyset$ by
a similar clique cutset argument.
It is not hard to see that every vertex $t\in T$
is either complete or anti-complete to $B$ as $G$ is $P_6$-free.
Let $T'\subseteq T$ be the set of those vertices that are complete to $A$.
By the definition of $T'$, any $t\in T'$ is complete to $\{v_2,v_3\}\cup S'_3(v_2)\cup S'_3(v_3)$.
Let $x\in S_5$ and $t\in T'$ be a neighbor of some vertex $y\in S_3(v_1)$.
Then $xytv_3\neq C_4$ implies that $tx\in E$.
Hence, $T'$ is complete to $S_5$.

Next we show that $T'$ is a clique.
Let $t$ and $t'$ be any two vertices in $T'$, and $p\in B$.
If $t$ is a neighbor of some vertex in $S_3(v_4)$
and $t'$ is a neighbor of some vertex in $S_3(v_1)$, then $v_0v_1t'ptv_4$ would induce a $C_6$,
unless $tt'\in E$. Now we assume that both $t$ and $t'$ are neighbors of some vertex in $S_3(v_4)$.
If $t$ and $t'$ have a common neighbor in $S_3(v_4)$, then $tt'\in E$ as $G$ is $C_4$-free.
So we may assume that there exist two distinct vertices $x,x'\in S_3(v_4)$
such that $xt,x't'\in E$ but $xt',x't\notin E$.
If $tt'\notin E$, then $C^*=xtpt'x'$ would be an induced $C_5$.
However, this contradicts Lemma \ref{dominating induced C_5}, since $v_1$ is anti-complete to $C^*$.
Therefore, $T$ is a clique and so $V'\cup T$ is a clique cutset of $G$.
Thus, $B=\emptyset$ and $A=T$.
Since $A$ is an arbitrary component of $S_2(v_2,v_3)$,
the above argument shows that $S_2(v_2,v_3)$ is dominated by $S_3(v_1)\cup S_3(v_4)$.
Note that for any vertex $x\in S_3(v_1)\cup S_3(v_4)$,
the neighbors of $x$ in $S_2(v_2,v_3)$ form a clique and hence have size at most $k$. This shows that
$|S_2(v_2,v_3)|\le 2k(k-2)\le 2k^2$. \qed

Now we consider the following cases.

{\bf Case 1.} There exists some $i$ such that $S_1(v_i)$ and $S_1(v_{i+2})$ are nonempty.

In this case $S_2=\emptyset$ by the property (P9).
Further, each nonempty $S_1(v_i)$ is a clique by (P1).
Hence, the size of $G$ is bounded.

{\bf Case 2.} There exists some $i$ such that $S_1(v_i)$ and $S_1(v_{i+1})$ are nonempty.

In this case $S_2=S_2(v_i,v_{i+1})$ by (P9).
Further, $S_1$ and $S_2$ are anti-complete to each other, hence by Lemma \ref{bound lemma},
the sizes of $S_1(v_i)$ and $S_2(v_i,v_{i+1})$ are bounded.

{\bf Case 3.} $S_1=\emptyset$. Then the size of $G$ is bounded by Lemma \ref{bound lemma}.

{\bf Case 4.} There is exactly one $S_1(v_i)$ that is nonempty.
We may assume that $S_1(v_0)\neq \emptyset$ and that $S_1(v_0)$ is not anti-complete to $S_2(v_2,v_3)$.
If $S_2(v_1,v_2)\neq \emptyset$ or $S_2(v_3,v_4)\neq \emptyset$, then each nonempty $S_2(v_i)$
would be a clique (and hence bounded) as $G$ is $C_4$-free.
So we assume that $S_2(v_1,v_2)=S_2(v_3,v_4)=\emptyset$.
By Lemma \ref{bound lemma}, $S_2(v_0,v_4)$ and $S_2(v_0,v_1)$ are bounded.
The remaining sets are $S_1(v_0)$ and $S_2(v_2,v_3)$.

\noindent {\bf Bounding the size of $S_1(v_0)$.}
Let $X\subseteq S_1(v_0)$ be the set of vertices that are not anti-complete to $S_2(v_2,v_3)$,
let $S'_1(v_0)=S_1(v_0)\setminus X$, and let $A$ be a component of  $S'_1(v_0)$.
As $G$ is $P_6$-free, we conclude that any vertex $x\in X\cup S_3(v_1)\cup S_3(v_4)$ is either
complete or anti-complete to $A$. If $A$ has a  neighbor in both $S_3(v_1)$ and $S_3(v_4)$,
then $A$ must be a clique and thus of size at most $k$.
Further, there are at most $k^2$ such components.

Hence, we may assume that $A$ is anti-complete  to $S_3(v_4)$.
Let $X'\subseteq X$ be the set of vertices that are complete to $A$.
We claim that $X'$ is a clique. Let $x_i\in X'$ ($i=1,2$) and $p\in A$.
If $x_1$ and $x_2$ have a common neighbor $y\in S_2(v_2,v_3)$,
then $x_1x_2\in E$ or $x_1px_2y$ would induce a $C_4$.
So, we assume that there exist $y_i\in S_2(v_2,v_3)$ ($i=1,2$)
such that $x_iy_i\in E$ but $x_iy_j\notin E$ for $i\neq j$.
Now $x_1y_1y_2x_2p$ is an induced $C_5$, and it is anti-complete to $v_1$,
which contradicts Lemma \ref{dominating induced C_5}.
Let $S'_3(v_1)\subseteq S_3(v_1)$ be the set of vertices that are complete to $A$.
By $C_4$-freeness of $G$ it is easy to see that $S'_3(v_1)$ is complete to $X'$.
Let $V'=\{v_0\}\cup S_3(v_0)\cup S'_3(v_1)$.
If $A$ is anti-complete to $X'$ or $S_5$, then
$G$ has a clique cutset $V'\cup S_5$ or $V'\cup X'$.
So, $A$ has a neighbor $x\in X'$ and $p\in S_5$ with $px\notin E$.
As $|X'|\le k^2$ and $|S_5|\le k$, there are at most $k^3$ such pairs of vertices.
Hence, there are at most $k^3$ such components,
otherwise by the pigeonhole principle an induced $C_4$ would arise.

Hence, it suffices to bound the size of $A$.
Let $R\subseteq S_5$ be the set of vertices that are not anti-complete to $A$
and have a non-neighbor in $X'$. Let $S'_5=S_5\setminus R$.
Note that $X'$ and $S'_5$ are complete to each other.
Let $T\subseteq A$ be the set of vertices that are neighbors of $R$.
Since any $r\in R$ has a non-neighbor $x\in X'$, the set $N_A(r)$ is a clique and hence
$|N_A(r)|\le k$. So, $|T|\le k^2$. Let $B$ be a component of $A\setminus T$.
Observe that any $t\in T$ is either complete or anti-complete to $B$.
If not, let $bb'\in E(B)$ with $bt\in E$ but $b't\notin E$. Let $r\in R$ be a neighbor
of $t$, let $x\in X'$ be a non-neighbor of $r$, and let $y\in S_2(v_2,v_3)$ be a neighbor of $x$.
If $ry\in E$ then $tryx$ would induce a $C_4$. But now $b'btrv_2y$ induces a $P_6$.

Let $T'\subseteq T$ be the set of vertices that are complete to $B$.
Note that by definition $V^*=S'_5\cup X'\cup V'$ is a clique.
Our goal is to show that
$V^*\cup T'$ is a clique. Let $t_i\in T'$. If $t_1$ and $t_2$ have a common neighbor in $R$,
then an induced $C_4$ would arise unless $t_1t_2\in E$.
So, we assume that there exist $r_i\in R$ ($i=1,2$)
such that $t_ir_i\in E$ but $t_ir_j\notin E$ for $i\neq j$.
Let $b\in B$. Now $t_1r_1r_2t_2b$ induces a $C_5$. Let $x_i\in X'$ be a non-neighbor of $r_i$
and $y_i\in S_2(v_2,v_3)$ be a neighbor of $x_i$. Note that for any $r\in S_5$, $r$ is either
complete or anti-complete to any edge between $S_1(v_0)$ and $S_2(v_2,v_3)$.
If $x_1=x_2$ or $y_1=y_2$,  $t_1r_1r_2t_2b$ would not be dominating.
Hence, $x_1\neq x_2$,  $y_1\neq y_2$ and $y_ix_j\notin E$ for $i\neq j$.
Now $x_1x_2y_2y_1$ induces a $C_4$. This proves that $T'$ is a clique.
By definition, $T'$ is complete to $X'\cup S_3(v_0)\cup \{v_0\}$.
Let $q\in S'_3(v_1)$ and $t\in T'$, and $r\in R$ be a neighbor of $t$. Since $qbtr$ does not
induce a $C_4$, we have $tq\in E$. Now suppose that $q\in S'_5$ and $q$ has a neighbor $b\in B$.
As $qbtr$ does not induce a $P_4$, we have $qt\in E$.
Hence, $T'$ is complete to $S'_3(v_1)\cup S'_5$.
We have shown that $T'$ is complete to $V^*$ and $T'$ is a clique. So, $V^*\cup T'$
is a clique cutset if $B\neq \emptyset$. Therefore, $A=T$ and has size at most $k^2$.
Thus, $|S_1(v_0)|\le k^2+k^2\times k+k^2\times k^3= k^2+k^3+k^5$.

\noindent {\bf Bounding the size of $S_2(v_2,v_3)$.}
Let $Y\subseteq S_2(v_2,v_3)$ be the set of vertices that
are not anti-complete to $S_1(v_0)\cup S_3(v_1)\cup S_3(v_4)$.
Let $A$ be a component $S'_2(v_2,v_3)=S_2(v_2,v_3)\setminus Y$.
As in previous case, we can show that any $y\in Y$ is either complete or anti-complete to $A$.
Let $Y'\subseteq Y$ be the set of vertices that are complete to $A$.
Since any vertex in $S_2(v_2,v_3)$ that is not anti-complete to $S_1(v_0)$
is a universal vertex in $S_2(v_2,v_3)$, we conclude that $Y'$ is a clique.
Let $S'_3(v_3)$ and $S'_3(v_2)$ be the subsets of $S_3(v_3)$ and $S_3(v_2)$ consisting of all vertices that
are complete to $A$, respectively. Let $V'=\{v_3,v_2\}\cup S'_3(v_2)\cup S'_3(v_3)$. If $A$ is anti-complete to
$S_5$ or $Y'$, then $V'\cup S_5$ or $V'\cup Y'$ would be a clique cutset.
Hence, $A$ corresponds to a pair of nonadjacent vertices $y\in Y'$ and $r\in S_5$ such that $r$
is not anti-complete to $A$. By property (P4),
each $y\in Y$ is a dominating vertex in $S_2(v_2,v_3)$, and so $|Y'|\le |Y|\le k$.
Since $|Y'|\le k$ and $|S_5|\le k$, there are at most $k^2$ components of $S'_2(v_2,v_3)$
by the pigeonhole principle and the fact that $G$ is $C_4$-free.

It suffices to bound the size of $A$.
We define $R\subseteq S_5$, $S'_5=S_5\setminus R$ and $T=N_A(R)$ as in the previous case.
Then $|T|\le k^2$.
Let $B$ be a component of $A\setminus T$. Note that any $t\in T$
is either complete or anti-complete to $B$.
Let $T'\subseteq T$ be the set of vertices that are complete to $A$.
By definition, $V^*=V'\cup S'_5\cup Y'$ is a clique.
Moreover, $T'$ is complete to $V^*\setminus S'_5$.
Let $b \in B$ be a neighbor of $q\in S'_5$, and let $t\in T'$. Then $tb\in E$.
Let $r\in R$ be a neighbor of $t$.
Since $btrq$ does not induce a $C_4$, we have $tq\in E$, as $rb\notin E$ by definition.
Hence, $T'$ is complete to vertices in $S'_5$ that are not anti-complete to $B$.
Finally, we show that $T'$ is a clique. Let $t_i\in T'$ for $i=1,2$.
Let $r_i\in R$ be a neighbor of $t_i$. If $r_1=r_2$, then $t_1t_2\in E$ or $t_1bt_2r_1$ would
induce a $C_4$. So $r_1\neq r_2$ and $r_it_j\notin E$ if $i\neq j$.
Suppose that $t_1t_2\notin E$. Then  $bt_1r_1r_2t_2$ induces a $C_5$.
Let $y_i\in Y'$ be a non-neighbor of $r_i$, and let $x_i\in S_1(v_0)$ be a neighbor of $y_i$ ($i=1,2$).
If $y_1=y_2$ or $x_1=x_2$, then $bt_1r_1r_2t_2$ is not dominating, contradicting
Lemma \ref{dominating induced C_5}.
Hence, $y_1\neq y_2$ and $y_ix_j\notin E$. Thus, $x_1x_2\notin E$.
Since $\{y_1,y_2\}$ is complete to $A$ and thus to $\{b,t_1,t_2\}$, the set
$\{y_1,y_2,r_1,r_2\}$ induces a disjoint union of two copies of $K_2$.
Moreover, $r_ix_i\notin E$ or $x_ir_it_iy_i$ would induce a $C_4$.
Since $bt_1r_1r_2t_2$ is dominating, we obtain that $r_1x_2\in E$ and $r_2x_1\in E$.
But then $\{y_1,y_2,x_1,x_2,r_1,r_2\}$ induces a $C_6$, a contradiction.
Hence, $A=T$ and so has size at most $k^2$.
Therefore, $|S_2(v_2,v_3)|\le k^2+k^2\times k^2=k^4+k^2$. \qed

\section{Obstructions to $3$-Coloring}

In this section we explicitly describe all the minimal non-$3$-colorable $(P_6,C_4)$-free graphs.
We note that \cite{maffray}, in conjunction with \cite{Bruce}, describe all minimal non-3-colorable
$P_5$-free graphs, and that \cite{Hoang1} describes all minimal non-$4$-colorable
$(P_5,C_5)$-free graphs.

\begin{figure}
\subfigure[$K_4$.]{
\label{f7a} 
\includegraphics[width=0.2\textwidth]{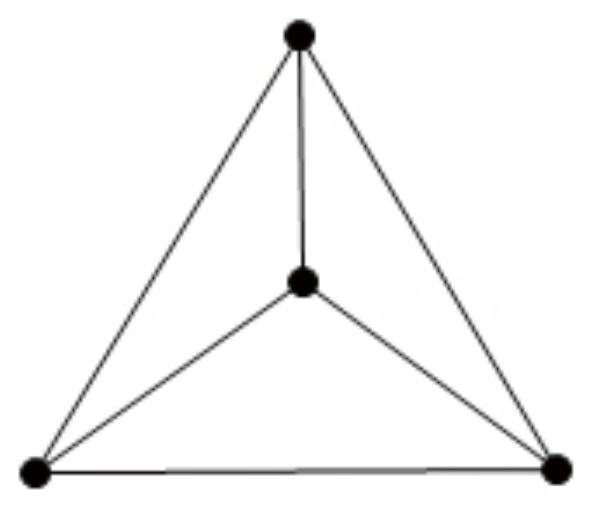}}
\hspace{0.1in}
\subfigure[$W_5$.]{
\label{f7b} 
\includegraphics[width=0.2\textwidth]{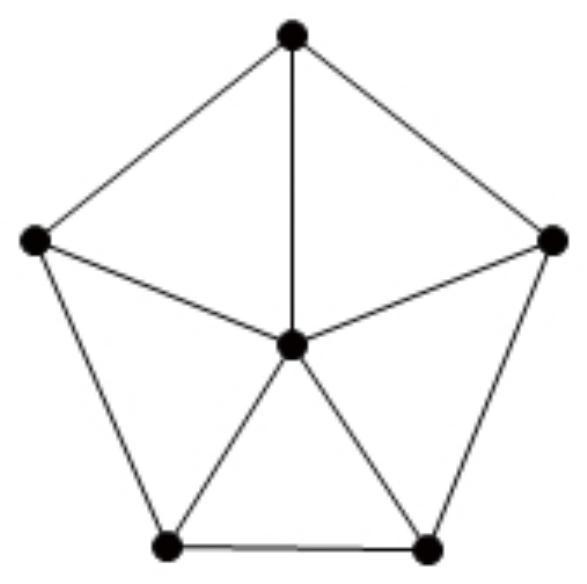}}
\hspace{0.1in}
\subfigure[The Hajos graph.]{
\label{f7a} 
\includegraphics[width=0.2\textwidth]{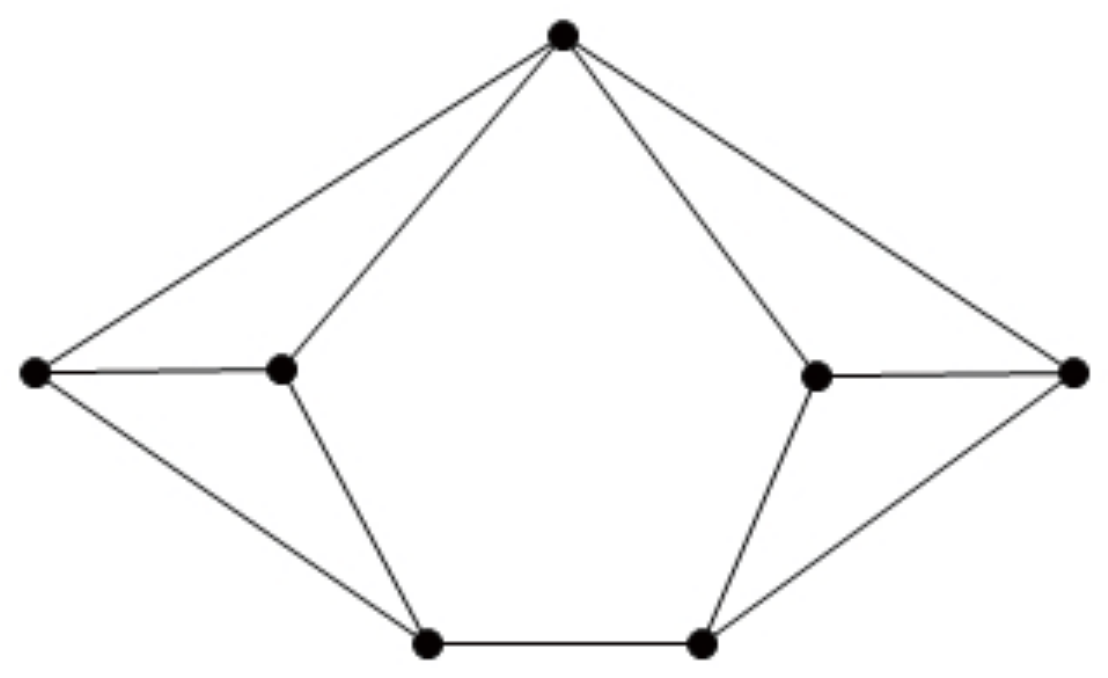}}
\hspace{0.1in}
\subfigure[$F$.]{
\label{f7a} 
\includegraphics[width=0.2\textwidth]{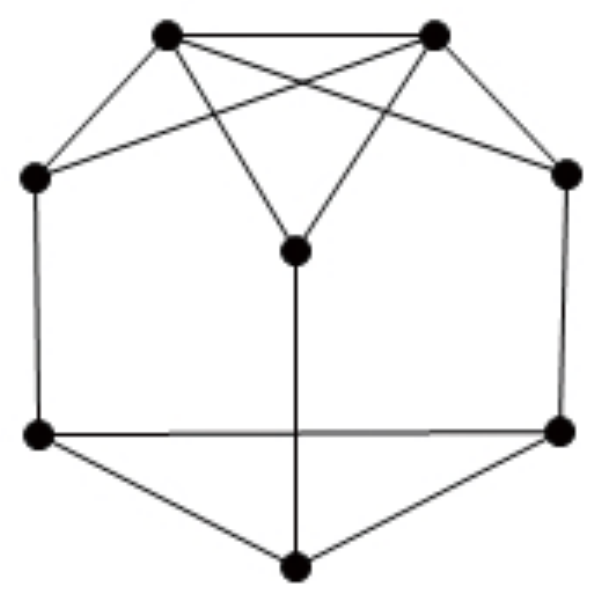}}
\vspace*{-10pt}
\caption{All minimal non-3-colorable $(P_6,C_4)$-free graphs.}
\label{3MO} 
\end{figure}

\begin{theorem}\label{3-coloring}
There are exactly four minimal non-3-colorable $(P_6,C_4)$-free graphs,
depicted in Figure \ref{3MO}.
\end{theorem}

\noindent {\bf Proof.} Let $G$ be a $(P_6,C_4)$-free minimal non $3$-colorable graph.
From the first few lines of the proof of Theorem \ref{finiteness} we know that $G$ has
$\delta(G)\ge k$, contains no clique cutset, is $K_4$-free, and contains an
induced $C=C_5=v_0v_1\ldots v_4$. We use the notation $S_p$, $S_1(v_i)$,
$S_2(v_i,v_{i+1})$, and $S_3(v_i)$ from Section 2. From Lemma \ref{dominating induced C_5},
we have $S_0=\emptyset$. It is easy to see that $|S_5|\le 1$. If $|S_5|=1$, then $G=W_5$.
So we may assume that $S_5=\emptyset$. If there exists an index $i$ such that
$S_3(v_i)\neq \emptyset$ and $S_3(v_{i+2})\neq \emptyset$,  then $G$ is the Hajos graph.
Hence, at most two $S_3(v_i)$'s are nonempty. Furthermore, each $S_3(v_i)$ is clique
and contains at most one vertex, since $G$ is $(C_4,K_4)$-free. Therefore, $|S_3|\le 2$.
We distinguish three cases.

{\bf Case 1.} $|S_3|=2$.

Without loss of generality, assume that $S_3(v_0)=\{x\}$ and $S_3(v_1)=\{y\}$.
$xy\notin E$ as $G$ is $K_4$-free. Also $S_1(v_3)=\emptyset$, otherwise let $t\in S_1(v_3)$
and then $tv_3v_2yv_0x=P_6$. Moreover, $x$ (respectively $y$) is complete to $S_2(v_3,v_4)$ (respectively $S_2(v_2,v_3)$).
Otherwise there exists some vertex $z\in S_2(v_3,v_4)$ with $xz\notin E$. Then $zv_3v_2yv_0x=P_6$.
Hence, $S_2(v_3,v_4)$ and $S_2(v_3,v_2)$ are cliques and each of them contains at most one vertex.
As $d(v_3)\ge 3$ and $S_1(v_3)=\emptyset$, at least one of them is nonempty.
Suppose first that $p\in S_2(v_3,v_4)$ and $q\in S_2(v_2,v_3)$.
Then $xp\in E$ and $yq\in E$. It follows from $S_1(v_3)=\emptyset$ and property (P7) that $S_1=\emptyset$.
Further, $S_2(v_1,v_2)=S_2(v_0,v_4)=\emptyset$ by (P10).
Hence we have $S_2=\{p,q\}$ by (P8), and therefore $N(x)=\{v_4,v_1,v_0,p\}$.
Since $G$ is a minimal obstruction, there exists a 3-coloring $\phi$ of $G-x$.
Note that we must have $\phi(v_4)=\phi(q)=\phi(v_1)$ and $\phi(p)=\phi(v_2)=\phi(v_0)$.
Consequently, we can extend $\phi$ to $G$ by setting $\phi(x)=\{1,2,3\}\setminus \{\phi(v_0),\phi(v_1)\}$.
This contradicts the fact that $G$ is not 3-colorable.
Therefore, exactly one of $S_2(v_3,v_4)$ and $S_2(v_3,v_2)$ is empty.
Without loss of generality, assume that $S_2(v_3,v_4)=\emptyset$ and let $z\in S_2(v_2,v_3)$.
Note that $N(v_3)=\{v_4,v_2,z\}$.
Let $\phi$ be a 3-coloring of $G-v_3$, and note that we must have $\phi(v_4)=\phi(v_1)=\phi(z)$.
Thus we can extend $\phi$ to $G$. This is a contradiction.

{\bf Case 2.} $|S_3|=1$. Without loss of generality, assume that $x\in S_3(v_0)$.

{\bf Case 2.1} $S_1(v_0)=\emptyset$.

We claim that in this case $S_2(v_2,v_3)=\emptyset$.
Otherwise we let $z\in S_2(v_2,v_3)$.
Note that $S_2(v_2,v_3)$ is independent and anti-complete to $x$ since $G$ is $(C_4,K_4)$-free.
By property (P4), the set $S_2(v_2,v_3)$ is anti-complete to $S_1$.
Since $\{v_2,v_3\}$ is not a clique cutset separating $S_2(v_2,v_3)$,
one of $S_2(v_3,v_4)$ and $S_2(v_1,v_2)$ is nonempty.
We assume by symmetry that $S_2(v_3,v_4)\neq \emptyset$ and let $w\in S_2(v_3,v_4)$.
By property (P7), $S_1=S_1(v_3)$.
Moreover, $x$ is anti-complete to $S_2(v_1,v_2)$ and $S_2(v_3,v_4)$.
Otherwise consider induced $C_5=C'=xv_1v_2v_3v_4$.
We define $S'_3$ with respect to $C'$ in the same way as we define $S_3$.
It is easy to check that $|S'_3|\ge 2$ and we are in Case 1.
Also, $S_2(v_0,v_4)=\emptyset$. Otherwise let $t\in S_2(v_0,v_4)$.
Since $xv_0twzv_2$ does not induce a $P_6$, $xt$ must be an edge,
and hence $\{x,v_0,v_4,t\}$ would induce a $K_4$.
Therefore, $N(x)=\{v_0,v_1,v_4\}$.
If $S_2(v_1,v_2)\neq \emptyset$, then in any 3-coloring $\phi$ of $G-x$
we would have $\phi(v_1)=\phi(v_4)$ and so $\phi$ can be extended to $G$.
This contradicts that $G$ is a minimal obstruction.
Hence, $S_2(v_1,v_2)=\emptyset$.
Note that $S_2=\{w,z\}$ since $G$ is $(C_4,K_4)$-free,
and hence $N(v_2)=\{v_3,z,v_1\}$.
Observe that in any 3-coloring $\phi$ of $G-v_2$ we have $\phi(z)=\phi(v_4)=\phi(v_1)$.
Consequently, we can extend $\phi$ to $G$, and this is a contradiction.
So the claim follows.
By (P7), one of $S_2(v_3,v_4)$ and $S_1(v_2)$ is empty,
and one of $S_2(v_1,v_2)$ and $S_1(v_3)$ is empty.
On the other hand, $S_2(v_3,v_4)\cup S_1(v_3)\neq \emptyset$
and $S_2(v_1,v_2)\cup S_1(v_2)\neq \emptyset$ as $\delta(G)\ge 3$.
This leads to the following two cases.

{\bf Case 2.1.a} $S_1(v_2)\neq \emptyset$ and $S_1(v_3)\neq \emptyset$ while $S_2(v_1,v_2)=S_2(v_3,v_4)=\emptyset$.

By (P7), the set $S_2(v_0,v_1)=S_2(v_0,v_4)=\emptyset$, and so $S_2=\emptyset$.
Since $\{v_3\}$ is not a clique cutset separating $S_1(v_3)$, we have $S_1(v_1)\neq \emptyset$.
Similarly, $S_1(v_4)\neq \emptyset$. Let $u_i\in S_1(v_i)$ for $i\neq 0$.
By (P1), each $S_1(v_i)$ is a clique, for $i\neq 0$.
Moreover, $|S_1(v_1)|+|S_1(v_3)|= 3$ and $|S_1(v_2)|+|S_1(v_4)|= 3$
as $\delta(G)\ge 3$ and $G$ is $K_4$-free.
If $|S_1(v_1)|=2$, then $|S_1(v_4)|=1$ and so $|S_1(v_2)|=2$.
Hence, $\{u_4,v_1,v_2\}\cup S_1(v_1)\cup S_1(v_2)$ induces a Hajos graph.
Therefore, $|S_1(v_1)|=|S_1(v_4)|=1$ and $|S_1(v_2)|=|S_1(v_3)|=2$.
Note that $x$ is anti-complete to $\{u_1,u_4\}$ or $G$ would contain either a $C_4$ or a $W_5$ as an induced subgraph.
Now $G$ has a 3-coloring: $\{v_1,u_3,u_2,v_4\}$, $\{v_0,v_3,u_1,u_2'\}$, $\{x,u_4,u'_3,v_2\}$
where $u'_2\in S_1(v_2)$ and $u'_3\in S_1(v_3)$.

{\bf Case 2.1.b} $S_2(v_1,v_2)\neq \emptyset$ and $S_2(v_3,v_4)\neq \emptyset$ while $S_1(v_2)=S_1(v_3)=\emptyset$.

Recall that $x$ is anti-complete to $S_2(v_1,v_2)$ and $S_2(v_3,v_4)$.
Let $y\in S_2(v_3,v_4)$ and $z\in S_2(v_1,v_2)$.
By (P8), $S_2=S_2(v_1,v_2)\cup S_2(v_3,v_4)$.
Since $\{v_3,v_4\}$ is not a clique cutset,
$S_2(v_3,v_4)$ has a neighbor in $S_1(v_1)$.
Similarly, $S_2(v_1,v_2)$ has a neighbor in $S_1(v_4)$.
However, this contradicts (P11).

{\bf Case 2.2} $S_1(v_0)\neq \emptyset$. Let $y\in S_1(v_0)$.

In this case $xy\in E$ by property (P12).
It follows from properties (P7) to (P9) that $S_2(v_1,v_2)=S_2(v_3,v_4)=\emptyset$.
If $S_1(v_0)$ is not anti-complete to $S_2(v_2,v_3)$, $G$ would contain $F$ as an induced subgraph and so $G=F$.
Hence, we may assume that $S_1(v_0)$ is anti-complete to $S_2(v_2,v_3)$.
Therefore, $S_2(v_2,v_3)=\emptyset$ or $\{v_2,v_3\}$ would be a clique cutset of $G$.
Since $\delta(G)\ge 3$, $S_1(v_2)\neq \emptyset$ and $S_1(v_3)\neq \emptyset$.
By (P9), $S_2=\emptyset$.
Let $p\in S_1(v_2)$ and $q\in S_1(v_3)$.
Note that $pq\notin E$, $py\in E$ and $qy\in E$.
Consider induced $C_5=C'=v_0v_1v_2py$.
We define $S'_3$ and $S'_p(v_0)$ in the same way we define $S_3$ and $S_p(v_0)$.
It is easy to see that $S'_3=S'_3(v_0)=\{x\}$.
By (P1), $S_1(v_0)$ is a clique and hence $S_1(v_0)=\{y\}$.
Now we are in Case 2.1 since $S'_1(v_0)=\emptyset$.

{\bf Case 3.} $|S_3|=0$, i.e., $V=C\cup S_1\cup S_2$.

We first claim that now $S_1\neq \emptyset$.
Assume that $S_1=\emptyset$ and thus $S_2\neq \emptyset$ or $G$ is 3-colorable.
Note that each $S_2(v_i,v_{i+1})$ is an independent set.
If there is exactly one nonempty $S_2(v_i,v_{i+1})$, then $G$ is 3-colorable.
If there are exactly three nonempty $S_2(v_i,v_{i+1})$'s,
then each of them is a clique by property (P2).
Since $G$ is $K_4$-free, each $S_2(v_i,v_{i+1})$ contains only one vertex.
Therefore, $G$ has eight vertices and it is easy to check that $G$ is 3-colorable.
Let us assume now that there are exactly two nonempty $S_2(v_i,v_{i+1})$.
If two $S_2(v_i,v_{i+1})$'s are complete to each other,
then we either find a $K_4$ or conclude that $|S_2|=2$ so that $G$ is 3-colorable.
If two $S_2(v_i,v_{i+1})$'s are anti-complete to each other, $G$ is also 3-colorable.
Therefore, we may assume that $S_1(v_0)\neq \emptyset$ and let $x\in S_1(v_0)$.
$S_2(v_3,v_4)=S_2(v_1,v_2)=\emptyset$ by (P7).
We claim that $S_1(v_3)\neq \emptyset$ and $S_1(v_4)\neq \emptyset$.
Otherwise we must have $S_2(v_2,v_3)\neq \emptyset$ and $S_2(v_0,v_4)\neq \emptyset$, and
$S_1(v_3)=S_1(v_4)=\emptyset$ since $d(v_3)\ge 3$ and $d(v_4)\ge 3$.
By properties (P7) and (P8), the set $S_2(v_0,v_1)=S_1(v_1)=\emptyset$.
This contradicts the fact that $\delta(G)\ge 3$.
By symmetry, $S_1(v_1)\neq \emptyset$ and $S_1(v_2)\neq \emptyset$.
Hence, $S_2=\emptyset$ and $S_1(v_i)$ is nonempty for each $i$.
Since $G$ is $K_4$-free, we have $5\le |S_1|\le 7$.
It is easy to check that $G$ is 3-colorable if $|S_1|\le 6$.
Thus $|S_1|=7$ and we may assume that $|S_1(v_0)|=|S_1(v_1)|=2$.
Let $u_i\in S_1(v_i)$ and $u'_0\in S_1(v_0)$, $u'_1\in S_1(v_1)$.
The subgraph induced by $\{u_3,u_1,v_1,v_0,u_0,u'_0,u'_1\}$
is isomorphic to the Hajos graph. \qed

\section{Obstructions to $4$-Coloring}

\begin{figure}
\subfigure[$G_{3,1}$.]{
\label{G31}
\includegraphics[width=0.2\textwidth]{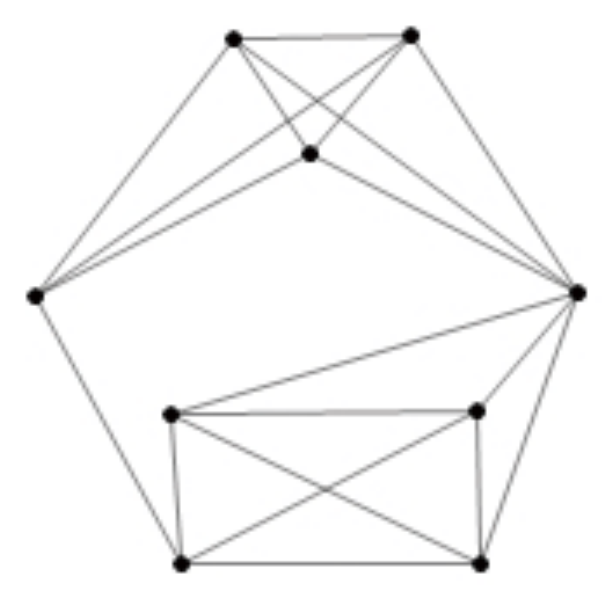}}
\hspace{0.1in}
\subfigure[$G_{2,2}$.]{
\label{G22}
\includegraphics[width=0.2\textwidth]{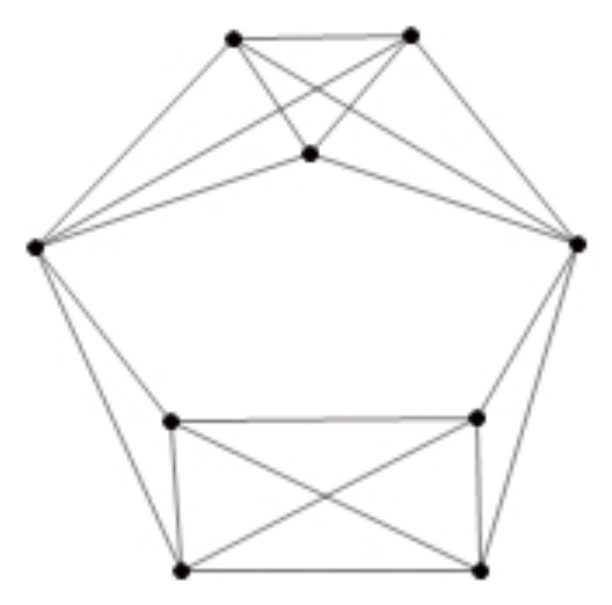}}
\hspace{0.1in}
\subfigure[$G_{2,1,1}$.]{
\label{G211}
\includegraphics[width=0.2\textwidth]{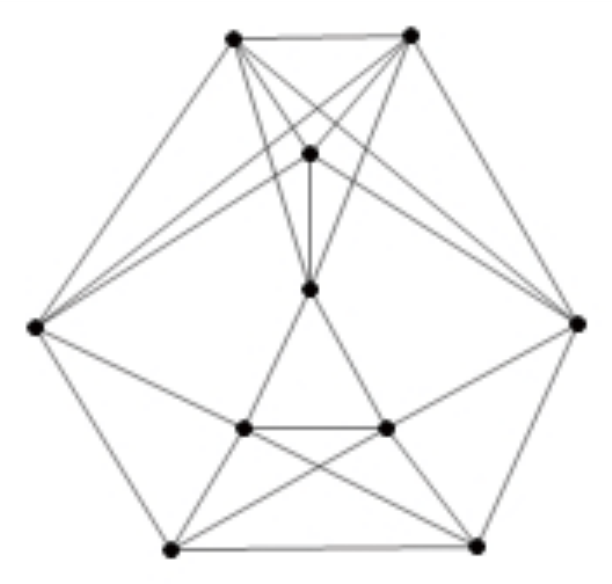}}
\hspace{0.1in}
\subfigure[$G_{1,1,1,1}$.]{
\label{G1111}
\includegraphics[width=0.2\textwidth]{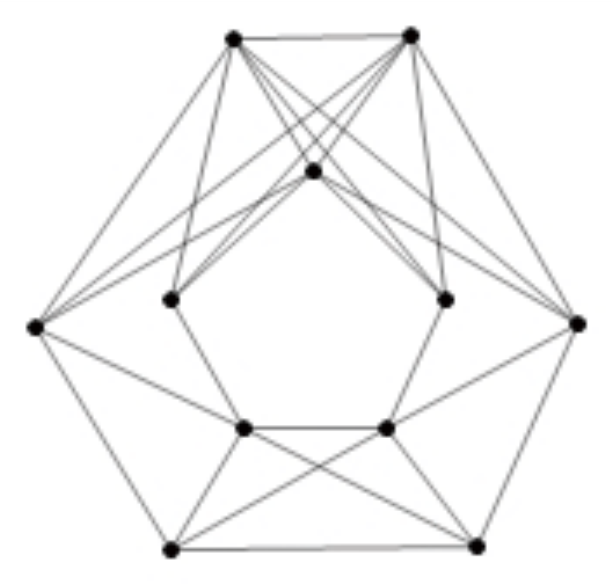}}
\subfigure[$F_1$.]{
\label{F1}
\includegraphics[width=0.2\textwidth]{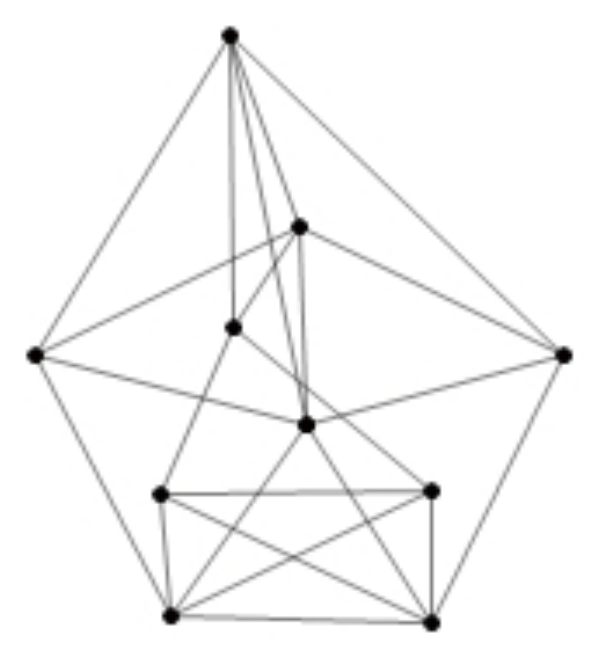}}
\hspace{0.1in}
\subfigure[$F_2$.]{
\label{F2}
\includegraphics[width=0.2\textwidth]{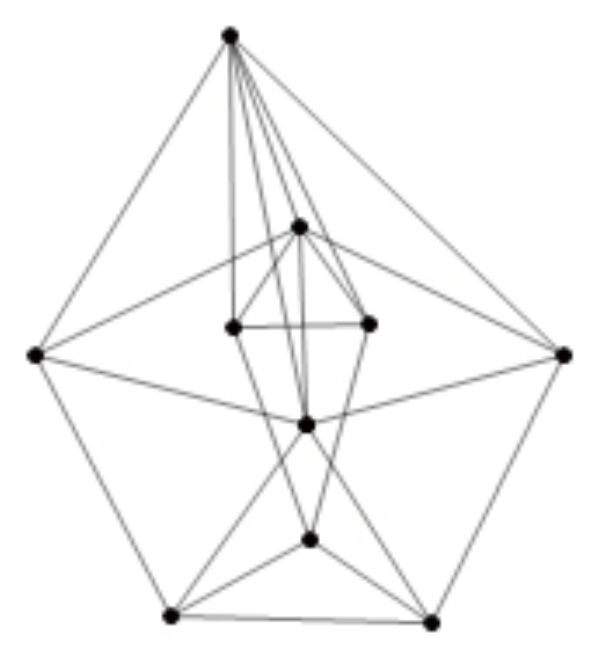}}
\hspace{0.1in}
\subfigure[$H_{1}$.]{
\label{H11}
\includegraphics[width=0.2\textwidth]{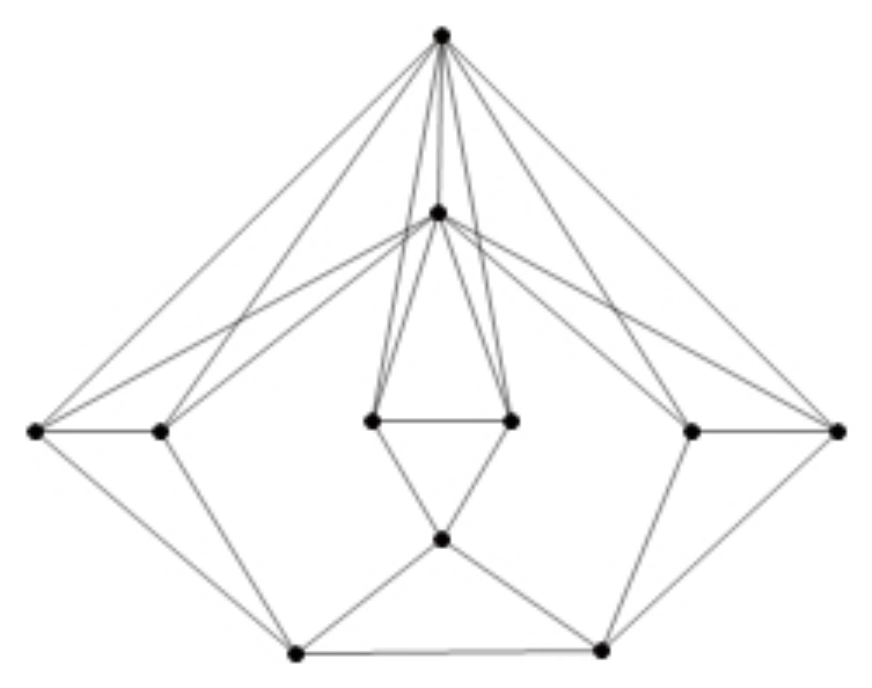}}
\hspace{0.1in}
\subfigure[$H_{2}$.]{
\label{H12}
\includegraphics[width=0.2\textwidth]{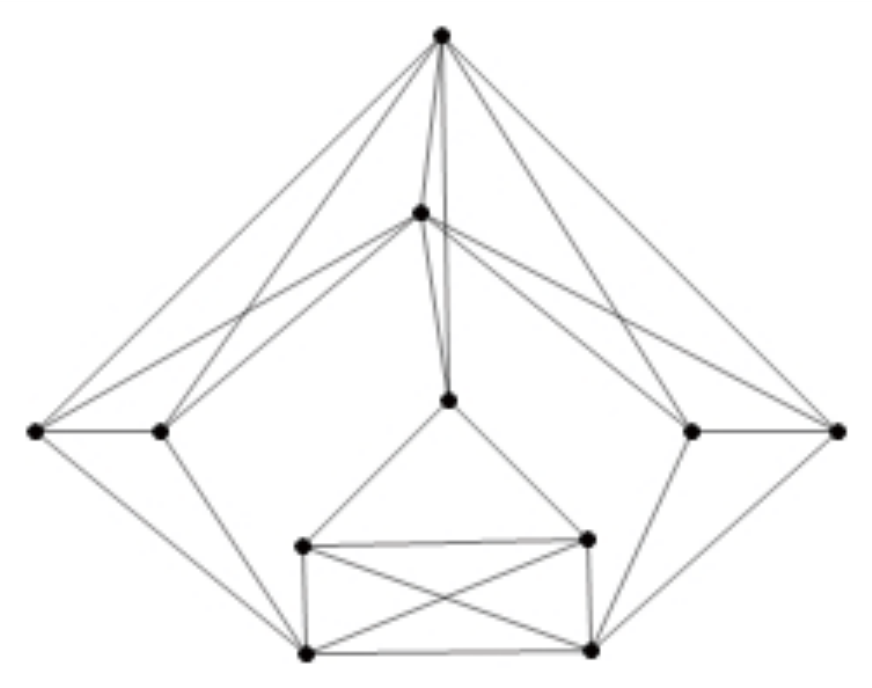}}
\hspace{0.1in}
\subfigure[$G_{P_4}$.]{
\label{GP4}
\includegraphics[width=0.2\textwidth]{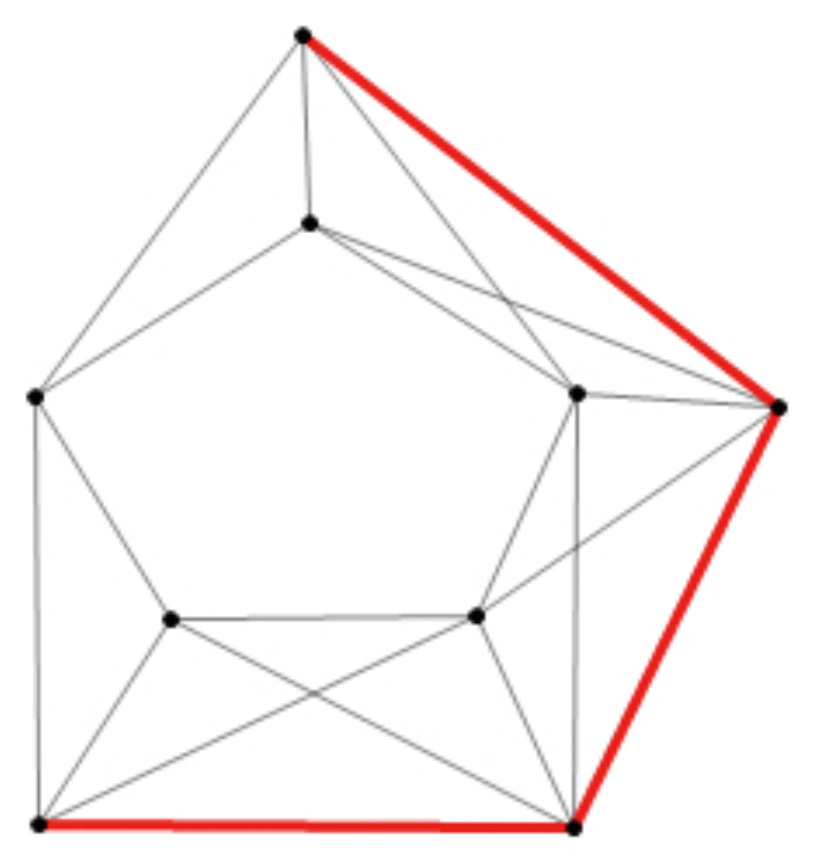}}
\vspace*{-10pt}
\caption{9 nontrivial minimal non-4-colorable $(P_6,C_4)$-free graphs.}
\label{4MO} 
\end{figure}

\begin{theorem}\label{4-coloring}
There are exactly $13$ minimal non-4-colorable  $(P_6,C_4)$-free graphs, depicted in Figure \ref{4MO}.
\end{theorem}


Our proof of Theorem \ref{4-coloring} has two parts. The first part deals with the case when $G$
contains an induced $W_5$. In the second part of the proof, we handle the case when $G$ has no
induced $W_5$. The technique we use is to choose some induced $C_5$ with a certain minimality
condition and derive some additional properties, valid for graphs without induced $W_5$.

\begin{lemma}\label{with W_5}
Let $G$ be a $(P_6,C_4)$-free minimal non-$4$-colorable graph with an induced $W_5$.
Then $G$ either is one of four minimal non-$3$-colorable graphs with an additional dominating
vertex or $G$ is $F_1$ or $F_2$ from Figure \ref{4MO}.
\end{lemma}
\noindent {\bf Proof.} If $G$ is perfect, then $G=K_5$. Hence, we assume that $G$ is imperfect and $K_5$-free.
Let $C=v_0\ldots v_4$ be an induced $C_5$. If $|S_5|\ge 2$, then $G$ is $W_5$ with an additional dominating vertex.
Hence we may assume that every induced $C_5$ has at most one 5-vertex. In particular, $|S_5|=1$.
Let $S_5=\{w\}$. Note that $S_5$ is complete to $S_3$. Hence, if there exists $i$ such that
$S_3(v_i)\neq \emptyset$ and $S_3(v_{i+2})\neq \emptyset$, then $G$ is the Hajos graph with an additional dominating vertex.
So there are at most two $S_3(v_i)$ are nonempty. Further, $|S_3(v_i)|\le 1$ as $G$ contains no $K_5$.
So $|S_3|\le 2$.

{\bf Case 1.} $|S_3|=2$. Let $x\in S_3(v_0)$ and $y\in S_3(v_1)$.
Then $xy\notin E$ as $G$ contains no $K_5$.
If $t\in S_1(v_3)$, then $tv_3v_4xv_1y$ would induce a $P_6$.
So, $S_1(v_3)=\emptyset$. Also, $x$ is complete to $S_2(v_3,v_4)$.
Otherwise let $z\in S_2(v_3,v_4)$ with $xz\notin E$. Then $zv_3v_2yv_0x$ would induce a $P_6$.
By symmetry, $y$ is complete to $S_2(v_2,v_3)$.
Note that $N(v_3)=\{v_2,v_4,w\}\cup S_2(v_3,v_4)\cup S_2(v_3,v_2)$. Now let $\phi$
be a 4-coloring of $G-v_3$. Note that $\phi(v_4)=\phi(v_1)$, $\phi(x)=\phi(y)$ and $\phi(v_0)=\phi(v_2)$.
As $v_4xv_0$ induces a triangle, we may assume that $\phi(v_4)=1$, $\phi(v_0)=2$, $\phi(x)=3$. Hence,
$\phi(w)=4$. Since $x$ is complete to $S_2(v_3,v_4)$ and $y$ is complete to $S_2(v_2,v_3)$, any vertex $t$
in $S_2(v_3,v_4)\cup S_2(v_3,v_2)$ has $\phi(t)\neq 3$. Hence, only colors 1, 2, 4 appear on $N(v_3)$
and so we can extend $\phi$ to $G$ by setting $\phi(v_3)=3$.

{\bf Case 2.} $|S_3|=0$. We claim that $S_1\neq \emptyset$. If not, $S_2\neq \emptyset$.
If there is exactly one nonempty $S_2(v_i,v_{i+1})$ then $S_2(v_i,v_{i+1})\cup \{w\}$ must be bipartite otherwise
a $K_5$ or $W_5$ with an additional dominating vertex would arise.
It is easy to see $G$ is 4-colorable.
Now suppose that there are exactly three nonempty $S_2(v_i,v_{i+1})$.
We may assume that $S_2(v_4,v_3)$, $S_2(v_3,v_2)$ and $S_2(v_2,v_1)$ are nonempty.
Observe that each $S_2(v_i,v_{i+1})$ is a clique now and thus contains at most two vertices.
Further, $|S_2(v_i,v_{i+1})|+|S_2(v_{i+1},v_{i+2})|\le 3$.
Let $p\in S_2(v_4,v_3)$, $r\in S_2(v_3,v_2)$ and $q\in S_2(v_2,v_1)$.
Suppose that $wr\in E$. Then the fact that $wrqv_1$ does not induce a $C_4$ implies that $wq\in E$.
By symmetry, $wp\in E$. Let $r'\in S_2(v_3,v_2)$. $wqr'v_3\neq C_4$ implies that $wr'\in E$.
Therefore, $w$ is either complete or anti-complete to $S_2$.
In the former case, $w$ is a dominating vertex and hence $G-w$ is a minimal non-$3$-colorable graph.
In the latter case, it is easy to check that $G$ is 4-colorable.

Suppose now that there are exactly two nonempty $S_2(v_i,v_{i+1})$.
If the two sets are complete to each other, then it is same as the above case.
So let us assume that  the two sets are anti-complete to each other.
Without loss of generality, assume that
$S_2(v_0,v_1)\neq \emptyset$ and $S_2(v_2,v_3)\neq \emptyset$.
Since $G-v_4$ is 4-colorable, both $\{w\}\cup S_2(v_2,v_3)$ and $\{w\}\cup S_2(v_0,v_1)$ are bipartite.
In fact, $T=\{w\}\cup S_2(v_0,v_1)\cup S_2(v_2,v_3)$ is also bipartite.
If not, let $Q$ be an induced odd cycle in $T$.
As $Q\nsubseteq \{w\}\cup S_2(v_0,v_1)$ and $Q\nsubseteq \{w\}\cup S_2(v_2,v_3)$, $Q$ contains a vertex in both
$S_2(v_0,v_1)$ and $S_2(v_3,v_4)$. As $S_2(v_0,v_1)$ is anti-complete to $S_2(v_2,v_3)$,
$Q$ must contain $w$ and $Q$ is not a triangle.
However, $Q-w$ is connected and hence is fully contained in $S_2(v_0,v_1)$
or $S_2(v_2,v_3)$. This is a contradiction.
We therefore can 4-color $G$ as following: $\phi(v_0)=\phi(v_2)=1$,
$\phi(v_1)=\phi(v_3)=2$, $\phi(v_4)=3$, $\phi(w)=4$,
and color one partite of $T$ with color 3 and the other with color 4.

Therefore, we may assume that $S_1(v_0)\neq \emptyset$.
Going through the same argument for Case 3 in Theorem \ref{3-coloring},
we conclude that $S_1(v_i)\neq \emptyset$ for each $i$ and $S_2=\emptyset$.
Moreover, $w$ is either complete or anti-complete to $S_1$ as $G$ is $C_4$-free.
In the former case, $w$ is a dominating vertex and hence $G-w$ is a minimal minimal non-$3$-colorable graph.
In the latter case, we let $u_i\in S_1(v_i)$ for each $0\le i\le 4$, and $wv_2u_2u_4u_1u_3$ induce a $P_6$.

{\bf Case 3.} $|S_3|=1$. Let $x\in S_3(v_0)$. We distinguish two cases.

{\bf Case 3.1} $S_1(v_0)=\emptyset$. We claim that $S_2(v_2,v_3)=\emptyset$.
Otherwise let $z\in S_2(v_2,v_3)$.
By (P7), we have $S_1(v_1)=S_1(v_4)=\emptyset$.
Note that $S_2(v_2,v_3)$ is bipartite and is anti-complete to $x$.
Since $\{v_2,v_3\}$ does not separate $S_2(v_2,v_3)$,
one of $S_2(v_3,v_4)$ and $S_2(v_1,v_2)$ is nonempty.
By symmetry, we assume that $p\in S_2(v_3,v_4)$.
By properties (P7) to (P9), we have $S_1=S_1(v_3)$ and $S_2(v_0,v_1)=\emptyset$.
In fact, $S_1(v_3)=\emptyset$ otherwise $\{v_3,w\}$ would separate $S_1(v_3)$.
Moreover, $x$ is anti-complete to $S_2(v_3,v_4)$.
If not, we may assume $xp\in E$ and consider induced $C_5=C'=C\setminus \{v_0\}\cup \{x\}$.
Observe that $w\in S'_5$ and $\{v_0,p\}\subseteq S'_3$, so we are in Case 1.
Going through the same argument in Case 2 we conclude that $w$ is either complete or anti-complete to $S_2$.
In the former case $w$ is a dominating vertex of $G$ and we are done.
Therefore, we assume $w$ is anti-complete to $S_2$. Note also that $2\le |S_2|\le 5$.
In the following we either find a minimal obstruction or show $G$ is 4-colorable.
Consider first that $S_2=S_2(v_4,v_3)\cup S_2(v_3,v_2)$.
If $S_2=\{p,z\}$, $G$ has a 4-coloring $\phi$:
$\{v_4,v_1,z\}$, $\{v_2,v_0,p\}$, $\{x,v_3\}$, $\{w\}$. If there exists $p'\in S_2(v_4,v_3)$ or
$z'\in S_2(v_4,v_3)$, then we can extend $\phi$ by adding $p'$ or $z'$ to $\{w\}$.
So, we assume that $S_2(v_4,v_0)\neq \emptyset$ and let $r\in S_2(v_0,v_4)$.
The fact that $v_2zprv_0x\neq P_6$ implies that $xr\in E$, hence $S_2(v_4,v_0)=\{r\}$ as $G$ is $K_5$-free.
If $S_2(v_4,v_3)=\{p,p'\}$,
then $\{w,x,v_0,v_3,v_4,p,p',r\}$ induces a graph that is not 4-colorable.
Note that $v_4$ is a dominating vertex in this subgraph,
and hence $G$ is the Hajos graph with an additional dominating vertex.
Thus, $S_2(v_4,v_3)=\{p\}$.  Note that $S_2(v_3,v_2)$ might contain a vertex $z'\neq z$ or not.
In either case, $G$ has a 4-coloring: $\{v_4,v_1,z\}$, $\{v_2,v_0,p\}$, $\{x,v_3\}$, $\{w,r,z'\}$.
Finally, we assume that $S_2(v_0,v_4)=\emptyset$
and let $r\in S_2(v_2,v_1)$. If $S_2(v_2,v_3)=\{z,z'\}$, then $S_2(v_4,v_3)=\{p\}$ and  $S_2(v_1,v_2)=\{r\}$
since $G$ is $K_5$-free.
$G$ has a 4-coloring: $\{v_4,v_1,z\}$, $\{v_2,v_0,p\}$, $\{x,v_3,r\}$, $\{w,z'\}$.
Hence, $S_2(v_2,v_3)=\{z,\}$. By $\delta(G)\ge 4$ we have $S_2(v_4,v_3)=\{p,p'\}$ and $S_2(v_1,v_2)=\{r,r'\}$.
In this case $G$ has a 4-coloring: $\{v_4,v_1,z\}$, $\{v_2,v_0,p\}$, $\{x,v_3,r\}$, $\{w,p',r'\}$.

Therefore, $S_2(v_2,v_3)=\emptyset$.
Consider first that $S_2(v_1,v_2)\neq \emptyset$ and $S_2(v_3,v_4)\neq \emptyset$ but $S_1(v_2)=S_1(v_3)=\emptyset$.
Then $S_2=S_2(v_1,v_2)\cup S_2(v_3,v_4)$ by (P7) and the fact that $S_2(v_2,v_3)=\emptyset$.
Since $\{v_3,v_4,w\}$ is not a clique cutset separating $S_2(v_3,v_4)$, $S_2(v_3,v_4)$ has a neighbor in $S_1(v_1)$.
Similarly, $S_2(v_1,v_2)$ has a neighbor in $S_1(v_4)$. However, this contradicts property (P11).
Hence, we must have $S_1(v_2)\neq \emptyset$ and $S_1(v_3)\neq \emptyset$ but $S_2(v_1,v_2)=S_2(v_3,v_4)=\emptyset$.
By (P7), we have $S_2=\emptyset$.
$S_1(v_1)\neq \emptyset$, since $\{v_3,w\}$ is not a clique cutset separating $S_1(v_3)$.
Similarly, $S_1(v_4)\neq \emptyset$. Let $u_i\in S_1(v_i)$ for $i\neq 0$.
Note that $w$ is either complete or anti-complete to $S_1$.
In the former case $w$ is a dominating vertex and we are done.
In the latter case we find an induced $P_6=wv_2u_2u_4u_1u_3$.

{\bf Case 3.2} $S_1(v_0)\neq \emptyset$. Let $y\in S_1(v_0)$.
$S_2(v_1,v_2)=S_2(v_3,v_4)=\emptyset$ by (P7).
Consider first that $S_2(v_2,v_3)=\emptyset$. Since $d(v_2)\ge 4$ and $d(v_3)\ge 4$, we have
$S_1(v_2)\neq \emptyset$ and $S_1(v_3)\neq \emptyset$. By properties (P7) to (P9), the set $S_2=\emptyset$.
Let $p\in S_1(v_3)$ and $q\in S_1(v_2)$.
Consider induced $C_5=C'=v_0v_1v_2qy$.
If $w$ is complete to $S_1$, $w$ is a dominating vertex in $G$ and we are done.
Hence, $w$ is anti-complete to $S_1$.
Note that $S_1(v_0)$ is a clique and thus contains at most two vertices.
Suppose first that $S_1(v_0)=\{y,y'\}$.
If $S_1(v_3)=\{p,p'\}$, then $\{v_4,v_0,v_3,w,x,p,p',y,y'\}$ induces a $G_{P_4}$
with respect to $C_5=v_0y'p'v_3w$.
Thus, $G=G_{P_4}$ but this contradicts that $G$ contains an induced $W_5$.
Hence, $S_1(v_3)=\{p\}$ and $S_1(v_2)=\{q\}$.
As $d(p)\ge 4$ and $d(q)\ge 4$,  both $S_1(v_1)$ and $S_1(v_4)$ are nonempty.
Let $u_i\in S_1(v_i)$ for each $i$, and so $G$ contains an induced $P_6=wv_2u_2u_4u_1u_3$.
Hence, $S_1(v_0)=\{y\}$.
If $|S_1(v_3)|=|S_1(v_2)|=3$, then $G=G_{3,1}$ which is $W_5$-free.
Thus we assume that $|S_1(v_3)|\le 2$. Note that $S_1(v_1)\neq \emptyset$ as $d(p)\ge 4$.
Let $t\in S_1(v_1)$, and so $wv_1tpyq=P_6$.

Therefore, $S_2(v_2,v_3)\neq \emptyset$. Let $z\in S_2(v_2,v_3)$.
As $\{v_2,v_3,w\}$ is not a clique cutset separating $S_2(v_2,v_3)$,
we may assume that $yz\in E$. If $wy\in E$, then the fact that $wyzv_3\neq C_4$ implies that $wz\in E$.
Hence $G$ is the graph $F$ with an additional dominating vertex.
If $wz\in E$, $G$ is the graph $F$ with an additional dominating vertex.
Therefore, $w$ is anti-complete to $\{y,z\}$.
By (P11), we have $S_1=S_1(v_0)$.
Further, $S_2(v_0,v_1)=\emptyset$ otherwise $\{v_0,v_1,x,w\}$ would be a clique cutset.
Similarly, $S_2(v_0,v_4)=\emptyset$. Hence, $S_2=S_2(v_2,v_3)$.
Note that $S_1(v_0)\cup S_2(v_2,v_3)$ contains no induced $C_5$,
since $v_1$ is anti-complete to $S_1(v_0)\cup S_2(v_2,v_3)$.
If $S_1(v_0)\cup S_2(v_2,v_3)$ is not bipartite,
it must contain a triangle, and hence $G=F_1$ or $G=F_2$.
Therefore, we assume that $S_1(v_0)\cup S_2(v_2,v_3)$ is triangle-free and
the edges between $S_1(v_0)$ and $S_2(v_2,v_3)$ form a matching.
As $d(y)\ge 4$ and $d(z)\ge 4$,
$y$ and $z$ have a neighbor $y'\in S_1(v_0)$ and $z'\in S_2(v_2,v_3)$, respectively.
Note $y'z'\notin E$ or $z'y'yz=C_4$. If $w$ is complete to $\{y',z'\}$,
$\{w,y,y',z,z'x,v_0,v_2,v_3\}$ would induce a $G_{3,1}$ with respect to $C_5=wy'yzz'$.
If $wy'\in E$, then $wz'\notin E$ and hence $v_1wy'yzz'=P_6$. Thus, $wy'\notin E$.
Similarly, $wz'\notin E$.
By (P7), the vertex $z$ is universal in $S_2(v_2,v_3)$, and so $z'$ cannot have
a neighbor different from $z$, as otherwise A $K_5$ would arise.
As $d(z')\ge 4$, $z'$ must have a neighbor $y''$ in $S_1(v_0)$. Note that $y''\notin \{y,y'\}$.
Applying the argument for $\{z,y\}$ to $\{z',y''\}$,
we conclude that $w$ is anti-complete to $\{z',y''\}$.
$y''$ is not complete to $\{y,y'\}$ or $K_5$ would arise.
If $y''y\in E$, then $y''yzz'=C_4$. If $y''y'\in E$, then $y''y\notin E$ and thus
$y''y'yzv_3w=P_6$.
As $d(y'')\ge 4$, $y''$ has a neighbor $y'''\in S_1(v_0)$.
$y'''\notin \{y,y',y''\}$.
Moreover, $y'''$ is not complete to $\{y,y'\}$. If $y'''y\in E$, then $y'''y'\notin E$ and thus $y'yy'''y''z'v_2=P_6$.
By symmetry, $y'''y'\notin E$. Now $y'''y''z'zyy'=P_6$. \qed

The following holds under the assumption that $G$ has no induced $W_5$.
\begin{ob}\label{o5}
Let $G$ be a $(P_6,C_4)$-free graph without an induced $W_5$.
Let $C=v_0v_1v_2v_3v_4$ be an induced $C_5$ of $G$. Then the following
properties hold.

(1) If both $S_1(v_{i-1})$ and $S_1(v_{i+1})$ are nonempty then $S_3(v_i)$ is anti-complete to $S_1(v_{i-1})$ and $S_1(v_{i+1})$.

(2) If both $S_2(v_{i-1},v_i)$ and $S_2(v_i,v_{i+1})$ are nonempty,
then $S_3(v_i)$ is complete to $S_2(v_{i-1},v_i)$ and $S_2(v_i,v_{i+1})$.

(3) Let $x\in S_3(v_{i-1})\cup S_3(v_{i+1})$. Suppose that
$pq\in E$ where $p\in S_1(v_i)$ and $q\in S_2(v_{i+2},v_{i+3})$.
Then $x$ is anti-complete to $\{p,q\}$.
\end{ob}

\begin{lemma}\label{without $W_5$}
Suppose that $G$ is a $(P_6,C_4)$-free minimal non-$4$-colorable graph without an induced $W_5$.
Then $G\in \{G_{3,1},G_{2,2},G_{2,1,1},G_{1,1,1,1},H_1,H_2,G_{P_4}\}$ (see Figure \ref{4MO}).
\end{lemma}

We postpone the lengthy proof of this lemma to the Appendix.

\section{The Complexity of $k$-Coloring}

We now apply our results to the questions of complexity of $k$-coloring $(P_6,C_4)$-free
graphs. Reference \cite{Short Cycle} gives a linear time algorithm for
$k$-coloring $(P_t,C_4)$-free graphs for any $k, t$. However, that algorithm depends
on Ramsey-type results, and end up using tree-decompositions with very high widths.
We offer more practical algorithms for $3$-coloring and $4$-coloring $(P_6,C_4)$-free
graphs. Our algorithms are linear time, once a clique cutset decomposition is given.
Moreover, our algorithms are certifying algorithms. Indeed, they are based on our
characterizations of minimal non-$k$-colorable $(P_6,C_4)$-free graphs, and when no
coloring is found, they exhibit a forbidden induced subgraph from Theorems \ref{3-coloring}
and \ref{4-coloring}.


The proof of Theorem \ref{3-coloring} can be easily turned into a linear time algorithm
for $3$-coloring $(P_6,C_4)$-free graphs without clique cutset.
We first test if $G$ is chordal. If so, we can
test whether or  not $G$ is 3-colorable. Otherwise we have an induced $C=C_\ell$ for
some $\ell\ge 4$. Up to this point every step can be done in linear time \cite{Golumbic}.
If $\ell=4$ or $\ell\ge 7$ then $G$ is not $(P_6,C_4)$-free.
If $\ell=5$ we follow the above proof, and it can be readily checked that every step
can be performed in linear time. The remaining case is $\ell=6$, and we can now
assume $G$ is also $C_5$-free. By Lemma \ref{C_6}, either $G$ is specific or $C$
is dominating. In the former case, a $k$-coloring of $G$ or a $K_4$ can be found
in linear time. Therefore, we assume that $C$ is dominating. We define $p$-vertices
and $S_p$ with respect to $C$. We either find that $G$ is not $(P_6,C_4)$-free or the
vertices of $G$ consist of $C\cup S_6\cup S_3$. Finally, in linear time we either find
a $K_4$ or conclude that $G$ has at most $13$ vertices, in which case a $3$-coloring
of $G$ can be obtained by brute force. A similar algorithm applies to the problem of
$4$-coloring $(P_6,C_4)$-free graphs. Thus we have the following result.

\begin{theorem}
There exist linear time certifying algorithms for $3$-coloring and $4$-coloring
$(P_6,C_4)$-free graphs, given a clique cutset decomposition of the input graph.
\end{theorem}

We note that a clique cutset decomposition can be obtained in time $O(mn)$
\cite{Tarjan}.

We now complement our results by proving most of the remaining problems of
$k$-coloring $(P_t,C_\ell)$-free graphs NP-complete (at least as long as $k \ge 3$
and $\ell > 3$).

Recently, Huang \cite{Huang MFCS} proved that the $5$-coloring problem for $P_6$-free
graphs is NP-complete, and that the $4$-coloring problem for $P_7$-free graphs is also
NP-complete. The proof used the following framework. We call a $k$-critical graph {\em nice}
if $G$ contains three independent vertices $\{c_1,c_2,c_3\}$ such that the clique number
$\omega(G-\{c_1,c_2,c_3\})=\omega(G)=k-1$. For example, any odd cycle of
length at least 7 is a nice 3-critical graph.

We give a reduction from 3-SAT, as in \cite{Huang MFCS}.
Let $I$ be any 3-SAT instance with variables $X=\{x_1,x_2,\ldots,x_n\}$
and clauses $\mathcal{C}=\{C_1,C_2,\ldots,C_m\}$,
and let $H$ be a nice $k$-critical graph with three specified independent vertices $\{c_1,c_2,c_3\}$.
We construct a new graph $G_{H,I}$ as follows.

$\bullet$ Introduce for each variable $x_i$ a {\em variable component} $T_i$ which is isomorphic to
$K_2$, labeled by $x_i\bar{x_i}$. Call these vertices {\em $X$-type}.

$\bullet$ Introduce for each variable $x_i$ a vertex $d_i$. Call these vertices {\em $D$-type}.

$\bullet$ Introduce for each clause $C_j=y_{i_1}\vee y_{i_2}\vee y_{i_3}$ a {\em clause component}
$H_j$ which is isomorphic to $H$, where $y_{i_t}$ is either $x_{i_t}$ or $\bar{x_{i_t}}$.
Denote three specified independent vertices in $H_j$
by $c_{i_tj}$ for $t=1,2,3$. Call $c_{i_tj}$ {\em $C$-type} and all remaining vertices {\em $U$-type}.

For any $C$-type vertex $c_{ij}$ we call $x_i$ or $\bar{x_i}$ its {\em corresponding literal vertex},
depending on whether $x_i\in C_j$ or $\bar{x_i}\in C_j$.

$\bullet$ Make each $U$-type vertex adjacent to each $D$-type and $X$-type vertices.

$\bullet$ Make each $C$-type vertex $c_{ij}$ adjacent to $d_i$ and its corresponding literal vertex.

We refer to \cite{Huang MFCS} for the proofs of the following two lemmas.
\begin{lemma}
Let $H$ be a nice $k$-critical graph.
Suppose $G_{H,I}$ is the graph constructed from $H$ and a 3-SAT instance $I$.
Then $I$ is satisfiable if and only if $G_{H,I}$ is $(k+1)$-colorable.
\end{lemma}

\begin{lemma}
Let $H$ be a nice $k$-critical graph.
Suppose $G_{H,I}$ is the graph constructed from $H$ and a 3-SAT instance $I$.
If $H$ is $P_t$-free where $t\ge 6$, then $G_{H,I}$ is $P_t$-free as well.
\end{lemma}

To obtain NP-completeness results for $(P_t,C_{\ell})$-free graphs,
we need an additional lemma.
\begin{lemma}
Let $\ell\ge 6$. If $H$ is $C_\ell$-free, then $G_{H,I}$ is $C_\ell$-free.
\end{lemma}
\noindent {\bf Proof.} Let $Q=v_1\ldots v_\ell$ be an induced $C_\ell$ in $G_{H,I}$.
Let $C_i$ (respectively $\bar{C_i}$) be the set of $C$-type vertices that connect to $x_i$ (respectively $\bar{x_i}$).
Let $G_i=G[\{T_i\cup\{d_i\}\cup C_i\cup \bar{C_i}\}]$.
Note that $G-U$ is disjoint union of $G_i$, $i=1,2,\ldots,n$.
If $Q\cap U=\emptyset$, then $Q\subseteq G_i$ for some $i$.
It is easy to see that $G_i$ is $C_\ell$-free as $\ell\ge 6$.
Thus, $Q\cap U\neq \emptyset$. Without loss of generality,
we assume that $v_1$ is a $U$-type vertex
where $v_1$ is in the $j$th clause component $H_j$.
If $v_2$ and $v_\ell$ are both in $H_j$, then $Q\subseteq H_j$, which contradicts our assumption that
$H_j=H$ is $C_\ell$-free.
If $v_2$ and $v_\ell$ are both in $X\cup D$,
then as $U$-type vertices are complete to $X$-type and $D$-type vertices,
all other vertices on $Q$ are of $C$-type. This is impossible since $C$ is independent.
The last case is $v_\ell$ is in $H_j$ and $v_2$ is in $X\cup D$. Similar to the second case, we have
$v_4,v_5,\ldots v_{l-1}$ are $C$-type vertices. This contradicts that $v_4v_5$ is an edge. \qed

The following theorem follows now directly from the above lemmas.
\begin{theorem}\label{main}
Let $\ell\ge 6$.
Then $k$-coloring is NP-complete for $(P_t,C_{\ell})$-free graphs whenever there
exists a $(P_t,C_{\ell})$-free nice $(k-1)$-critical graph.
\end{theorem}

We apply Theorem \ref{main} to derive a series of hardness results on $(P_t,C_{\ell})$-free
graphs for various values of $k$ and $t$.
\begin{figure}[htbp]
\centering
\includegraphics[width=0.4\textwidth]{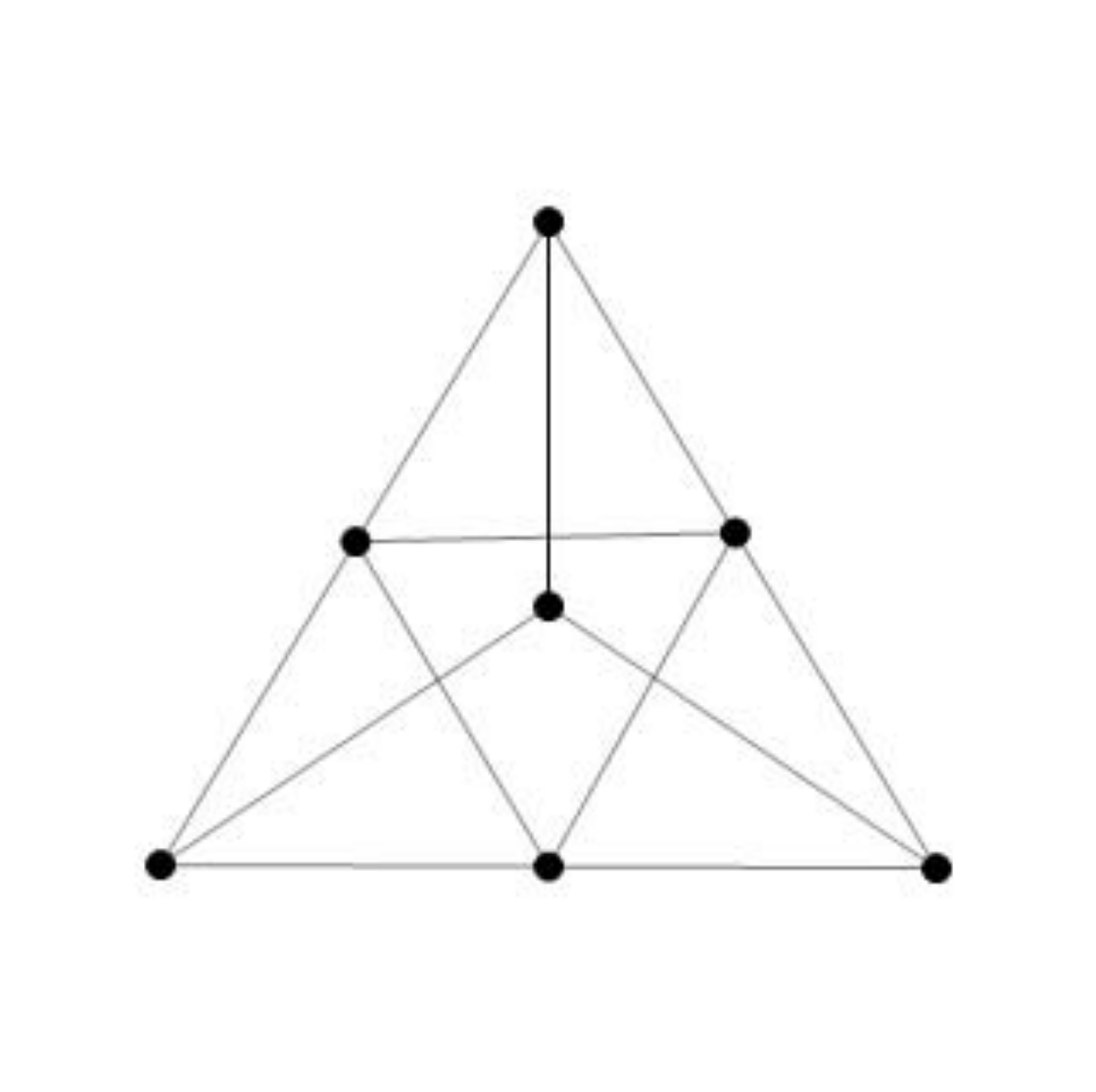}
\vspace*{-10pt}
\caption{$G_1$.}
\label{G1}
\end{figure}
\begin{theorem}
Let $k\ge 5$, $t\ge 6$ and $\ell\ge 6$ be fixed integers.
Then $k$-coloring is NP-complete for $(P_t,C_{\ell})$-free graphs.
\end{theorem}
\noindent {\bf Proof.}
It is easy to check that the graph $G_1$ shown in Figure \ref{G1}
is a nice $4$-critical $(P_6,C_{\ell})$-free graph for any fixed $\ell\ge 6$.
Applying Theorem \ref{main} with $G_1$ will complete our proof. \qed

\begin{theorem}
$4$-coloring is NP-complete for $(P_t,C_{\ell})$-free graphs when $t\ge 7$ and $\ell\ge 6$ with $\ell\neq 7$;
and 4-coloring is NP-complete for $(P_t,C_{\ell})$-free graphs when $t\ge 9$ and $\ell\ge 6$ with $\ell \neq 9$.
\end{theorem}
\noindent {\bf Proof.}
It is easy to check that $C_7$ is a nice $3$-critical $(P_t,C_{\ell})$-free graph for any
$t\ge 7$ and $\ell\ge 6$ except $\ell=7$,
and that $C_9$ is a nice $3$-critical $(P_t,C_{\ell})$-free graph for any $t\ge 9$ and $\ell\ge 6$ except $\ell=9$.
Applying Theorem \ref{main} with $C_7$ and $C_9$ will complete the proof. \qed

We shall use a different
reduction to prove the next result.

\begin{theorem}\label{C5}
$4$-coloring is NP-complete for $(P_7,C_5)$-free graphs.
\end{theorem}

\noindent {\bf Proof.} We reduce NOT-ALL-EQUAL 3-SATISFIABILITY with positive literals
only (NAE 3-SAT PL for short) to our problem.  The NAE 3-SAT PL is NP-complete \cite{NAE}
and is defined as follows. Given
a set $X=\{x_1,x_2,\ldots,x_n\}$ of logical variables,
and a set $\mathcal{C}=\{C_1,C_2,\ldots, C_m\}$ of three-literal clauses over $X$
in which all literals are positive, does there exist
a truth assignment for $X$ such that each clause contains at least one true
literal and at least one false literal?
Given an instance $I$ of NAE 3-SAT PL we construct a graph $G_I$ as follows.

$\bullet$ For each variable $x_i$ we introduce a single vertex named as $x_i$. Call these vertices $X$-type.

$\bullet$ For each variable $x_i$ we introduce a "truth assignment" component $F_i$
where $F_i$ is isomorphic to $P_4$ whose vertices are labeled by $d_ie'_ie_id'_i$.

$\bullet$ For each clause $C_j=x_{i_1}\vee x_{i_2}\vee x_{i_3}$
we introduce two copies of $C_7$ denoted by $H_j$ and $H'_j$.
Choose three independent vertices of $H_j$ and name them as $c_{i_1j}$, $c_{i_2j}$ and $c_{i_3j}$.
Choose three independent vertices of $H'_j$ and name them as $c'_{i_1j}$, $c'_{i_2j}$ and $c'_{i_3j}$.
Call these vertices $C$-type and $C'$-type, respectively.
The remaining vertices in clause components are said to be of $U$-type.

$\bullet$ Make each $U$-type vertex adjacent to each $X$-type vertex and each vertex in $F_i$ for $1\le i\le n$.

$\bullet$ Make each $C$-type vertex $c_{ij}$ adjacent to $x_i$ and $d_i$
and make each $C'$-type vertex $c'_{ij}$ adjacent to $x_i$ and $d'_i$.


This completes the construction of $G_I$.
It is easy to see that $d_i$ and $d'_i$ have no common neighbor
in $G-U$ and same for $e_i$ and $e'_i$.

\noindent {\bf Claim 1.} {\em The instance $I$ is satisfiable if and only if $G_I$ is 4-colorable.}

\noindent {\bf Proof.} Suppose first that $G_I$ is 4-colorable and $\phi$ is a 4-coloring of $G_I$.
Without loss of generality, we may assume that
the two adjacent $U$-type vertices in $H_1$ receive color 1 and 2, respectively.
Now as $U$ is complete to $X\cup F$,
it follows that each $x_i$ and each vertex in $F_i$ receives color 3 or 4. Further, $\phi(d_i)\neq \phi(d'_i)$
for each $i$. We define a truth assignment as follows.

$\bullet$ We set $x_i$ to be TRUE if $\phi(x_i)=\phi(d_i)$ and to be FALSE if $\phi(x_i)\neq \phi(d_i)$.

We show that every clause $C_j$ contains at least one true literal and one false literal.
Suppose $x_{i_1}$, $x_{i_2}$, and $x_{i_3}$ are all TRUE.
Then it implies that $\phi(d'_{i_j})\neq \phi(x_{i_j})$ for all $j=1,2,3$.
As a result, $c'_{i_j}$ must be colored with color 1 or 2 under $\phi$.
Moreover, all $U$-type vertices in $H'_j$ are colored with 1 or 2 under $\phi$.
This contradictions the fact that $H'_j=C_7$ is not 2-colorable.
If $x_{i_1}$, $x_{i_2}$, and $x_{i_3}$ are all FALSE we would reach a similar contradiction.
Conversely, suppose that every clause $C_j$ contains at least one true literal and one false literal.
We define a 4-coloring $\phi$ as follows.

$\bullet$ Set $\phi(x_i)=3$ if $x_i$ is TRUE and $\phi(x_i)=4$ if $x_i$ is FALSE.

$\bullet$ We color vertices in $F_i$ alternately with color 3 and 4 starting from setting $\phi(d_i)=3$.

$\bullet$ Let $C_j=x_{i_1}\vee x_{i_2}\vee x_{i_3}$ be a clause.
Without loss of generality, we may assume that $x_{i_1}$ is TRUE and $x_{i_2}$ is FALSE.
It follows from the definition of $\phi$ that $\phi(x_{i_1})=\phi(d_{i_1})=3$.
Hence, we can color $c_{i_1j}$ with color 4, so that $H_j-c_{i_1j}$ can be colored with colors 1 and 2.
Similarly, we can 4-color $H'_j$. \qed

\noindent {\bf Claim 2.} {\em $G_I$ is $C_5$-free.}

\noindent {\bf Proof.} Let $Q=v_1\ldots v_5$ be an induced $C_5$ in $G_I$.
Let $C_i$ (respectively $C'_i$) be the set of $C$-type (respectively $C'$-type) vertices that are adjacent to $x_i$.
Let $G_i$ be the subgraph of $G_I$ induced by $\{x_i\}\cup C_i\cup C'_i\cup F_i$.
Note that $G-U$ is disjoint union of $G_i$.
Suppose first that $Q\cap U=\emptyset$. Note that both $e_i$ and $e'_i$ have degree 2 in $G_i$.
If $Q$ contains $e_i$ or $e'_i$, then $Q$ contains $F_i$ as an induced subgraph and thus the
fifth vertex of $Q$ would be a common neighbor of $d_i$ and $d'_i$, a contradiction.
So $Q\cap \{e_i,e'_i\}=\emptyset$. If $Q\cap \{d_i,d'_i\}=\emptyset$,
then $Q$ is a star which is impossible. Without loss of generality, we assume that $d_i\in Q$. If $d'_i$ is also
in $Q$, then there would be a common neighbor of $d_i$ and $d'_i$. So $d'_i\notin Q$.
Then the two neighbors of $d_i$ on $Q$ must be of $C$ or $C'$-type,
and so the other two vertices have to be of $X$-type, which is not possible.
Hence, $Q\cap U\neq \emptyset$. Suppose $v_1$ is of $U$-type and from $H_j$.
If both $v_2$ and $v_5$ are of $X$-type or $F$-type, then $v_3$ and $v_4$ have to be $C$ or $C'$-type.
But this is a contradiction as $C\cup C'$ is independent. If both $v_2$ and $v_5$ are in $H_j$, then
$Q\subseteq H_j$, which is impossible as $H_j=C_7$ is $C_5$-free. So we assume that $v_2$ is of
$X$-type or $F$-type and $v_5$ is in $H_j$. Then $v_5$ must be $C$ or $C'$-type. Moreover, $v_4$ must
be of $C$ or $C'$-type as it is not adjacent to $v_1$ or $v_2$.
This is impossible since $v_4v_5$ is an edge. \qed

\noindent {\bf Claim 3.} {\em $G_I$ is $P_7$-free.}

\noindent {\bf Proof.}
Let $P$ be an induced $P_7$ in $G_I$. We first consider the case $P\cap U\neq \emptyset$.
Let $u\in P$ be an $U$-type vertex and $u$ is in some clause component $H_j$.
For any vertex $x$ on $P$ we denote by $x^-$ and $x^+$ the left and right neighbor of $x$ on $P$, respectively.
Suppose first that $u$ is an endvertex of $P$. Then $u^+$ is in $X\cup F$ or $H_j$.
If $u^+$ is in $H_j$, then $P\subseteq H_j$, which is a contradiction since $H_j=C_7$ is $P_7$-free.
So $u^+$ is in $X\cup F$. If $u^{++}$ is in $C\cup C'$, then $|P|=3$, a contradiction.
So $u^{++}$ is in $U$. But now $u^{+++}$ must be in $C\cup C'$, and thus $|P|=4$, a contradiction.
Hence, $u$ must have degree 2 on $P$. If $u^-$ and $u^+$ are both in $H_j$, then $P\subseteq H_j$, a contradiction.
If $u^-$ and $u^+$ are both in $X\cup F$, then $u^{--}$ and $u^{++}$ are both of $C$- or $C'$-type.
Hence, $|P|\le 5$ since $C\cup C'$ is independent.
So we may assume that $u^+$ is in $H_j$ and $u^-$ is in $X\cup F$.
Now $u^+$ must be of $C$- or $C'$-type and hence an endvertex of $P$.
Therefore, $|P|\le 2+4-1=5$.

We have shown that $P\cap U=\emptyset$. So $P\subseteq G_i$ for some $i$.
Now we show that $|P\cap C_i|=1$. Otherwise assume that $|P\cap C_i|=2$.
Let $c_1$ and $c_2$ be the vertices in $P\cap C_i$.
If $c_1$ and $c_2$ are not at distance 2 on $P$,
then $x_i$ and $d_i$ are not on $P$ otherwise $P$ would not be induced. However,
$x_i$ and $d_i$ are the only neighbors of $C$-type vertices in $G_i$, a contradiction.
So, $c_1$ and $c_2$ must be at distance 2 on $P$.
If they are connected by $d_i$, then $x_i\notin P$ and vice versa.
But now $|P|=3$, since $C_i\cup C'_i$ is independent.
Therefore, $|P\cap C|\le 1$ and similarly $|P\cap C'|\le 1$.
So, we must have $F_i\cup \{x_i\}\subseteq P$, and thus
$P=C_7$, a contradiction.
\qed

The following result is a direct corollary of Theorem \ref{C5}.
\begin{theorem}
Let $k\ge 4$ and $t\ge 7$. Then $k$-coloring is NP-complete for $(P_t,C_5)$-free graphs.
\end{theorem}

\section{Conclusions}

We have undertaken a first systematic study of the $k$-coloring problem for graphs without
an induced cycle $C_\ell$ and an induced path $P_t$. We have shown that while for many
values of $k$, $t$ and $\ell$ these problems are NP-complete, the case of $(P_6,C_4)$-free
graphs offers much structure to be exploited. In particular, we have shown that there are for
each $k$ only finitely many non-$k$-colorable $(P_6,C_4)$-free graphs.

For $k=3$ and $k=4$, we were able to describe these minimal obstructions explicitely, and
so obtained certifying polynomial time (linear time if a clique cutset decomposition is given)
algorithms for coloring $(P_6,C_4)$-free graphs.
However, for larger $k$, we do not know certifying $k$-coloring algorithms for $(P_6,C_4)$-free graphs.

Our hardness results come close to classifying the complexity all cases of $k$-coloring for
$(P_t,C_{\ell})$-free graphs. There seem to be two stubborn cases about which not much can
be said with the current tools, when $k=3$ or $\ell=3$. (But note \cite{P_7 1,P_7 2}.) Beyond
these cases, our results leave only the following remaining open problems.
\begin{problem}\label{p1}
What is the complexity of $k$-coloring $(P_6,C_5)$-free graphs for $k\ge 4$?
\end{problem}
\begin{problem}\label{p3}
What is the complexity of $4$-coloring $(P_6,C_6)$-free graphs?
\end{problem}
\begin{problem}
What is the complexity of $4$-coloring $(P_t,C_7)$-free graphs for $t=7$ and $t=8$?
\end{problem}
In \cite{Huang MFCS} Huang conjectured that $4$-coloring is polynomial time solvable for
$P_6$-free graphs. If the problems in Problem \ref{p1} for $k=4$ or Problem \ref{p3} are polynomial,
this would add evidence to the conjecture.

We are grateful to Daniel Paulusma for very valuable advice and suggestions.

\newpage

\section*{Appendix}

\noindent {\bf Proof of Lemma \ref{without $W_5$}.}
By Lemma \ref{with W_5}, we may assume that no induced $C_5$ has a 5-vertex.
Let $C=v_0\ldots v_4$ be an induced $C_5$ such that $|S_3|$ is as small as possible.
As the graph $G_{3,1}$ is a minimal obstruction, we obtain that $|S_3|\le 7$.
Suppose first $|S_3|=0$. It is easy to check that
either $G$ contains a $K_5$ or is 4-colorable if $S_1=\emptyset$. Hence, we may assume that $S_1(v_0)\neq \emptyset$.
Going through the same argument as in Case 2 of Lemma \ref{with W_5}, we conclude that each $S_1(v_i)\neq \emptyset$
for each $i$.  If two $S_1(v_i)$ have size at least 3, then $G$ either contains $K_5$ or $G_{3,1}$.
Now suppose that $|S_1(v_0)|=3$. Thus $|S_1(v_2)|=|S_1(v_3)|=1$ or $K_5$ arises. If $|S_1(v_1)|=|S_1(v_4)|=2$, then
$G=G_{2,2}$. Otherwise one of $S_1(v_1)$ and $S_1(v_4)$ has size 1 in which case it is easy to check $G$ is 4-colorable.
Now we assume that each $|S_1(v_i)|\le 2$. If all but one $S_1(v_i)$ have size 2, then $G=G_{P_4}$. Otherwise,
there are at least two $|S_1(v_i)|=1$. It is easy to check $G$ is 4-colorable. Therefore, $1\le |S_3|\le 7$.

{\bf Case 1.} $|S_3|=1$. Let $x\in S_3(v_0)$. Suppose first that $S_1=\emptyset$.
If $S_2(v_2,v_3)=\emptyset$, then both $S_2=S_2(v_1,v_2)$ and $S_2(v_3,v_4)$ have at least
two vertices as $d(v_2)\ge 4$ and $d(v_3)\ge 4$.
By (P8), $S_2=S_2(v_1,v_2)\cup S_2(v_3,v_4)$.
As $d(x)\ge 3$, we have that $x$ is not anti-complete to $S_2$ and hence complete to $S_2$ by (P10).
Now $G$ contains $G_{3,1}$ as an induced subgraph and so $G=G_{3,1}$.
Thus $S_2(v_2,v_3)\neq \emptyset$. By (P8), one of $S_2(v_0,v_1)$ and $S_2(v_0,v_4)$
is empty, say $S_2(v_0,v_1)$. As $d(v_1)\ge 4$ we have $S_2(v_1,v_2)\neq \emptyset$ and thus
$S_2(v_0,v_4)=\emptyset$. As $d(v_4)\ge 4$ we have $S_2(v_3,v_4)\neq \emptyset$. By (P10),
$x$ must be anti-complete to $S_2$. But now $d(x)=3$ contradicting $\delta(G)\ge 4$.

Therefore, $S_1\neq \emptyset$. Suppose first that $S_1(v_0)\neq \emptyset$.
Going through the same argument as in Case 2 of Lemma \ref{with W_5}
we conclude that $S_1(v_i)\neq \emptyset$ for each $i$ and $S_2=\emptyset$.
It is easy to check that either $G\in \{G_{3,1},G_{2,2},G_{P_4}\}$ or $G$ is 4-colorable.
So, $S_1(v_0)=\emptyset$. We first show that $S_1(v_2)$ is anti-complete to
$S_2(v_0,v_4)$. If not, let $x\in S_1(v_2)$ be adjacent to $y\in S_2(v_0,v_4)$.
By (P11), $S_1=S_1(v_2)$. Further, $S_2(v_0,v_1)=S_2(v_4,v_3)=\emptyset$
, and one of $S_2(v_1,v_2)$ and $S_2(v_3,v_2)$ is empty by properties (P7) to (P9).
As $\delta(G)\ge 4$ there are at least two 3-vertices adjacent to $v_1$ or $v_3$.
This is impossible as $|S_3|=1$. Now if $S_1(v_2)\neq \emptyset$, then $S_1(v_4)\neq \emptyset$ as
$v_2$ does not separate $S_1(v_2)$ and by (P9) we have $S_2=\emptyset$.
As $d(v_3)\ge 4$ $S_1(v_3)\neq \emptyset$ and hence $S_1(v_1)\neq \emptyset$. Now by Observation
\ref{o5}, we have $x$ is anti-complete to $S_1$, contradicting $\delta(G)\ge 4$.
Therefore, $S_1(v_2)=\emptyset$. Similarly, $S_1(v_3)=\emptyset$.
Now we may assume that $S_1(v_1)\neq \emptyset$. Then $S_2(v_2,v_3)=\emptyset$.
As $\delta(G)\ge 4$, both $S_2(v_1,v_2)$ and $S_2(v_3,v_4)$ have at least two vertices
and so $S_2=S_2(v_1,v_2)\cup S_2(v_3,v_4)$.
By (P10), $x$ must be anti-complete to $S_2$ or $G=G_{3,1}$.
By Observation \ref{o5} and $d(x)\ge 4$ we have $S_1(v_4)=\emptyset$.
But now $\{v_1,x\}$ is a clique cutset separating $S_1(v_1)$.

{\bf Case 2.} $|S_3|=2$. We distinguish two subcases.

{\bf Case 2.1} There exists some $i$ such that $S_1(v_i)\neq \emptyset$ and $S_1(v_{i+1})\neq \emptyset$.
Without loss of generality, assume that $x\in S_1(v_0)$ and $y\in S_1(v_1)$.
By (P9), $S_2=S_2(v_0,v_1)$.
Suppose first that $S_2(v_0,v_1)\neq \emptyset$. Then $S_1(v_2)=S_1(v_4)=\emptyset$.
As $d(v_2)\ge 4$ and $d(v_4)\ge 4$ we have $|S_3(v_3)|=2$.
Note that $S_1(v_3)$ is a clique since $S_1(v_0)\neq \emptyset$ and is complete to $S_3(v_3)$.
So, $|S_1(v_3)|\le 1$ or $K_5$ would arise. Note that $S_2(v_0,v_1)$ is bipartite. If $S_1(v_3)=\emptyset$,
then it is easy to check that $G$ is 4-colorable.
So we assume that $S_1(v_3)=\{w\}$.
If $w$ has two neighbors in $S_2(v_0,v_1)$, then $G=G_{2,1,1}$. Thus $w$ has at most one neighbor in $S_2(v_0,v_1)$.
If $|S_1(v_0)|=3$ or $|S_1(v_1)|=3$, then $G=G_{3,1}$. Thus $|S_1(v_i)|\le 2$ for $i=0,1$. Now it is easy to check
that $G$ is 4-colorable.
Therefore, $S_2(v_0,v_1)=\emptyset$ and thus $S_2=\emptyset$. We consider two subcases.

{\bf Case 2.1.a} There exists some $i$ such that $|S_3(v_i)|=2$. Suppose that $i=0$ (or $i=1$).
As $d(v_2)\ge 4$ and $d(v_3)\ge 4$, we have that $|S_1(v_i)|\ge 2$ for $i=2,3$. By (P12),
$x$ is complete to $S_3(v_0)$ and hence $S_1(v_0)=\{x\}$. If $|S_1(v_1)|\ge 2$, then $G=G_{2,2}$.
So assume that $S_1(v_1)=\{y\}$. If $|S_1(v_i)|=3$ for some $i\in \{2,3\}$, then $G=G_{3,1}$. If $|S_1(v_4)|\ge 2$
then $G=G_{2,2}$. Hence, $|S_1(v_3)|=|S_1(v_2)|=2$ and $|S_1(v_4)|\le 1$.
Let $u_i\in S_1(v_i)$ for nonempty $S_1(v_i)$ and let $u'_i\in S_1(v_i)$ with $u'_i\neq u_i$ for $i=2,3$.
Let $S_3(v_0)=\{z,z'\}$.
If $u_4$ exists, then $S_3(v_0)$ is anti-complete to $S_1\setminus \{x\}$ by Observation \ref{o5}.
Thus $G$ has a 4-coloring: $\{v_1,v_4,x\}$, $\{v_0,v_2,u_4,u_3\}$, $\{v_3,y,u_2,z\}$, $\{u'_2,u'_3,z'\}$.
If $u_4$ does not exist, then $y$ may or may not be adjacent to $S_3$. In either case,
$G$ has a 4-coloring:
$\{v_1,v_4,x\}$,  $\{v_0,v_3,u_2,y\}$,  $\{v_2,u_3,z\}$,  $\{z',u'_2,u'_3\}$.

Now suppose that $i=4$ (or $i=2$). As $d(v_i)\ge 4$ we have $|S_1(v_i)|\ge 2$ for $i=1,2$.
We may assume that $S_1(v_4)=\emptyset$ or we are in the case $i=0$.
Since $\{v_1\}$ does not separate $S_1(v_1)$, $S_1(v_3)\neq \emptyset$.
By Observation \ref{o5}, $S_3$ is anti-complete to $S_1$.
Note that $|S_1(v_3)|+|S_1(v_0)|\le 3$ otherwise $G=G_{3,1}$ or $G=G_{2,2}$.
Let $u_i,u'_i\in S_1(v_i)$ for $i=2,3$ and $S_3(v_4)=\{z,z'\}$.
If each $S_1(v_i)$ has size less than 3, then $G$ has a 4-coloring:
$\{v_0,v_3,y\}$, $\{v_2,v_4,u'_1,x\}$, $\{v_1,u_2,u_3,z\}$, $\{u'_2,u'_3,z'\}$.
So assume without loss of generality that $|S_1(v_2)|=3$ and hence $S_1(v_0)=\{x\}$.
It is easy to check $G$ is also 4-colorable.

Finally, suppose that $i=3$. As $d(v_i)\ge 4$ we have $|S_1(v_i)|\ge 2$ for $i=0,1$. If $S_1(v_3)=\emptyset$,
then as $G$ has no clique cutset, $S_1(v_j)\neq \emptyset$ for $j=1,4$, and we are in the case $i=4$.
So $S_3(v_1)=\{z\}$. Note that $|S_1(v_0)|=|S_1(v_1)|=2$ or $G=G_{3,1}$.
Moreover, each of $S_1(v_1)$ and $S_1(v_4)$ has size at most 1 or $G=G_{2,2}$.
Now it is easy to check that $G$ is 4-colorable.

{\bf Case 2.1.b} Each $S_3(v_i)$ has at most one vertex. Let $N$ be the set of $v_i$ such that
$S_3(v_i)\neq \emptyset$. Then there are six possible cases.

Suppose first that $N=\{v_0,v_1\}$.
Let $t\in S_3(v_0)$ and $r\in S_3(v_1)$.
Since $xtv_4v_3v_2r\neq P_6$, we have $rt\in E$ or $rx\in E$.
Similarly, the fact that $yrv_2v_3v_4t\neq P_6$ implies that $rt\in E$ or $yt\in E$.
If $rt\notin E$, then $xr$ and $yt$ are edges and so $txry=C_4$. Hence, $rt\in E$.
As $d(v_3)\ge 4$, $|S_1(v_3)|\ge 2$. Similarly,
both $S_1(v_2)$ and $S_1(v_4)$ are nonempty. By Observation \ref{o5}, $t$ (respectively $r$)
is anti-complete to $S_1\setminus S_1(v_0)$ (respectively $S_1\setminus S_1(v_1)$).
Note that $|S_1(v_3)|=2$ or $G=G_{3,1}$.
If $S_1(v_0)$ has two vertices, then $\{v_0,v_1,t,r,y\}\cup S_1(v_0)\cup S_1(v_3)$
induces a $G_{P_4}$ with respect to $tv_1u_1v_3v_0$ and $u'_3u'_0v_0r$ where $u_i,u'_i\in S_1(v_i)$ for each $i$.
Hence, $S_1(v_0)=\{x\}$. Similarly, $S_1(v_1)=\{y\}$. If $|S_1(v_2)|=3$, then $G=G_{2,2}$.
So $|S_1(v_i)|\le 2$ for $i=2,4$. Now it is easy to check $G$ is 4-colorable.

Now suppose that $N=\{v_1,v_2\}$. Let $t\in S_3(v_2)$ and $r\in S_3(v_1)$. As $\delta(G)\ge 4$
we have $|S_1(v_4)|\ge 2$ and $|S_1(v_3)|\ge 1$. By Observation \ref{o5}, we have $ty\notin E$.
Since $tv_3v_4wyr\neq P_6$, we have $rt\in E$, where $w\in S_1(v_4)$.
We may assume that $S_1(v_2)=\emptyset$
or we are in the case $N=\{v_0,v_1\}$. Note that $r$ is anti-complete to $S_1(v_0)$
as $S_1(v_3)\neq \emptyset$ and $G$ is $C_4$-free. Since $d(x)\ge 4$ we have $|S_1(v_0)|+|S_1(v_3)|=4$.
If $|S_1(v_4)|=3$, then $|S_1(v_1)|=1$. Also, $|S_1(v_3)|\le 2$ or $G=G_{3,1}$. Now $G$ is 4-colorable.
So, $|S_1(v_4)|=2$. As $\delta(G)\ge 4$ we have  $|S_1(v_1)|=2$.
If $|S_1(v_3)|\ge 2$ then $\{v_1,v_2,v_3,u'_3,u'_1,u_3,u_1,r,t\}$ induces a $G_{P_4}$
with respect to $v_1v_2v_3u'_3u'_1$ and $u_3u_1rt$ where $u_i,u'_i\in S_1(v_i)$.
So, $|S_1(v_3)|=1$ and then $|S_1(v_0)|=3$. Now $S_1(v_0)\cup \{v_0,u_3,r,v_1,u_1,u'_1\}$
induces a $G_{2,2}$ with respect to induced $K_5-e=S_1(v_0)\cup \{v_0,u_3\}$ and $K_4=rv_1u_1u'_1$.
This completes the proof of $N=\{v_1,v_2\}$.

Let $N=\{v_2,v_3\}$. As $\delta(G)\ge 4$, both $S_1(v_1)$ and $S_1(v_4)$ are nonempty, and
$S_1(v_0)$ has at least two vertices. If one of $S_1(v_2)$ and $S_1(v_3)$ is nonempty,
we are in one of previous cases. But now $\{v_0\}$ is a clique cutset separating $S_1(v_0)$.

Let $N=\{v_0,v_3\}$ and let $r\in S_3(v_0)$, $t\in S_3(v_3)$. As $\delta(G)\ge 4$ we have $S_1(v_i)\neq \emptyset$
for $i\neq 4$. Let $u_i\in S_1(v_i)$. By $G$ is $C_4$-free, we have $r$ (respectively $t$)
is anti-complete to $S_1(v_1)$ (respectively $S_1(v_2)$).
Then as $d(u_1)\ge 4$ and $d(u_2)\ge 4$, we have $|S_1(v_1)|+|S_1(v_3)|=4$ and $|S_1(v_2)|+|S_1(v_0)|=4$.
If $|S_1(v_0)|=3$, then $|S_1(v_2)|=|S_1(v_2)|=1$, and so $|S_1(v_1)|=3$. Now $G=G_{3,1}$.
So each $S_1(v_i)$ has size 2. But now $\{r,v_4,t,v_3,v_0\}\cup S_1(v_0)\cup S_1(v_1)$ induces a $G_{P_4}$.

Let $N=\{v_0,v_2\}$. As in the case where $N=\{v_0,v_3\}$, we obtain that each $S_1(v_i)\neq \emptyset$.
Moreover, each $S_1(v_i)$ has size 2 except $S_1(v_1)$. Hence, $G=G_{P_4}$.

The case $N=\{v_2,v_4\}$ is the same as $N=\{v_0,v_3\}$. This completes the proof of Case 2.1.

{\bf Case 2.2} One of $S_1(v_i)$ and $S_1(v_{i+1})$ is empty for each $i$.
Hence, there are at most two nonempty
$S_1(v_i)$. We consider following three cases.

Suppose first that there are exactly two $S_1(v_i)$ that are nonempty.
Without loss of generality, we assume that $S_1(v_0)$ and $S_1(v_2)$ are nonempty.
By (P9), we have $S_2=\emptyset$. As $d(v_1)\ge 4$ and $d(v_4)\ge 4$,
we have $|S_3(v_0)|=2$ but this contradicts $d(v_3)\ge 4$.

Now we suppose that $S_1(v_0)\neq \emptyset$ while $S_1(v_i)=\emptyset$ for $i\neq 0$.
Let $x\in S_1(v_0)$. Note that $S_2(v_2,v_1)=S_2(v_3,v_4)=\emptyset$.

We first claim that $S_3$ is not anti-complete to $S_1$.
If not, $x$ has a neighbor $y\in S_2(v_2,v_3)$ or $\{v_0\}$ would be a clique cutset.
Further, one of $S_2(v_0,v_4)$ and $S_2(v_0,v_1)$ is empty, say $S_2(v_0,v_4)$.
Since $d(v_4)\ge 4$, we have $S_3(v_1)=S_3(v_2)=\emptyset$. Also, $S_3(v_0)=\emptyset$ by our assumption.
If $S_3(v_4)\neq \emptyset$,  then $|S_2(v_0,v_1)|\ge 1$ since $d(v_1)\ge 4$.
Since $\{v_0,v_1\}$ does not separate $S_2(v_0,v_1)$,
$S_2(v_0,v_1)$ has a neighbor $t$ in $S_3(v_4)$ and hence $ty\in E$ by the property (P10).
Now $tyxv_0=C_4$. So it must be the case that $|S_3(v_3)|=2$. Then $|S_2(v_0,v_1)|\ge 1$ since $d(v_1)\ge 4$
and so $\{v_0,v_1\}$ is a clique cutset separating $S_2(v_0,v_1)$.

Hence, $S_3$ is not anti-complete to $S_1$, and thus $|S_3(v_3)\cup S_3(v_2)|\le 1$.
By $d(v_2)\ge 4$ and $d(v_3)\ge 4$ we have that $S_2(v_2,v_3)\neq \emptyset$.
We first consider the case that $S_1(v_0)$ is anti-complete to $S_2$.
If $S_3(v_3)\neq \emptyset$, then $G$ would have a clique cutset separating $S_1(v_0)$. So,
$S_3(v_3)=S_3(v_2)=\emptyset$. Further, there is no $S_3(v_i)$ having size 2 or $G$
would have a clique cutset. Let $S_3=\{r,t\}$. If $r\in S_3(v_4)$ and $t\in S_3(v_0)$,
then $rt\notin E$ or clique cutset would arise. Thus, $S_2(v_0,v_1)\neq \emptyset$ as $d(v_1)\ge 4$.
As $\{v_2,v_3\}$ is not a clique cutset, $r$ is not anti-complete to $S_2(v_2,v_3)$ and hence
complete to $S_2(v_2,v_3)$ and $S_2(v_0,v_1)$. As $d(v_2)\ge 2$, $|S_2(v_2,v_3)|\ge 2$ and thus
$S_2(v_0,v_1)=\{q\}$. As $d(q)\ge 4$ we have $qt\in E$ and thus $rqtv_4=C_4$.
By symmetry, it is impossible for $r\in S_3(v_0)$ and $t\in S_3(v_1)$.
Finally, it is impossible for $S_3(v_4)$ and $S_3(v_1)$ to be nonempty
by properties (P7) to (P9) and $\delta(G)\ge 4$.
Therefore, we may assume that $x$ has a neighbor $y$ in $S_2(v_2,v_3)$.
Without loss of generality, we assume that $S_3(v_2)=\emptyset$.
Next we distinguish two cases by properties (P7) to (P9).

{\bf (I)} $S_2(v_0,v_1)=\emptyset$. Then $S_3=S_3(v_0)\cup S_3(v_1)$ by $d(v_1)\ge 4$.
If $|S_3(v_0)|=2$, then $|S_2(v_2,v_3)|\ge 2$ by $d(v_3)\ge 4$. If $x$ has a different
neighbor $y'$ in $S_2(v_2,v_3)$, then $G=G_{2,1,1}$. If there is an edge other than $xy$ between
$S_1(v_0)$ and $S_2(v_2,v_3)$, then $G=G_{1,1,1,1}$. Hence, $S_2(v_2,v_3)\setminus \{y\}$
is anti-complete to $S_1$ and thus $\{v_2,v_3\}$ is a clique cutset by (P4) to (P6).
If $|S_3(v_1)|=2$, then we are in the case $S_3$ is anti-complete to $S_1$.
Now let $t\in S_3(v_1)$ and $r\in S_3(v_0)$. By $\delta(G)\ge 4$ we have $|S_2(v_2,v_3)|\ge 2$
and $|S_2(v_0,v_4)|\ge 1$. As $\{v_0,v_4,r\}$ does not separate $S_2(v_0,v_4)$,
$t$ is not anti-complete to $S_2(v_0,v_4)$ and hence complete to $S_2(v_0,v_4)$ and $S_2(v_2,v_3)$.
Thus $S_2(v_0,v_4)=\{q\}$. As $d(q)\ge 4$ we have $qr\in E$ and thus $rt\in E$ or $qrv_1t=C_4$.
But now it is easy to see $G$ contains $G_{3,1}$ as an induced subgraph.

{\bf (II)} $S_2(v_0,v_4)=\emptyset$. So, $S_3(v_i)=\emptyset$ for $i=1,2$.
Note that it is impossible that $|S_3(v_3)|=2$ by our assumption.
If $|S_3(v_4)|=2$, then we are in the case where $S_3$ is anti-complete to $S_1$.
If $|S_3(v_0)|=2$, then the only edge between $S_1(v_0)$ and $S_2(v_2,v_3)$ is $xy$
or $G\in \{G_{1,1,1,1},G_{2,1,1}\}$. As $G$ has no clique cutset, $S_1(v_0)=\{x\}$
and $S_2(v_0,v_1)=\emptyset$. Note that $S_2(v_2,v_3)$ is bipartite and thus $G$ is 4-colorable.
If $|S_3(v_3)|=|S_3(v_0)|=1$, then $S_2(v_0,v_1)\neq \emptyset$ by $d(v_1)\ge 4$ and thus
$\{v_0,v_1\}\cup S_3(v_0)$ would be a clique cutset. If $|S_3(v_4)|=|S_3(v_0)|=1$,
then it is same as the third case in {\bf (I)}. Finally, $|S_3(v_3)|=|S_3(v_4)|=1$.
Let $r\in S_3(v_4)$ and $t\in S_3(v_3)$.
Then $|S_2(v_0,v_1)|\ge 2$ as $d(v_1)\ge 4$. As $G$ has no clique cutset, $r$
is not anti-complete to $S_2(v_0,v_1)$ and thus complete to $S_2$.
Thus, $S_2(v_0,v_1)=\{p,p'\}$ and $S_2(v_2,v_3)=\{y\}$. Since
$tv_3yxv_0p\neq P_6$, we have $ty\in E$ and so $rt\in E$ or $yrv_4t=C_4$.
Now $G=G_{3,1}$.

Finally, we assume that $S_1=\emptyset$.
Consider first that $|S_3(v_0)|=2$. If $S_2(v_2,v_3)=\emptyset$, then
both $S_2(v_1,v_2)$ and $S_2(v_3,v_4)$ contain at least two vertices since $d(v_2)\ge 4$ and $d(v_3)\ge 4$.
As $G$ has no clique cutset, there exists $t\in S_3(v_0)$ that is complete to $S_2$
by (P9). But now $G=G_{3,1}$. So, let $x\in S_2(v_2,v_3)$. Then
one of  $S_2(v_1,v_2)$ and $S_2(v_3,v_4)$ is nonempty, say $y\in S_2(v_3,v_4)$.
If $S_2=S_2(v_1,v_2)\cup S_2(v_3,v_4)$, then $|S_2(v_3,v_2)|=2$ and $|S_2(v_3,v_4)|=1$.
Now $G$ is 4-colorable. Hence, either $S_2(v_1,v_2)\neq \emptyset$ or $S_2(v_0,v_4)\neq \emptyset$.
In the former case, $S_3$ is anti-complete to $S_2$ or $C_4$ occurs and thus $G$ is 4-colorable.
In the latter case, we have $|S_2(v_3,v_2)|=2$, $|S_2(v_3,v_4)|=1$ and $|S_2(v_0,v_4)|\le 2$.
Note that any $t\in S_2(v_0,v_4)$ is not complete to $S_1(v_0)$ or $K_5$ would occur, and hence
$G$ is 4-colorable.
Hence, no $S_3(v_i)$ has size 2. Suppose that $|S_3(v_0)|=|S_3(v_2)|=1$.
If $S_2(v_3,v_4)=\emptyset$, then $S_2(v_0,v_4)\neq \emptyset$ and so $S_3(v_0)\cup \{v_0,v_4\}$
is a clique cutset. So, $S_2(v_3,v_4)\neq \emptyset$. Now as $d(v_0)\ge 2$ and $d(v_2)\ge 2$
we have three $S_2(v_i,v_{i+1})$ are nonempty, contradicting the property (P7).

So, there must be the case that $|S_3(v_0)|=|S_3(v_1)|=1$. Let $r\in S_3(v_0)$ and $t\in S_3(v_1)$.
If $S_2(v_2,v_3)=\emptyset$, then $|S_2(v_1,v_2)|\ge 1$ and $|S_2(v_3,v_4)|\ge 2$. As $G$
has no clique cutset,  $r$ is not anti-complete to $S_2(v_3,v_4)$ and thus complete to $S_2$.
So, $|S_2(v_1,v_2)|= 1$ and $|S_2(v_3,v_4)|= 2$. Let $q\in S_2(v_1,v_2)$. Note that $qt\in E$ as $d(q)\ge 4$.
Hence, $rt\in E$ or $qrv_1t=C_4$. Now $G=G_{3,1}$.
Therefore, $S_2(v_2,v_3)\neq \emptyset$. By symmetry, $S_2(v_4,v_3)\neq \emptyset$.
Let $p\in S_2(v_3,v_2)$ and $q\in S_2(v_4,v_3)$.
If $S_2=S_2(v_3,v_4)\cup S_2(v_3,v_2)$, then $G$ has a 4-coloring $\phi$: $\{v_0,v_2,q\}$,
$\{v_1,v_4,p\}\}$, $\{r,p'\}$, $\{v_3,t\}$ if $S_2(v_3,v_2)=\{p,p'\}$.
If $S_2(v_4,v_3)=\{q,q'\}$, then $G$ has a 4-coloring
by replacing $\{r,p'\}$, $\{v_3,t\}$ in $\phi$ with $\{t,q'\}$, $\{v_3,r\}$.
Hence we assume by symmetry that $S_2(v_1,v_2)\neq \emptyset$.
Let $s\in S_2(v_1,v_2)$. By (P9) and $C_4$-freeness of $G$,
we have $r$ is anti-complete to $S_2$ and thus $rt\in E$ since $d(r)\ge 4$.
Suppose $S_2(v_2,v_3)=\{p\}$. If $S_2(v_1,v_2)=\{s,s'\}$, then
$t$ is not complete to $\{s,s'\}$, say $ts'\notin E$, since $G$
is $K_5$-free. Hence $G$ has a 4-coloring: $\{r,q,v_2\}$, $\{v_0,v_3,s\}$,
$\{v_4,v_1,p\}$, $\{t,s',q'\}$ where $q'\in S_2(v_3,v_4)$.
Finally, suppose that $S_2(v_2,v_3)=\{p,p'\}$. Then $S_2(v_3,v_4)=\{q\}$ and $S_2(v_1,v_2)=\{s\}$.
If $t$ is complete to $\{p,p',s\}$, then $K_5$ would arise.
Otherwise it is easy to check that $G$ is 4-colorable. This completes the proof of Case 2.

In the remaining of the proof,
we shall frequently consider some induced $C_5=C'$ with $C'\neq C$
or $C_5=C_{t}$ by modifying $C$ with respect to some vertex $t\notin C$.
We can then define $p$-vertices with respect to $C'$ and $C_{t}$ as well.
We adapt those definitions by using the notation $S'_p$ and $S^{t}_p$.
For example, $S'_1$ is the set of 1-vertices with respect to $C'$,
and $S^{t}_1$ is the set of 1-vertices with respect to $C_t$, and so on.
Let $s=(s_1,\ldots, s_5)$ be an integer vector. We say that $C$ is of {\em type $s$}
if $S_3(v_i)$ has size $s_i$ for each $0\le i\le 4$.

{\bf Case 3.} $|S_3|=3$. There are four possible configurations.

{\bf $C$ is of type (2,1,0,0,0)}. Let $S_3(v_0)=\{x,x'\}$ and $S_3(v_1)=\{y\}$.
We may assume that $xy\notin E$. If $t\in S_1(v_3)$ then
$tv_3v_4xv_1y=P_6$. So, $S_1(v_3)=\emptyset$.
Let $C_x=C\setminus \{v_0\}\cup \{x\}$ and $C_y=C\setminus \{v_0\}\cup \{y\}$.
As $xy\notin E$, we have $S^x_3\cap S_2\neq \emptyset$ and $S^y_3\cap S_2\neq \emptyset$.
Let $p\in S^x_3\cap S_2$ and $q\in S^y_3\cap S_2$.
Note that $xp\in E$ and $qy\in E$ by definition of $p$ and $q$.
Suppose first that $p\in S_2(v_1,v_2)$. Then $py\notin E$ or $pyv_0x=C_4$.
If $q\in S_2(v_2,v_3)$, then $qpv_1y=C_4$. So, $q\in S_2(v_0,v_4)$.
By (P8), $S_2(v_3,v_4)=S_2(v_3,v_2)=\emptyset$.
Now $d(v_3)=2$ as $S_1(v_3)=\emptyset$.
Therefore, $p\in S_2(v_3,v_4)$. Then $py\in E$ or $pyv_1x=C_4$.
If $q\in S_2(v_0,v_4)$, then $qx\notin E$ or $qxv_1y=C_4$.
Also, $qp\in E$ and so $pqv_0x=C_4$. Thus $q\in S_2(v_2,v_3)$.
Now by (P7) to (P9) and the fact that $xp,qy\in E$,
we have $S_2=S_2(v_2,v_3)\cup S_2(v_3,v_4)$ and $S_1=\emptyset$.
Thus, $2\le |S_2|\le 3$.
Suppose first that $S_2(v_2,v_3)=\{q,q'\}$.
Then $G$ has a 4-coloring: $\{v_1,v_4,q\}$, $\{v_0,v_2,p\}$,
$\{y,x,v_3\}$, $\{x',q'\}$.
Now suppose that $S_2(v_3,v_4)=\{p,p'\}$.
If $x'y\notin E$ then $N(y)=\{v_0,v_1,v_2,q\}$.
Since {\bf $G$ is a minimal obstruction}, $G-y$ has a 4-coloring $\phi$.
Note that $\phi(v_1)=\phi(v_4)=\phi(q)$. Hence, $\phi$ can be extended to
$G$, a contradiction. So, $x'y\in E$ and so $x'p\notin E$ or $x'pqy=C_4$.
Now $G$ has a 4-coloring: $\{v_1,v_4,q\}$, $\{v_0,v_2,p'\}$,
$\{y,x,v_3\}$, $\{x',p\}$.

{\bf $C$ is of type (2,0,1,0,0)}. Let $S_3(v_0)=\{x,x'\}$ and $S_3(v_2)=\{y\}$.
We first claim that $S_1(v_3)=\emptyset$.
Otherwise let $t\in S_1(v_3)$. Suppose that $S_1(v_1)\neq \emptyset$.
Let $p\in S_1(v_1)$. Then $tp\in E$.
By (P9), $S_2=\emptyset$.
Let $C'=v_3tpv_1v_2$. Note that $x,x'\notin S'_3$ as $\{x,x'\}$
is anti-complete to $\{t,v_3,v_2\}$. So, $|S'_3\cap S_1|\ge 2$
by the minimality of $C$. It is straightforward to check that
$|S'_3\cap (S_1(v_1)\cup S_1(v_3))|\ge 2$ and thus
$|S_1(v_1)|+|S_1(v_3)|\ge 4$. So, $|S_1(v_1)|+|S_1(v_3)|= 4$ by $G$ is $K_5$-free.
As $d(v_2)\ge 4$ we have $S_1(v_2)\neq \emptyset$. Let $q\in S_1(v_2)$.
Since $\{v_2,y\}$ is not a clique cutset separating $S_1(v_2)$,
$S_1(v_0)\cup S_1(v_4)\neq \emptyset$. Suppose that $t'\in S_1(v_4)$.
Then $t'q\in E$. Let $C''=v_4t'qv_2v_3$. Similar as above we have
that $|S_1(v_2)|+|S_1(v_4)|= 4$. If $S_1(v_0)\neq \emptyset$ then $G=G_{P_4}$.
If $|S_1(v_1)|=3$ or $|S_1(v_4)|=3$ then $G=G_{3,1}$. So, each $S_1(v_i)$ has size 2.
But now $\{x,x',v_0,v_1,v_4\}\cup S_1(v_1)\cup S_1(v_4)$ induce a $G_{2,2}$.
Thusm $S_1(v_4)=\emptyset$ and so $S_1(v_0)\neq \emptyset$. Since $S_1(v_0)$
is complete to $S_3(v_0)$, we have $\{x,x',v_0,v_1\}\cup S_1(v_0)\cup S_1(v_1)\cup S_1(v_3)$
induces a $G_{2,2}$ or $G_{3,1}$.
So, $S_1(v_1)=\emptyset$. Let $p\in S_1(v_0)$. $pt\in E$.
Note that $S_1(v_0)=\{p\}$ or $K_5$ would arise. Also, $S_2=\emptyset$.
Let $C'=v_0ptv_3v_4$. As $y$ is anti-complete to $\{v_0,v_4,p\}$, $y\notin S'_3$
and so $S'_3\cap S_1\neq \emptyset$. Let $t'\in S'_3\cap S_1$ and it is easy to see that
$t'\in S_1(v_3)$. Thus, $S_1(v_3)=\{t,t'\}$ or $\{x,x'\}\cup C'\cup S_1(v_3)$
induces a $G_{3,1}$. Now by (P11), we have $S_1(v_3)$ is anti-complete to
$S_2(v_0,v_1)$. Hence, $\{t,t'\}$ is complete to $y$ as $d(t)\ge 4$ and $d(t')\ge 4$.
Now $G$ contains $G_{2,1,1}$ as an induced subgraph.
So far, we have showed that $S_1(v_1)=S_1(v_0)=\emptyset$ if $S_1(v_3)\neq \emptyset$.
As $\{v_3,y\}$ is not a clique cutset separating $S_1(v_3)$, we may assume that
$t$ has a neighbor $p\in S_2(v_0,v_1)$. By Observation \ref{o5} (3), $y$ is anti-complete
to $\{p,t\}$. Then the fact that $yv_3tpv_0x(x')\neq P_6$ implies that $p$ is complete to $\{x,x'\}$
and so $\{p,v_0,v_1,x,x'\}$ induces a $K_5$.

Therefore, $S_1(v_3)=\emptyset$. Next we claim that $S_2(v_3,v_2)\neq \emptyset$.
If not, we have $S_2(v_3,v_4)\neq \emptyset$ and $S_2(v_1,v_2)\neq \emptyset$
as $d(v_2)\ge 4$ and $d(v_3)\ge 4$. So, $S_2=S_2(v_1,v_2)\cup S_2(v_3,v_4)$ and
$S_1=S_1(v_1)\cup S_1(v_4)$. Suppose that $q\in S_2(v_1,v_2)$ is adjacent a vertex $t\in S_1(v_4)$.
By Observation \ref{o5} (3), $\{x,x'\}$ is anti-complete to $\{q,t\}$.
Hence, $\{x,x'\}$ is anti-complete to $S_2$. Also, $S_1(v_1)=\emptyset$.
Let $C'=v_4v_3v_2qt$. Note that $x,x'\notin S'_3$ and so $|S'_3\cap (S_1\cup S_2)|\ge 2$.
Note that $S'_3\cap S_2(v_3,v_4)=\emptyset$. If $|S'_3\cap S_1(v_4)|\ge 2$
or $|S'_3\cap S_2(v_4)|\ge 2$, then $\{x,x',v_0,v_1,v_4\}\cup S_1(v_4)\cup S_2(v_4)$ induces
a $G_{3,1}$ or $G_{2,2}$. Thus, there exists a vertex $q\in S'_3\cap S_2(v_1,v_2)$ with $q'\neq q$.
Now $qy\in E$ as $xv_4tqv_2y\neq P_6$. Since $q'q\in E$, $q'y\notin E$ or $K_5$ would arise.
But then $yv_2q'tv_4x=P_6$.
Therefore, $S_1(v_4)$ is anti-complete to $S_2(v_1,v_2)$. Since $\{v_1,v_2,y\}$ is not a clique
cutset separating $S_2(v_1,v_2)$, $S_2(v_1,v_2)$ is not anti-complete to $\{x,x'\}$.
Thus, we may assume that $x'$ is complete to $S_2$ by (P9).
Now note that $S_1(v_1)$ is anti-complete to $S_2(v_3,v_4)$ by Observation \ref{o5} (3).
If $t\in S_1(v_4)$ and $t'\in S_1(v_1)$, then $\{x,x'\}$ is anti-complete to $\{t,t'\}$
by Observation \ref{o5} (1). Also, $ty\notin E$. As $d(t)\ge 4$, we have $|S_1(v_1)|+|S_1(v_4)|=4$.
Then $\{x,x',v_0,v_1,v_4\}\cup S_1(v_1)\cup S_1(v_4)$ induces a $G_{2,2}$ or $G_{3,1}$.
So, if $S_1(v_4)\neq \emptyset$, then $S_1(v_1)=\emptyset$ and thus $\{v_4,x,x'\}$ would be a clique cutset.
Hence, $S_1(v_4)=\emptyset$. Since $x'$ is complete to $S_2$, we have that $2\le |S_2|\le 3$.
Moreover, $py\notin E$ or $pyv_1x'=C_4$. Thus $y$ is anti-complete to $S_2(v_3,v_4)$.
Next we show that $S_1(v_1)$ is a clique. Let $t\in S_1(v_1)$ and $A$ be the component of $S_1(v_0)$
containing $t$. Since $\{v_1,x,x'\}$ is not a clique cutset separating $A$,
$A$ is not anti-complete to $y$ and hence complete to $y$. Further, since $v_0x'pv_3yt\neq P_6$, we have
$x't\in E$ and thus $A$ is complete to $x'$. Hence, $A$ is a clique. By $G$ is $C_4$-free,
$S_1(v_1)=A$ and $|A|\le 2$ by $G$ is $K_5$-free.
If $S_2(v_3,v_4)=\{p\}$ then $xp\in E$ as $d(p)\ge 4$.
Thus, $S_2(v_1,v_2)=\{q\}$ or $K_5$ would arise. Note that $qy\notin E$ or $G=G_{P_4}$ with respect
to $xqv_2v_3p$. Now $S_1=\emptyset$ or if $t\in S_1(v_1)$ then $tx'qv_2y$ and $v_1$ induce a $W_5$.
Now $G$ has a 4-coloring: $\{v_0,p,q,y\}$, $\{v_4,v_1\}$, $\{x,v_3\}$, $\{x',v_2\}$.
So, we assume that $S_2(v_4,v_3)=\{p,p'\}$ and thus $S_2(v_1,v_2)=\{q\}$.
Now $x$ is anti-complete to $S_2$ or $K_5$ would arise. If $|S_1(v_1)|=2$,
then $G=G_{3,1}$ .
So, $S_1(v_1)$ contains at most one vertex. If $S_1(v_1)=\{t\}$, then $qy\notin E$ or $qytx'=C_4$.
Now $G$ has a 4-coloring: $\{v_1,v_4\}$, $\{x',v_3\}$, $\{t,v_2,p,v_0\}$, $\{x,y,p',q\}$.
So, $S_1(v_1)=\emptyset$. Also, $qy\in E$ as $d(q)\ge 4$.
Now $\{x',v_4,p,p',v_3\}=K_5-e$ and $\{v_1,q,v_2,y\}=K_4$ induce a $G_{2,2}$.

Therefore, let $p\in S_2(v_2,v_3)$. As $\{v_2,v_3,y\}$ is not a clique cutset separating $S_2(v_2,v_3)$
the following three cases are possible. First we suppose that $S_2(v_1,v_2)\neq \emptyset$.
Let $q\in S_2(v_1,v_2)$. By Observation \ref{o5} (2), $y$ is complete to $S_2(v_1,v_2)\cup S_2(v_2,v_3)$.
Thus, $S_2(v_2,v_3)=\{p\}$ and $S_2(v_2,v_3)=\{q\}$.
Further, $S_1=S_1(v_2)$ and so $S_1(v_2)=\emptyset$ or $\{v_2,y\}$ would be a clique cutset.
Suppose that $S_2(v_3,v_4)\neq \emptyset$. Note that $\{x,x'\}$ is anti-complete to $S_2$.
If $S_2(v_3,v_4)=\{r\}$, then $G$ has a 4-coloring:
$\{v_1,v_4,p\}$, $\{v_0,v_3,q\}$, $\{x,r,v_2\}$, $\{x',y\}$.
So, $S_2(v_3,v_4)=\{r,r'\}$. Then $y$ is not complete to $\{r,r'\}$, say $yr'\notin E$
or $\{r,r',p,v_3,y\}$ would induce a $K_5$. Then $G$ has a 4-coloring:
$\{v_1,v_4,p\}$, $\{v_0,v_3,q\}$, $\{x,r,v_2\}$, $\{x',y,r'\}$.
So, we may assume that $S_2(v_3,v_4)=\emptyset$.  If $S_2(v_0,v_1)\neq \emptyset$,
then $G$ has a 4-coloring as above. Suppose that $r\in S_2(v_0,v_1)$.
The fact that $v_3pqrv_0x\neq P_6$ implies that $xq\in E$ or $xr\in E$.
Similarly, $x'q\in E$ or $x'r\in E$.
Also, the fact that $xq$ (respectively $x'q$) is an edge implies that $xr$ (respectively $x'r$) is an edge,
 since $G$ is $C_4$-free. Hence, $q$ is not complete to $\{x,x'\}$, say $qx'\notin E$ or $\{x,x',v_0,v_1,r\}$
would induce a $K_5$. As $qx'\notin E$, $x'r\in E$. Hence, $xr\notin E$ and so $xq\in E$.
Now $xx'rq=C_4$.

Therefore, we may assume that $S_2(v_1,v_2)= \emptyset$. Suppose that $S_2(v_3,v_4)\neq \emptyset$.
Let $q\in S_2(v_3,v_4)$. Note that $S_1=\emptyset$ since $S_1(v_3)=\emptyset$.
Suppose that $r\in S_2(v_0,v_4)$. Note that $r$ is not complete to $\{x,x'\}$, say $xr\notin E$.
As $v_2pqrv_0x\neq P_6$, we have $xq\in E$. But now $xqrv_0=C_4$.
So, $S_2(v_0,v_4)= \emptyset$. Thus $2\le |S_2|\le 3$.
Now as $yv_2pqv_4v_0\neq P_6$, we have $yp\in E$ or $yq\in E$.
Also, if $yq\in E$ then $yp\in E$ or $yv_2pq=C_4$.
So, $yp\in E$ and thus $S_2(v_2,v_3)=\{p\}$.
Suppose first that $S_2(v_3,v_4)=\{q\}$.
Note that $q$ is not complete to $\{x,y\}$ or $\{x',y\}$.
Thus $G$ has a 4-coloring: $\{v_1,v_4,p\}$, $\{q,y,v_0\}$, $\{x,v_2\}$, $\{x',v_3\}$
if $qy\notin E$, and otherwise we move $q$ from $\{q,y,v_0\}$ to $\{x,v_2\}$.
Now suppose that $S_2(v_3,v_4)=\{q,q'\}$.
Then we may assume that $qy\notin E$ or $K_5$ would arise.
Also, $\{q,q'\}$ is not anti-complete to $\{x,x'\}$.
If $q'y\notin E$, then $q$ and $q'$ are in the same place thus we may assume that $qx\notin E$.
Now $G$ has a 4-coloring: $\{v_1,v_4,p\}$, $\{q',y,v_0\}$, $\{q,x,v_2\}$, $\{x',v_3\}$.
Otherwise $q'y\in E$ and so $q'$ is anti-complete to $\{x,x'\}$.
Then $G$ has a 4-coloring: $\{v_1,v_4,p\}$, $\{q,y,v_0\}$, $\{q,x,v_2\}$, $\{x',v_3\}$.

Now we may assume that $S_2(v_1,v_2)=S_2(v_3,v_4)=\emptyset$ and $p$ has a neighbor
$t\in S_1(v_0)$. Then $t$ is complete to $\{x,x'\}$. Also $S_1=S_1(v_0)$ by (P11).
Let $C'=v_0v_4v_3p$. Clearly, $y\notin S'_3$ as $y$ is anti-complete to $\{v_4,v_0,t\}$.
Thus $S'_3\cap (S_1\cup S_2)\neq \emptyset$. It is easy to check that
$S'_3\cap S_1\subseteq S_1(v_0)\cap S'_3(t)$ and $S'_3\cap S_2(v_2,v_3)\subseteq S_1(v_0)\cap S'_3(p)$.
Let $r\in S'_3$. If $r\in S_1(v_0)$, then $rt\in E$ and so $\{x,x',v_0,r,t\}$ induces a $K_5$.
Hence, $r\in S_2(v_2,v_3)$ and now $\{x,x'v_0,v_1,v_4,t,v_3,v_2,p,r\}$ induces a $G_{2,1,1}$.

{\bf $C$ is of type (1,0,1,1,0)}. Let $x\in S_3(v_0)$, $y\in S_3(v_2)$ and $z\in S_3(v_3)$.
We first show that $S_1(v_0)=\emptyset$. If $yz\in E$, then if $t\in S_1(v_0)$ we have
$G=G_{2,2}$. If $yz\notin E$, then $yv_2zv_4v_0t=P_6$. Next we claim that $yz\in E$.
Otherwise let $C_y=C\setminus \{v_2\}\cup \{y\}$ and $C_z=C\setminus \{v_3\}\cup \{z\}$.
As $yz\notin E$, we have that $S^y_3\cap (S_1\cup S_2)\neq \emptyset$ and $S^z_3\cap (S_1\cup S_2)\neq \emptyset$.
Let $p\in S^y_3\cap (S_1\cup S_2)$ and $q\in S^z_3\cap (S_1\cup S_2)$.
Then $py,qz\in E$ by definition of $p$ and $q$. Since $G$ is $C_4$-free, $pz,qy\notin E$.
We consider the case $p\in S_2(v_0,v_1)$ first. If $q\in S_2(v_0,v_4)$,
then $x$ is complete to $S_2(v_0,v_1)\cup S_2(v_0,v_4)$ by Observation \ref{o5} (2).
Note that $S_1=\emptyset$. Further, $S_2(v_1,v_2)=S_2(v_3,v_4)=S_2(v_2,v_3)=\emptyset$
by (P7) to (P9). As $G$ is $K_5$-free, $S_2=\{p,q\}$. Hence, $G$ has a 4-coloring:
$\{x,y,z\}$, $\{p,v_4,v_2\}$, $\{q,v_1,v_3\}$, $\{v_0\}$.
Thus $q\in S_2(v_1,v_2)$. Now $pqv_2y=C_4$ as $qy\notin E$.
Hence, $p\in S_2(v_3,v_4)$. If $q\in S_2(v_1,v_2)$, then $S_2(v_0,v_1)=S_2(v_0,v_4)=\emptyset$
and so $N(v_0)=\{v_1,v_4,x\}$, a contradiction. Thus, $q\in S_2(v_0,v_4)$.
But now $pqzv_3=C_4$.
Therefore $yz\in E$. Recall that $S_1(v_0)=\emptyset$.
As $d(v_0)\ge 4$, we may assume that there exists a vertex $p\in S_2(v_0,v_1)$.
Since $G$ has no clique cutset, the following four cases are possible.

{\bf Case a.} $S_2(v_0,v_4)\neq \emptyset$. Let $q\in S_2(v_0,v_4)$.
By Observation \ref{o5} (2), $x$ is complete to $S_2(v_0,v_1)\cup S_2(v_0,v_4)$
and so $S_2(v_0,v_1)=\{p\}$ and $S_2(v_0,v_4)=\{q\}$.
Note that $S_1=\emptyset$. If $S_2=\{p,q\}$, then $G$
has a 4-coloring: $\{q,v_1,v_3\}$, $\{p,v_2,v_4\}$, $\{y,x\}$, $\{z,v_0\}$.
Now by symmetry, we may assume that $r\in S_2(v_1,v_2)$. Then $z$
is anti-complete to $S_2$ and so $G$ has a 4-coloring by adding $r$
to $\{z,v_0\}$ if $S_2(v_1,v_2)=\{r\}$. So, let $S_2(v_1,v_2)=\{r,r'\}$.
As $G$ is $K_5$-free, $y$ is not complete to $\{r,r'\}$, say $yr'\notin E$.
Then $yp\in E$ since  $yv_2r'pqv_4\neq P_6$.
But now $pr'v_2y=C_4$.

{\bf Case b.} $S_2(v_1,v_2)\neq \emptyset$. Let $q\in S_2(v_1,v_2)$.
We may assume that $S_2(v_0,v_4)= \emptyset$. Suppose that $r\in S_2(v_2,v_3)$.
Then $yr,yq\in E$ by Observation \ref{o5} (2). Hence, $zr\notin E$ or $\{v_2,v_3,y,z,r\}$
would induce a $K_5$. So, $zq\notin E$ or $v_3rqz=C_4$. But then $zv_3rpqv_0=P_6$.
So, $S_2(v_2,v_3)=\emptyset$. Hence $2\le |S_2|\le 3$.
Also, $S_1=S_1(v_1)$ and $S_1(v_1)$ is a clique and thus $|S_1(v_1)|\le 2$.
Suppose that $t\in S_1(v_1)$. Then $tyzv_4v_0p\neq P_6$ implies that $yp\in E$ and so $yq\in E$
or $ypqv_2=C_4$. Since $v_3v_2qpv_0x\neq P_6$, we have either $xp\in E$ or $xq\in E$.
In any case, we have an induced $C_4$ as $t$ is complete to $\{x,y\}$.
So, $S_1(v_1)=\emptyset$.
If $S_2=\{p,q\}$, then $G$ has a 4-coloring: $\{v_0,v_3,q\}$, $\{v_2,v_4,p\}$,
$\{x,y\}$, $\{z,v_1\}$.
Suppose that $S_2(v_0,v_1)=\{p,p'\}$.
Now since $v_3v_2qpv_0x\neq P_6$, we have either $xp\in E$ or $xq\in E$.
If $xq\in E$, then $x$ is complete to $\{p,p'\}$ by $G$ is $C_4$-free and hence
$\{x,v_0,v_1,p,p'\}$ induces a $K_5$. So, $xq\notin E$ and thus $xp\in E$.
Replacing the argument for $\{p',q\}$, we have $xp'\in E$ and so $K_5$ would arise.
If $S_2(v_1,v_2)$ contains two vertices, we would derive a similar contradiction.

{\bf Case c.} $S_2(v_0,v_1)$ is not anti-complete to $S_1(v_3)$. We now may assume that
$S_2(v_0,v_4)=S_2(v_1,v_2)=\emptyset$. Without loss of generality, we assume that $p$
has a neighbor $q\in S_1(v_3)$. So, $S_1=S_1(v_3)$ by (P11).
Moreover, $S_2(v_2,v_3)=\emptyset$ or $\{v_2,v_3,y,z\}$ would be a clique cutset.
Now $px\in E$ since  $v_2v_3qpv_0x\neq P_6$ and so $p$ is the only
neighbor of $q$ in $S_2(v_0,v_1)$ or $K_5$ would arise.
Let $C'=v_1v_2v_3qp$. Note that $v_0,v_4,x\notin S'_3$ and $y,z\in S'_3$.
Thus, $S'_3\cap (S_1\cup S_2)\neq \emptyset$ by the minimality of $C$.
Let $t\in S'_3\cap (S_1\cup S_2)$. It is easy to check that $t\in S_1(v_3)$ or $t\in S_2(v_0,v_1)$.
If $t\in S_2(v_0,v_1)$, then $t$ must be complete to $\{v_1,p,q\}$, contradicting the fact that
$p$ is the only neighbor of $q$. Hence, $t\in S_1(v_3)$ and $t$ must be complete to $\{v_3,q,p\}$.
By (P12), $z$ is complete to $\{q,t\}$.
Now $G=H_2$.

{\bf Case d.} Now we may assume that $py\in E$. If $S_2(v_2,v_3)\neq \emptyset$, then $S_2(v_3,v_4)=\emptyset$
and so $\{v_2,v_3,y,z\}$ would be a clique cutset since $S_1(v_0)=\emptyset$.
So, $S_2(v_2,v_3)=\emptyset$. Suppose that $S_1(v_3)\neq \emptyset$. Then $S_1(v_1)\neq \emptyset$
or $\{v_3,y,z\}$ would be a clique cutset. But then $S_2=\emptyset$ by (P7) to (P9), a contradiction.
So, $S_1(v_3)=\emptyset$. Thus, $S_1=S_1(v_1)$ and $S_2=S_2(v_0,v_1)\cup S_2(v_3,v_4)$.
Next we claim that $xp\notin E$. Otherwise $xp\in E$. Let $C'=xpyzv_4$. It is easy to check
that $v_1,v_3\in S'_3$ but $v_0,v_2\notin S'_3$. So, $S'_3\cap (S_1\cup S_2)\neq \emptyset$.
Let $t\in S'_3\cap (S_1\cup S_2)$.
As $S_1(v_1)$ is anti-complete to $\{v_4,z,p\}$, $t\notin S_1(v_1)$ and so $t\in S_2$.
If $t\in S_2(v_0,v_1)$, then $t$ is complete to $\{x,p,y\}$. If $t\in S_2(v_3,v_4)$,
then as $py\in E$ we have $yt\in E$ and so $xt\notin E$ or $xtyv_1=C_4$.
Hence, $t$ is complete to $\{v_4,z,y\}$.  If $S'_3\cap S_2=\{t\}$, then $|S'_3|=3$
and we are in one of previous two cases. Thus, there exists another vertex $t'\neq t$
with $t'\in S'_3\cap S_2$. If $t,t'\in S_2(v_0,v_1)$, then $\{v_0,v_1,t,t',p\}$ would induce a $K_5$.
If $t,t'\in S_2(v_4,v_3)$, then $\{y,z,t,t',v_3\}$ would induce a $K_5$.
Hence, $t\in S_2(v_0,v_1)$ and $t'\in S_2(v_4,v_3)$. But now $G=G_{3,1}$.
Therefore, $xp\notin E$. Now let $C''=v_0pyv_3v_4$. As $xp\notin E$,
$x\notin S''_3$. Also, $v_2\notin S''_3$ but $z,v_1\in S''_3$.
Hence, $S''_3\cap (S_1\cup S_2)\neq \emptyset$.
By the same argument as above, we either find an induced $K_5$ or $G_{2,2}$ or
we are in one of previous two cases.

{\bf $C$ is of type (1,1,0,0,1)}.  Let $x\in S_3(v_0)$, $y\in S_3(v_1)$ and $z\in S_3(v_4)$.
We first suppose that $S_2(v_2,v_3)=\emptyset$. As $\delta(G)\ge 4$, we have the following two cases.
Suppose first that $S_2(v_1,v_2)$ and $S_2(v_3,v_4)$ are nonempty but $S_1(v_2)=S_1(v_3)=\emptyset$.
By (P9), we may assume that $S_1(v_4)=\emptyset$.
Now $x$ is complete to $S_2$ or $\{v_1,v_2,y\}$ would be a clique cutset.
Also, $x$ is complete to $\{y,z\}$, otherwise considering $C_y=C\setminus \{v_1\}\cup \{y\}$
or $C_z=C\setminus \{v_4\}\cup \{z\}$ will obtain by the minimality of $C$
that $S_2(v_0,v_1)\cup S_2(v_0,v_4)\cup S_2(v_2,v_3)\neq \emptyset$
which contradicts our assumption and (P8).
But now $G$ contains $G_{3,1}$ as an induced subgraph.
So, $S_1(v_2)$ and $S_1(v_3)$ are nonempty and $S_2(v_2,v_1)\cup S_2(v_3,v_4)=\emptyset$.
Thus, $S_2=\emptyset$ and hence $x$ is complete to $\{y,z\}$ by the minimality of $C$.
Let $p\in S_1(v_3)$ and $q\in S_1(v_2)$. Suppose that $t\in S_1(v_0)$.
Let $C'=xtpv_3v_4$. Then $v_1,v_2,y\notin S'_3$. Hence, $S'_3\cap S_1\neq \emptyset$.
Let $r\in S'_3\cap S_1$. It is easy to check that $r\in S_1(v_0)\cup S_1(v_3)$.
We claim that $S'_3\cap S_1(v_0)\neq \emptyset$. Otherwise $r\in S_1(v_3)$.
Then $r\in S'_3(p)$ as $r$ is anti-complete to $\{v_4,x\}$. If $S'_3=\{v_0,z,r\}$,
then we are in the case {\bf $C$ is of type (1,0,1,1,0)}.
So, $|S'_3\cap S_1(v_3)|\ge 2$ and thus $G=G_{2,2}$.
Therefore, we may assume that $r\in S_1(v_0)$.
If $S_1(v_3)=\{p,p'\}$, then $C_5=xtpv_3z$ and $P_4=v_4v_0rt$ induce a $G_{P_4}$.
So, $S_1(v_3)=\{p\}$ and $S_1(v_2)=\{q\}$. Now let $C''=tqv_2v_3p$.
Clearly, $x,v_0,v_1,v_4\notin S''_3$ and so $S''_3\cap S_1\neq \emptyset$.
Let $s\in S''_3\cap S_1$
Clearly, $s\in S_1(v_0)\cup S_1(v_2)\cup S_1(v_3)$.
As $s\notin \{p,q\}$, we have $s\in S_1(v_0)$. Hence, $s=r$.
By the minimality of $C$, $y$ and $z$ must be in $S''_3$.
This implies that $y$ is complete to $\{t,q,v_2\}$. So, $ry\in E$ or $rqyx=C_4$.
But then $\{x,v_0,y,r,t\}$ induces a $K_5$.
We have shown that $S_1(v_0)=\emptyset$. As $G$ has no clique cutset,
$S_1(v_1)\neq \emptyset$ and $S_1(v_4)\neq \emptyset$.
Let $u_i\in S_1(v_i)$ for $i=1,4$.
Note that $x$ is anti-complete to $S_1$ by Observation \ref{o5} (1).
If $|S_1(v_1)|\ge 2$, say $u_1,u'_1\in S_1(v_1)$, then $G=G_{P_4}$
with respect to $xyu_1u_4z$ whose 3-vertices are $v_4v_0v_1u'_1$.
Hence, $S_1(v_i)=\{u_i\}$. Note that $pz\notin E$ or $zpu_1u_4=C_4$.
Thus, $z$ is anti-complete to $S_1(v_3)$ and $y$ is anti-complete to $S_1(v_2)$.
By $\delta(G)\ge 4$, we must have $|S_1(v_2)|=|S_1(v_3)|=3$. It is easy to check
$G$ is 4-colorable. 

Therefore, we may assume that $p\in S_2(v_2,v_3)$. By (P7) to (P9),
there are at most two nonempty $S_1(v_i)$.
If there exists $i$ such that $S_1(v_i)\neq \emptyset$
and $S_1(v_{i+1})\neq \emptyset$, then $i=2$ as $S_2(v_2,v_3)\neq \emptyset$.
Thus $\{v_2,y\}$ is a clique cutset separating $S_1(v_2)$.

{\bf Case a.} $S_1(v_i)\neq \emptyset$ for some $i$. As $S_2(v_2,v_3)\neq \emptyset$, $S_1(v_1)=S_1(v_4)=\emptyset$.
So, $i\in \{0,2,3\}$. Suppose first that $i=2$ (or $i=3$) and let $t\in S_1(v_2)$.
As $\{v_2,y\}$ is not a clique cutset, $S_1(v_2)$ is not anti-complete to $S_2(v_0,v_4)$.
We may assume that $t$ has a neighbor $q\in S_2(v_0,v_4)$.
By Observation \ref{o5} (3), $y$ is anti-complete to $\{q,t\}$
and thus anti-complete to $S_2(v_0,v_4)\cup S_2(v_2,v_3)$.
Let $C'=qtv_2v_3v_4$. Clearly, $x,v_0,v_1\notin S'_3$.
If $z\in S'_3$, then $z\in S'_3(v_4)$ and if
$y\in S'_3$, then $z\in S'_3(t)$.
As $x\notin S'_3$, $|S'_3\cap (S_1\cup S_2)|\ge 1$. Let $r\in S'_3\cap (S_1\cup S_2)$.
If $r\in S_1(v_2)=S_1$, then $r\in S'_3(t)$. Also, $qx\in E$ $pv_2tqv_0x\neq P_6$.
Hence, $q$ is the only neighbor of $t$ in $S_2$ and so $r\notin S_2(v_0,v_4)$. Clearly,
$r\notin S_2(v_2,v_3)$. If $r\in S_2(v_3,v_4)$, then $r$ must be complete to $\{q,v_3,v_4\}$
and hence $r\in S'_3(v_4)$. So, $S'_3=S'_3(v_4)\cup S'_3(t)$.
Now as $|S'_3|\ge 3$ either we are in one of previous cases or $G$ contains $G_{3,1}$ as an induced subgraph.

Therefore, we may assume that $i=0$. Let $C_x=C\setminus \{v_0\}\cup \{x\}$.
If $x$ is not complete to $\{y,z\}$, then by the minimality of $C$, we have
$S^x_3\cap (S_2(v_1)\cup S_2(v_4))\neq \emptyset$, which contradicts (P7).
Hence, $xy,xz\in E$. Suppose that $p$ has a neighbor $q\in S_1(v_0)$.
Let $C'=v_0v_1v_2pq$.
Note that $z$ is not complete to $\{p,q\}$ by Observation \ref{o5} (3) and hence $z\notin S'_3$.
Also, $v_3,v_4\notin S'_3$.
Thus, $S'_3\cap (S_1\cup S_2)\neq \emptyset$. Let $t\in S'_3\cap (S_1\cup S_2)$.
If $t\in S_1(v_0)$, $t$ is complete to $\{v_0,p,q\}$ and then $G=H_1$.
Clearly, $t\notin S_2(v_0,v_1)\cup S_2(v_0,v_4)$.
If $t\in S_2(v_2,v_3)$, $t$ is complete to $\{q,p,v_2\}$ and then $G$ is not 4-colorable
and $G=H_2$.
We have shown that $S_1(v_0)$ is anti-complete to $S_2(v_2,v_3)$.
As $\{v_2,v_3\}$ does not separate $S_2(v_2,v_3)$,
$S_2(v_2,v_3)$ is not anti-complete to $\{y,z\}$.
Without loss of generality, assume that $py\in E$.
As before we can show that $S_1(v_0)$ is complete to $\{y,z\}$
and thus a clique. So, $S_1(v_0)=\{q\}$ or $K_5$ would arise.
Also, $pz\in E$ or $pzxy=C_4$. As $d(p)\ge 4$, there exists $p'\in S_2(v_2,v_3)$
with $pp'\in E$. Note that in any 4-coloring $\phi$ of $G$,
$\phi(p')=\phi(y)=\phi(z)$. So if $p'$ is not anti-complete to $\{y,z\}$
$G$ is not 4-colorable. Specifically, if $p'y\in E$ then
$\{y,p,p',v_3,v_2\}=K_5-e$ and $\{z,x,v_0,q\}=K_4$ induces a $G_{3,1}$.
If $p'z\in E$, then $\{q,v_0,x,z,y\}=K_5-e$ and $\{v_2,v_3,p,p'\}=K_4$ induce a $G_{2,2}$.
Thus, we assume that $p'$ is anti-complete to $\{y,z\}$.
As $d(p')\ge 4$, there exists $p''\in S_2(v_2,v_3)$ with $p'p''\in E$.
Moreover, $pp''\notin E$ or $K_5$ would arise, and $p''y\notin E$ or $p''ypp'=C_4$.
Then the fact that $p''p'pyqz\neq P_6$ implies that $zp''\in E$, and thus $v_4zp''p'py=P_6$.

{\bf Case b.} $S_1=\emptyset$. Recall that $p\in S_2(v_2,v_3)$.
We first show that $x$ is complete to $\{y,z\}$. Otherwise suppose $xy\notin E$.
Since $yv_1xv_4v_3p\neq P_6$,  we have $yp\in E$ and so $zp\notin E$.
Since $p$ is an arbitrary vertex in $S_2(v_2,v_3)$, we have
that $y$ is complete to $S_2(v_2,v_3)$, and $z$ is anti-complete to $S_2(v_2,v_3)$.
Hence, $xz\in E$ by symmetry. Let $C_x=C\setminus \{v_0\}\cup \{x\}$.
As $xy\notin E$, $S^x_3\cap S_2\neq \emptyset$.
Let $q\in S^x_3\cap S_2$. $xq\in E$.
Suppose that $r\in S_2(v_0,v_4)$. Then $S_2(v_1,v_2)=\emptyset$. Hence, $q\in S_2(v_3,v_4)$.
By (P7) to (P9) and the fact that $yp\in E$, we have $yr\in E$ and thus $yrqp=C_4$.
So, $S_2(v_0,v_4)=\emptyset$. If $q\in S_2(v_1,v_2)$ then $pq\in E$. Note that $qy\notin E$
or $qyv_0x=C_4$. Then $pqv_1y=C_4$. Thus, $q\in S_2(v_3,v_4)$.
As $xq\in E$, $S_2(v_1,v_2)=\emptyset$ by (P9). Thus, $S_2=S_2(v_4,v_3)\cup S_2(v_2,v_3)$
and $2\le |S_2|\le 3$. If $S_2=\{p,q\}$, then $G$ has a 4-coloring: $\{x,v_3\}$, $\{v_0,v_2,q\}$, $\{y,z\}$,
$\{v_1,v_4,p\}$. Suppose now that $S_2(v_4,v_3)=\{q,q'\}$. As $v_2yv_0xqq'\neq P_6$, we have $xq'\in E$.
As $\{x,v_4,z,q,q'\}$ does not induce a $K_5$, $z$ is not complete to $\{q,q'\}$, say $zq'\notin E$.
Then $G$ has a 4-coloring by adding $q'$ to $\{y,z\}$. Finally, $S_2(v_2,v_3)=\{p,p'\}$.
Then $G$ has a 4-coloring $\{x,y, v_3\}$, $\{v_0,v_2,q\}$, $\{p',z\}$, $\{v_1,v_4,p\}$
as $z$ is anti-complete to $S_2(v_2,v_3)$.

Therefore, $xy,xz\in E$. Next we show that $S_2(v_1,v_2)=S_2(v_3,v_4)=\emptyset$.
By symmetry, we may assume that $S_2(v_1,v_2)\neq \emptyset$. Let $q\in S_2(v_1,v_2)$.
Then $S_2(v_0,v_4)=\emptyset$. As $v_4v_3pqv_1y\neq P_6$, we have $py\in E$ or $qy\in E$.
Suppose first that $py\in E$. Then $qy\in E$ or $pqv_1y=C_4$.
Let $C'=ypv_3v_4v_0$. Clearly, $v_1\notin S'_3$ as $v_1$ is anti-complete to $\{v_4,v_3,p\}$, and
$x\in S'_3(v_0)$, $z\in S'_3(v_4)$ and $v_2\in S'_3(p)$. If $S'_3=\{x,z,v_2\}$ then
we are in the case {\bf $C$ is of type (1,0,1,1,0)}. So, let $r\in S'_3\setminus \{x,z,v_2\}$.
Clearly, $r\notin S_2(v_0,v_1)\cup S_2(v_1,v_2)$.
If $r\in S_2(v_2,v_3)$ then $r\in S'_3(p)$ and $G=G_{2,2}$.
So, $r\in S_2(v_3,v_4)$. As $py\in E$, $pz\notin E$ and then
$rz\in E$ since $v_1v_2prv_4z\neq P_6$. Hence, $y$ and $z$
are complete to $S_2(v_1,v_2)$ and $S_2(v_3,v_4)$, respectively.
So, $S_2(v_1,v_2)=\{q\}$ and $S_2(v_3,v_4)=\{r\}$.
If $S_2(v_2,v_3)=\{p,p'\}$ and $p'y\in E$, then $p'\in S'_3(p)$ and thus $G=G_{2,2}$.
Note $x$ is anti-complete to $S_2$. Now $G$ has a 4-coloring:
$\{x,q,v_3\}$, $\{v_0,r,v_2\}$, $\{v_1,z,p\}$, $\{v_4,y,p'\}$.

Now we have shown that $py\notin E$ and thus $qy\in E$.
Since $p$ is an arbitrary vertex in $S_2(v_2,v_3)$, we may assume that
$y$ is anti-complete to $S_2(v_2,v_3)$. Also, replacing any $q'\in S_2(v_1,v_2)$
we obtain $yq'\in E$ and so $S_2(v_1,v_2)=\{q\}$ or $K_5$ would arise.
If $S_2(v_3,v_4)\neq \emptyset$, then $z$ is anti-complete to $S_2(v_1,v_2)$
and complete to $S_2(v_3,v_4)$ by symmetry. Thus, $S_2(v_3,v_4)=\{r\}$ and
$G$ has a 4-coloring: $\{x,q,v_3\}$, $\{v_0,r,v_2\}$, $\{y,z,p\}$, $\{v_4,v_1,p'\}$,
where $p'$ might be another vertex in $S_2(v_2,v_3)$.
If $S_2(v_0,v_1)\neq \emptyset$, then $y$ is complete to $S_2(v_0,v_1)\cup S_2(v_1,v_2)$ and thus
$S_2(v_0,v_1)=\{r\}$. Also, $xr\notin E$ or $\{x,y,v_0,v_1,r\}$ would induce a $K_5$.
Note that $z$ is anti-complete to $S_2$ and thus
$G$ has a 4-coloring:
$\{x,r,v_2\}$, $\{v_0,q,v_3\}$, $\{y,z,p\}$, $\{v_4,v_1,p'\}$,
where $p'$ might be another vertex in $S_2(v_2,v_3)$.
Finally, we have $S_2=\{q\}\cup S_2(v_2,v_3)$.
If $S_2(v_2,v_3)=\{p,p'\}$ and $z$ is complete to $\{p,p'\}$, then
$\{p,p,v_3,v_2,z\}=K_5-e$ and $\{x,y,v_0,v_1\}$ induce a $G_{2,2}$.
Otherwise in case of $S_2(v_2,v_3)=\{p,p'\}$, we may assume $p'z\notin E$ and thus
$G$ has a 4-coloring:
$\{x,v_2\}$, $\{v_0,q,v_3\}$, $\{y,z,p'\}$, $\{v_4,v_1,p'\}$

Therefore, $S_2(v_1,v_2)=S_2(v_3,v_4)=\emptyset$.
As $\{v_2,v_3\}$ is not a clique cutset,
$S_2(v_2,v_3)$ is not anti-complete to $\{y,z\}$.
By symmetry, we may assume that $py\in E$.
Let $C'=ypv_3v_4v_0$. $v_1\notin S'_3$.
Clearly, $x\in S'_3(v_0)$, $z\in S'_3(v_4)$ and $v_2\in S'_3(p)$.
If $S'_3=\{x,z,v_2\}$ then we are in the case {\bf $C$ is of type (1,0,1,1,0)}.
So, let $t\in S'_3\setminus \{x,z,v_2\}$.
Note that $t\notin S_2(v_0,v_1)$ as $S_2(v_0,v_1)$ is anti-complete to $\{v_3,v_4,p\}$.
If $t\in S_2(v_2,v_3)$, $t\in S'_3(p)$ and thus $G=G_{2,2}$.
So, $t\in S_2(v_0,v_4)$ and $t\in S'_3(v_0)$, namely $t$
is complete to $\{v_0,v_4,y\}$. Thus, $xt\in E$.
By (P9), $y$ is complete to $S_2$ and hence $2\le |S_2|\le 3$.
Note that $zt\notin E$ or $\{v_0,v_4,x,t,z\}$ would induce a $K_5$.
Now if $S_2(v_2,v_3)=\{p,p'\}$ then $p'\in S'_3(p)$ and $G=G_{2,2}$.
So, $S_2(v_2,v_3)=\{p\}$. If $S_2=\{p,t\}$, then $G$ has a 4-coloring $\phi$:
$\{x,v_3\}$, $\{v_0,v_2\}$, $\{p,t,z,v_1\}$, $\{v_4,y\}$.
If $S_2(v_0,v_4)=\{t,t'\}$ then $t'x\notin E$ or $\{v_0,v_4,x,t,t'\}$ would induce a $K_5$.
Then $G$ has a 4-coloring by adding $t'$ to $\{x,v_3\}$ in $\phi$. This completes the proof of Case 3.

Note that if $S_3(v_i)$ has two vertices then $S_3(v_i)$ is not complete to $S_3(v_{i+1})$ as $G$ is $K_5$-free.
Moreover if $S_3(v_{i+1})$ also has two vertices, then there is at most one edge
between $S_3(v_i)$ and $S_3(v_{i+1})$ as $G$ is $(K_5,C_4)$-free.

{\bf Case 4.} $|S_3|=4$. There are five possible configurations for $S_3$.

{\bf $C$ is of type (2,2,0,0,0)}.
Let $S_3(v_0)=\{x,x'\}$ and $S_3(v_1)=\{y,y'\}$. As $G$ is $(K_5,C_4)$-free, we may assume
that $x$ is anti-complete to $\{y,y'\}$ and $y$ is anti-complete to $\{x,x'\}$.
Let $C'=C\setminus \{v_0\}\cup \{x\}$. Note that $y,y'\notin S'_3$.
It is easy to check that
$S'_3\cap (S_1\cup S_2)\subseteq S'_3\cap (S_2(v_1,v_2)\cup S_2(v_3,v_4))$.
Hence, $|S'_3\cap (S_2(v_1,v_2)\cup S_2(v_3,v_4))|\ge 2$ by the minimality of $C$.
Suppose that  $p\in S'_3\cap S_2(v_3,v_4)$. Note that $px\in E$. Then as $x'xpv_3v_2y\neq P_6$, we have
$x'p\in E$. Further, $S'_3\cap S_2(v_3,v_4)$ is a clique and thus $|S'_3\cap S_2(v_3,v_4)|\le 1$
or $K_5$ would arise. Next we show that $|S'_3\cap S_2(v_1,v_2)|\le 1$. If not, let $p,p'$
be two vertices in $S'_3\cap S_2(v_1,v_2)$. Then $\{p,p',y,y',x,x',v_0,v_1\}$ contains a $W_5$.
Therefore, we may assume $q\in S_2(v_1,v_2)$ and $p\in S_2(v_3,v_4)$. Moreover, $x$ is complete to $\{p,q\}$
by definition. As shown above, we obtain that $x'p\in E$. So, $\{x,x'\}$ is complete to $S_2(v_1,v_2)\cup S_2(v_3,v_4)$
and $S_2(v_3,v_2)=\emptyset$ by (P9).
Thus, $S_2(v_1,v_2)=\{q\}$ and $S_2(v_1,v_2)=\{p\}$.
If $t\in S_1(v_3)$, then $tv_3v_4xv_1y=P_6$. So, $S_1(v_3)=\emptyset$ and now $N(v_3)=\{v_2,v_4,p\}$
which contradicts that $\delta(G)\ge 4$.

{\bf $C$ is of type (1,1,0,2,0)}. Let $S_3(v_3)=\{x,x'\}$, $S_3(v_0)=\{z\}$ and $S_3(v_1)=\{y\}$.
Note that $yz\notin E$ or $G=G_{2,2}$. Let $C'=C\setminus \{v_0\}\cup \{z\}$.
As $y\notin S'_3$ we have $S'_3\cap (S_2(v_0,v_1)\cup S_2(v_3,v_4))\neq \emptyset$.
Let $p$ be such a vertex. If $p\in S_2(v_3,v_4)$, then $p$ is not complete to $\{x,x'\}$,
say $xp\notin E$. Now $yv_1zpv_3x=P_6$. Therefore, $p\in S_2(v_1,v_2)$.
Let $C''=C\setminus \{v_1\}\cup \{y\}$. By symmetry, we obtain that there exists $q\in S''_3\cap S_2(v_0,v_4)$.
Note that $pz\in E$ and $qy\in E$ by definition of $p$ and $q$. Further, $qz\notin E$
or $qzv_1y=C_4$. If $xp\notin E$, then $xq\notin E$ by (P9) and thus
$qv_0zpv_2x=P_6$. Thus $xp\in E$ and now $zv_4xp=C_4$.

{\bf $C$ is of type (2,1,0,0,1)}.
Let $S_3(v_0)=\{x,x'\}$, $S_3(v_1)=\{z\}$ and $S_3(v_4)=\{y\}$.
As $G$ is $K_5$-free, each of $\{y,z\}$ is not complete to $\{x,x'\}$. We may assume
that $zx\notin E$. If $t\in S_1(v_2)$ then $tv_2v_1x(x')v_4y=P_6$. Thus, $S_1(v_2)=\emptyset$.
Similarly  $S_1(v_3)=\emptyset$. Let $C'=C\setminus \{v_1\}\cup \{z\}$.
By the minimality of $C$, we have $S'_3\cap (S_2(v_0,v_4)\cup S_2(v_2,v_3))\neq \emptyset$.
We first show that $S'_3\cap  S_2(v_2,v_3)=\emptyset$. Otherwise let $p\in S'_3\cap S_2(v_2,v_3)$.
Note that $pz\in E$, and $py\notin E$ or $pyv_0z=C_4$.
As $xv_1zpv_3y\neq P_6$, we have $yx\in E$ and so $x'y\notin E$.
Moreover, $x'z\in E$ since $x'v_1zpv_3y\neq P_6$, and so $yxx'z=P_4$.
Let $C''=C\setminus \{v_4\}\cup \{y\}$.
Then there exists $q\in S''_3\cap (S_2(v_0,v_1)\cup S_2(v_2,v_3))$.
It is clear that $qy\in E$ by definition of $q$. As $py\notin E$ and $qy\in E$, we have $q\in S_2(v_2,v_3)$
by (P9). Note that $p\neq q$. Also $qz\notin E$ or $qzv_0y=C_4$.
Then $pq\in E$ since $qyxx'zp\neq P_6$.
Let $C_x=C\setminus \{v_0\}\cup \{x\}$.
Then there exists $r\in S^x_3\cap (S_2(v_1,v_2)\cup S_2(v_3,v_4))$
by the minimality of $C$. If $r\in S_2(v_1,v_2)$, then $rxyq=C_4$. Thus, $r\in S_2(v_3,v_4)$.
Symmetrically considering $C_{x'}=C\setminus \{v_0\}\cup \{x'\}$ we obtain that there exists
$r'\in S_2(v_1,v_2)$. However, this contradicts (P9), since $xr\in E$.
Therefore, $S'_3\cap S_2(v_0,v_4)\neq \emptyset$. Symmetrically considering $C''=C\setminus \{v_4\}\cup \{y\}$
we can conclude that $S''_3\cap S_2(v_0,v_1)\neq \emptyset$. Hence, $S_2(v_2,v_3)=\emptyset$.
Since $d(v_2)\ge 4$ and $d(v_3)\ge 4$, we have $S_2(v_1,v_2)\neq \emptyset$ and $S_2(v_3,v_4)\neq \emptyset$.
This contradicts  (P8).

{\bf $C$ is of type (2,0,1,0,1)}.
Let $S_3(v_0)=\{x,x'\}$, $S_3(v_2)=\{z\}$ and $S_3(v_4)=\{y\}$.
We may assume that $xy\notin E$. If $t\in S_1(v_2)$ then $tv_2v_1xv_4y=P_6$. So, $S_1(v_2)=\emptyset$.
Let $C_x=C\setminus \{v_0\}\cup \{x\}$ and
$C_y=C\setminus \{v_4\}\cup \{y\}$. Then there exists $p\in S^y_3\cap (S_2(v_0,v_1)\cup S_2(v_2,v_3))$.
and $q\in S^x_3\cap (S_2(v_2,v_1)\cup S_2(v_4,v_3))$. Note that $py\in E$ and $qx\in E$ by definition.
We first claim that $S^y_3\cap S_2(v_3,v_2)=\emptyset$.
If not, suppose that $p\in S_2(v_2,v_3)$. Note that $pz\in E$ or $zv_2pyv_0x=P_6$.
If $q\in S_2(v_3,v_4)$, then $qy\in E$ or $yv_4qp=C_4$. But then $qyv_0x=C_4$.
So, $q\in S_2(v_1,v_2)$. Now $S_2(v_0,v_1)=S_2(v_3,v_4)=\emptyset$
by the fact that $py,qx\in E$ and (P9).
Moreover, $S_2(v_0,v_4)=\emptyset$. By Observation \ref{o5} (2), $z$ is complete to $S_2$
and hence $S_2(v_3,v_2)=\{p\}$ and $S_2(v_1,v_2)=\{q\}$.
Note that $S_1=\emptyset$ as $S_1(v_2)=\emptyset$.
Thus, $G$ has a 4-coloring: $\{v_1,v_4,p\}$, $\{v_0,v_3,q\}$, $\{v_2,x,y\}$, $\{x',z\}$.
Therefore, $p\in S_2(v_0,v_1)$. Suppose first that $q\in S_2(v_3,v_4)$.
Then $S_2(v_1,v_2)=S_2(v_2,v_3)=\emptyset$ by (P8).
Thus, $d(v_2)=3$, a contradiction. Hence, $q\in S_2(v_1,v_2)$. Note that $px,qy\notin E$.
If $x'y\in E$ then $px'\in E$ or $yx'v_1p=C_4$, and so $\{v_1,p,y,v_4,x\}\cup \{v_0\}$ induces a $W_5$.
So, $x'y\notin E$. Hence, $|S^y_3\cap S_2(v_0,v_1)|\ge 2$ by the minimality of $C$ and the above argument.
Let $p$ and $p'$ be two vertices in  $S^y_3\cap S_2(v_0,v_1)$, and then $\{p,p',x,x',y,v_0,v_1,v_4\}$
contains a $W_5$.

{\bf $C$ is of type (2,0,0,1,1)}.
Let $S_3(v_0)=\{x,x'\}$, $S_3(v_3)=\{z\}$ and $S_3(v_4)=\{y\}$.
We may assume that $xy\notin E$. If $t\in S_1(v_2)$ then $tv_2v_1xv_4y=P_6$. So $S_1(v_2)=\emptyset$.
Let $C_x=C\setminus \{v_0\}\cup \{x\}$ and
$C_y=C\setminus \{v_4\}\cup \{y\}$. Then there exists $q\in S^y_3\cap (S_2(v_0,v_1)\cup S_2(v_2,v_3))$.
and $p\in S^x_3\cap (S_2(v_2,v_1)\cup S_2(v_4,v_3))$ by minimality of $C$.
$px,qy\in E$ by definition of $p$ and $q$.
Suppose first that $p\in S_2(v_3,v_4)$. $py\notin E$ or $pyv_0x=C_4$.
If $q\in S_2(v_2,v_3)$ then $pqyv_4=C_4$. So $q\in S_2(v_0,v_1)$.
As $S_2(v_3,v_4)\neq \emptyset$ and $S_2(v_0,v_1)\neq \emptyset$, we have $S_2(v_1,v_2)=S_2(v_2,v_3)=\emptyset$.
Now $d(v_2)=3$ since $S_1(v_2)=\emptyset$, a contradiction.
Thus $p\in S_2(v_1,v_2)$.  $p$ is anti-complete to $\{y,z\}$ since $G$ is $C_4$-free. $zv_2pxv_0y$
implies that $yz\in E$. If $S^x_3\cap S_2=\{p\}$, then we are in the case $C$ is of type (2,0,1,0,1).
So we let $p'\in S^x_3\cap S_2(v_1,v_2)$ with $p'\neq p$. $p'x\in E$.
So $x'$ is not complete to $\{p,p'\}$, say $x'p\notin E$. $x'xpv_2v_3y$ implies that $x'y\in E$.
Now we consider $q$. If $q\in S_2(v_2,v_3)$ then $S_2(v_0,v_1)=S_2(v_3,v_4)=\emptyset$ by the
fact that $xp,qy\in E$ and (P9). Also, $S_2(v_0,v_4)=\emptyset$.
Now $S_2=\{p,p',q\}$ and $S_1=\emptyset$. $G$ has a 4-coloring:
$\{v_1,v_4,q\}$, $\{x,y,v_2\}$, $\{v_0,v_3,p'\}$, $\{z,p,x'\}$.
Thus $q\in S_2(v_0,v_1)$. Then $qx'\in E$ or $x'yqv_1=C_4$. But now $\{v_1,v_4,x,y,q\}\cup \{v_0\}$
induces a $W_5$.

{\bf $C$ is of type (1,1,1,1,0)}.
Let $S_3(v_0)=\{x\}$, $S_3(v_1)=\{y\}$, $S_3(v_2)=\{z\}$ and $S_3(v_3)=\{w\}$.
Note that $\{x,y,z,w\}$ does not induce a $P_4$ or $G=G_{P_4}$.
So, there are at most two edges in $\{x,y,z,w\}$.
We shall consider two subcases.

{\bf Case a.} There is at most one edge in $\{x,y,z,w\}$. Suppose that $yz\notin E$.
Without loss of generality, assume $xy\notin E$.
Let $C_y=C\setminus \{v_1\}\cup \{y\}$. As $x,z\notin S^y_3$
we have $|S^y_3\cap S_2|\ge 2$. If $|S^y_3\cap S_2(v_0,v_4)|\ge 2$ or
$|S^y_3\cap S_2(v_3,v_2)|\ge 2$, then $|S^y_3\cap S_2|\ge 3$ or we are in
one of previous four cases.
Thus, $S^y_3\cap S_2(v_3,v_2)\neq \emptyset$ and $S^y_3\cap S_2(v_0,v_4)\neq \emptyset$
or $K_5$ would arise. Also, $y$ is complete to $S_2(v_0,v_4)$ and $S_2(v_3,v_2)$
and hence $S_2=S_2(v_3,v_2)\cup S_2(v_0,v_4)$ by (P7) to (P9).
But now $C_z=C\setminus \{v_2\}\cup \{z\}$ has $|S^z_3|<4$, which contradicts the minimality of $C$.
So, it must be the case that $yz\in E$ and $xy,zw\notin E$. Consider $C_y$ and $C_z$ as above.
Let $p\in S^y_3\cap (S_2(v_0,v_4)\cup S_2(v_2,v_3))$ by the minimality of $C$.
Suppose that $S^y_3\cap S_2(v_0,v_4)=\emptyset$.
Then $p\in S_2(v_2,v_3)$. Note that $py\in E$ by definition of $p$, and $pz\notin E$ or $pyzv_3=C_4$.
So $S^y_3\cap S_2(v_2,v_3)=\{p\}$ or $K_5$ would arise.
Now $|S^y_3|=4$ and we are in one of previous four cases.
So, we may assume that $p\in S_2(v_0,v_4)$.
By symmetry, there exists a vertex $q\in S^z_3\cap S_2(v_4,v_3)$.
by definition of $p$ and $q$, $py, qz\in E$. Now $pyzq=C_4$.

{\bf Case b.} There are two edges in $\{x,y,z,w\}$.
Suppose first that $xy,wz\in E$ but $yz\notin E$.
Define $C_y$ and $C_z$ as above. As $y\notin S^z_3$ and $z\notin S^y_3$,
we have $S^y_3\cap S_2\neq \emptyset$ and $S^z_3\cap S_2\neq \emptyset$.
We claim that $S^y_3\cap S_2(v_2,v_3)\neq \emptyset$.
Otherwise, let $p\in S^y_3\cap S_2(v_0,v_4)$. Note that $py\in E$, and $px\in E$ or $v_4pyx=C_4$.
Also, $S^y_3\cap S_2(v_0,v_4)$ is a clique
and hence $S^y_3\cap S_2(v_0,v_4)=\{p\}$ or $K_5$ would arise.
Now $S^y_3=\{x,p,v_1,w\}$ with $x,p\in S^y_3(v_0)$ and so we are in one of four previous cases.
Hence, the claim holds. Similarly, $S^z_3\cap S_2(v_0,v_1)\neq \emptyset$.
Let $p\in S^y_3\cap S_2(v_2,v_3)$ and $q\in S^z_3\cap S_2(v_0,v_1)$.
Note that $py,qz\in E$. Also, $qy\notin E$ or $qyv_2z=C_4$.
As $yxv_4wzq\neq P_6$, we have $qx\in E$. Also, $qy\notin E$ or $\{v_0,v_1,q,x,y\}$
would induce a $K_5$. Then $\{v_2,z,q,x,y\}\cup \{v_1\}$ induces a $W_5$.

Now we consider the case $xy,yz\in E$ but $zw\notin E$.
Let $C_z=C\setminus \{v_2\}\cup \{z\}$ and $C_w=C\setminus \{v_3\}\cup \{w\}$.
As $zw\notin E$, we have that $S^z_3\cap S_2\neq \emptyset$ and $S^w_3\cap S_2\neq \emptyset$.
We claim that $S^z_3\cap S_2(v_3,v_4)\neq \emptyset$. If not, there exists $p\in S^z_3\cap S_2(v_0,v_1)$.
Note that $pz\in E$, and so $py\in E$ or $v_0yzp=C_4$. So, $S^z_3\cap S_2(v_0,v_1)=\{p\}$ or $K_5$ would arise.
Hence, $S^z_3=\{x,y,v_2,p\}$ with $y,p\in S^z_3(v_1)$ and we are in one of previous four cases.
So, the claim holds and let $p\in S^z_3\cap S_2(v_3,v_4)$. Note that $pz\in E$ and $py\notin E$.
Also, $pw\notin E$ or $pwv_2z=C_4$, and $px\notin E$ or $pxv_1z=C_4$.
Let $q\in S^w_3\cap S_2$. $qw\in E$. If $q\in S_2(v_0,v_4)$, then $pq\in E$ and thus $wv_3pq=C_4$.
So, $q\in S_2(v_1,v_2)$. Also, $qz\notin E$ or $qzv_3w=C_4$, and $qx\notin E$ or $qxv_4w=C_4$.
Hence, $x$ is anti-complete to $S_2(v_1,v_2)\cup S_2(v_3,v_4)$.
As $qwv_4xyz\neq P_6$, we have $qy\in E$. Note that $S_2(v_3,v_2)=\emptyset$ by the fact $wp\notin E$
and Observation \ref{o5} (2). Thus, $S_2=S_2(v_1,v_2)\cup S_2(v_3,v_4)$.
So, $S_1=S_1(v_1)\cup S_1(v_4)$.
Now consider $C^*=xyzpv_4$.
Note that $v_0\in S^*_3(x)$, $v_1\in S^*_3(y)$, and $v_3\in S^*_3(p)$.
but $v_2,q,w\notin S^*_3$. By the minimality of $C$, we have $S^*_3\cap (S_1\cup S_2)\neq \emptyset$.
Let $r\in S^*_3\cap (S_1\cup S_2)$. If $r\in S_2(v_3,v_4)$, then $r$ must be in $S^*_3(p)$
as $r$ is anti-complete $\{x,y\}$. Thus $G=G_{2,2}$.
Moreover, any vertex $t\in S_2(v_1,v_2)$ is anti-complete to $\{x,v_4,p\}$, and any vertex $t\in S_1(v_4)$
is anti-complete to $\{p,y,z\}$. Therefore, $r\in S_1(v_1)$.
If $r$ is complete to $\{x,y,z\}$, then there exists
$r'\in S^*_3\cap S_1(v_1)$ with $r'\neq r$ otherwise $|S^*_3|=4$ and we are in one of four pervious cases.
Note that $r'$ must be complete to $\{p,y,z\}$. Hence, in any case there exists
a vertex $r\in S_1(v_1)$ that is complete to $\{p,y,z\}$ but $px\notin E$.
Now $wv_3prv_1x=P_6$.


{\bf Case 5.} $|S_3|=5$. There are five possible configurations for $S_3$.

{\bf $C$ is of type (2,2,0,0,1)}. $S_3(v_0)=\{x,x'\}$, $S_3(v_1)=\{y,y'\}$, $S_3(v_4)=\{w\}$.
We may assume that $y$ is anti-complete to $\{x,x'\}$ and $x$ is anti-complete to $\{y,y'\}$.
If $t\in S_1(v_3)$ then $tv_3v_4xv_1y=P_6$. So, $S_1(v_3)=\emptyset$.
Let $C'=C\setminus \{v_1,v_4\}\cup \{y,w\}$ be an induced $C_5$.
Let $p\in S'_3\cap S_2$ by $x\notin S'_3$ and the minimality of $C$.
Suppose first that $p\in S_2(v_0,v_1)$. Note that $S_2(v_3,v_4)\cup S_2(v_2,v_3)\neq \emptyset$
by $d(v_3)\ge 4$. Let $q\in S_2(v_3,v_4)\cup S_2(v_2,v_3)$.
Without loss of generality,
we assume that $q\in S_2(v_2,v_3)$.
Since $qv_3v_4xv_1y(y')\neq P_6$, we have $q$ is complete to $\{y,y'\}$.
As $q$ is an arbitrary vertex in $S_2(v_3,v_4)$,
we have $S_(v_2,v_3)$ is complete to $\{y,y'\}$ and so $S_2(v_2,v_3)=\{q\}$. Note that $S_2(v_3,v_2)=\emptyset$
by (P8) and so $N(v_3)=\{v_2,v_4,q,w\}$. Now as {\bf $G$ is a minimal obstruction},
$G-v_3$ has a 4-coloring $\phi$. Note that $\phi(q)=\phi(v_1)=\phi(v_4)$ and therefore we can extend
$\phi$ to $G$, a contradiction. As $x,x'\notin S'_3$, there exists two different vertices $p$ and $q$
in $S'_3\cap S_2$. If $p,q\in S_2(v_0,v_4)$, then $\{p,q,v_0,v_4,w\}$ induces a $K_5$.
Note that $S_2(v_2,v_3)$ is complete to $\{y,y'\}$ and $S_2(v_2,v_3)$ contains
at most one vertex.  Hence, we may assume that $p\in S_2(v_0,v_4)$ and $S_2(v_2,v_3)=\{q\}$.
By the fact that $yq\in E$ and (P10), we have $S_2(v_3,v_4)=\emptyset$.
Hence, we derive a similar contradiction as above.

{\bf $C$ is of type (0,1,0,2,2)}. $S_3(v_3)=\{x,x'\}$, $S_3(v_4)=\{y,y'\}$, $S_3(v_1)=\{w\}$.
We may assume that $y$ is anti-complete to $\{x,x'\}$ and $x$ is anti-complete to $\{y,y'\}$.
Let $C'=C\setminus \{v_1,v_4\}\cup \{y,w\}$ be an induced $C_5$. Then $x,x'\notin S'_3$
and hence $S'_3\cap (S_1\cup S_2)$ contains at least two vertices.
Let $p$ and $q$ be such two vertices. Let $t\in S_1(v_0)$.  If $t$ is not anti-complete to $\{y,y'\}$, say
$ty\in E$, then $tyv_4xv_2v_1=P_6$. Thus $p,q\notin S_1(v_0)$. Now suppose that $q\in S_2(v_0,v_4)$.
Then $q$ is complete to $\{w,y\}$. Note that $qy'\notin E$. Then the fact that $y'yqwv_2x\neq P_6$
implies that $qx\in E$ and thus $qxv_2w=C_4$.
Thus, $p,q\notin S_2(v_0,v_4)$. If $p,q\in S_2(v_0,v_1)$, then $\{p,q,v_0,v_1,w\}$
would induce a $K_5$. Now let $p,q\in S_2(v_2,v_3)$. If $\{p,q\}$ is complete to $y$ or $w$, then $G=G_{3,1}$
otherwise $G$ would contain an induced $W_5$. Hence, we may assume that $py\in E$ and $qw\in E$.
Thus $pw,qy\notin E$. By (P10), $S_2(v_0,v_1)=S_2(v_0,v_4)=\emptyset$.
Also, $S_1(v_1)=\emptyset$ since $S_2(v_3,v_2)\neq \emptyset$.
By $d(v_1)\ge 4$ we have $S_2(v_1,v_2)\neq \emptyset$ and thus
$S_2(v_2,v_3)=\{p,q\}$. Now let $C_y=C\setminus \{v_4\}\cup \{y\}$. Then $|S^y_3\cap S_2(v_2,v_3)|\ge 2$.
But this is impossible since $qy\notin E$. Therefore, $p\in S_2(v_0,v_1)$ and $q\in S_2(v_2,v_3)$.
By definition of $q$, $py\in E$ and hence $S_2(v_1,v_2)=\emptyset$ by (P10).
Moreover, $S_2(v_4,v_0)=\emptyset$  by (P7).
Now consider $C_x=C\setminus \{v_3\}\cup \{x\}$ and thus $|S^x_3|<5$ contradicting the minimality of $C$.

{\bf $C$ is of type (2,1,1,0,1)}. $S_3(v_0)=\{x,x'\}$, $S_3(v_1)=\{y\}$, $S_3(v_2)=\{z\}$, $S_3(v_4)=\{w\}$.
If $yz\in E$ and one of $\{x,x'\}$ is complete to $\{y,w\}$, then $G=G_{P_4}$. Hence, either
$yz\notin E$ or no vertex in $\{x,x'\}$ is complete to $\{y,w\}$.
Let $C'=C\setminus \{v_1,v_4\}\cup \{y,w\}$ be an induced $C_5$.
Thus $|S'_3\cap (S_1\cup S_2)|\ge 2$. Let $p,q\in S'_3$.
Note that $p,q\in S_2(v_0,v_1)\cup S_2(v_0,v_4)\cup S_2(v_2,v_3)$.
If $\{p,q\}\subseteq S_2(v_0,v_1)$ or $\{p,q\}\subseteq S_2(v_0,v_4)$, then $K_5$ would arise.
Next we show that $\{p,q\}\nsubseteq S_2(v_2,v_3)$. If not, then both $p$ and $q$ are adjacent to
exactly one of $\{y,w\}$. If $pw\in E$, then the fact that $zv_2pwv_0x(x')\neq P_6$ implies that $pz\in E$.
We may assume that $xy\notin E$. If $py\in E$, Since $wv_3pyv_1x\neq P_6$, we have $wx\in E$.
Thus, $x'w\notin E$. As $wv_3pyv_1x'\neq P_6$, we have $x'y\in E$. Now $zy\in E$ since $wxx'yv_2z\neq P_6$.
Hence, $pz\in E$ or $ypv_3z=C_4$. We have showed if $p\in S_2(v_3,v_2)$ then $pz\in E$.
Therefore, $pq\notin E$ or $\{p,q,v_2,v_3,z\}$ would induce a $K_5$.
Further, $y$ or $w$ cannot be complete to $\{p,q\}$.
Thus, we may assume that $py\in E$ and $qz\in E$.
By previous argument we have that $\{y,x,x',w\}$ induces a $P_4$ and hence $qwxx'yp=P_6$.

Therefore, three cases remain. If $p\in S_2(v_0,v_1)$ and $q\in S_2(v_0,v_4)$, then $pq\in E$ by (P1) to (P3).
By Observation \ref{o5} (2), we have $\{x,x'\}$ is complete to $\{p,q\}$ and thus $\{x,x',v_0,p,q\}$
induces a $K_5$. If $p\in S_2(v_0,v_4)$ and $q\in S_2(v_2,v_3)$, then $p$ is complete to $\{y,w\}$
by definition. By (P9), we have $yq\in E$ and $wq\notin E$.
We may assume that $xy\notin E$. Thus $wxx'y=P_4$ as shown above.
Also $px\notin E$ or $pxv_1y=C_4$ and hence $px'\notin E$ or $px'xw=C_4$.
Now we have $wxx'yp$ is an induced $C_5$ with $v_0$ being a 5-vertex.
Finally, let $p\in S_2(v_0,v_1)$ and $q\in S_2(v_2,v_3)$. By definition,
$p$ is complete to $\{y,w\}$. By (P9), $qw\in E$ and $qy\notin E$.
Moreover, $qz\in E$, and $pz\in E$ or $zqwp=C_4$.
If $x$ is complete to $\{y,w\}$, then $px\in E$ or $xwpy=C_4$ and thus
$\{x,y,v_0,v_1,p\}$ would induce a $K_5$. Hence, none of $\{x,x'\}$ is complete to
$\{y,w\}$. Therefore, $yz\in E$ or $|S'_3\cap (S_1\cup S_2)|\ge 3$, which is impossible
by previous argument. Now $\{p,v_1,v_2,q,w,v_0,y,z,v_3\}$ induces a $G_{P_4}$
with respect to $C^*=wv_0yzv_3$ and $S^*_3=\{q,v_2,v_1,p\}$ for which $qv_2v_1p$
induces a $P_4$.

{\bf $C$ is of type (1,1,2,0,1)}. $S_3(v_0)=\{x\}$, $S_3(v_1)=\{y\}$, $S_3(v_2)=\{z,z'\}$, $S_3(v_4)=\{w\}$.
Note that $xw\notin E$ or $G=G_{2,2}$. We may assume that $yz\notin E$.
Let $C'=C\setminus \{v_1,v_4\}\cup \{y,w\}$.
Hence, $|S'_3\cap (S_1\cup S_2)|\ge 2$. Let
$p,q\in S'_3\cap (S_1\cup S_2)$. If $p\in S_1(v_0)$, then $p$ is complete to $\{x,y,w\}$
and hence $pxv_4w=C_4$. If $|S'_3\cap (S_1\cup S_2)|\ge 2$ or $|S'_3\cap (S_1\cup S_2)|\ge 2$
then $K_5$ would arise. Next we show that $\{p,q\}\nsubseteq S_2(v_2,v_3)$. If not, let $q,p\in S_2(v_2,v_3)$.
Note that $p$ is not complete to $\{z,z'\}$, say $zp\notin E$. If $pw\in E$ then $zv_2pwv_0x=P_6$.
Hence, $y$ is complete to $\{p,q\}$. But now $\{v_1,v_2,v_3,y,z,z',p,q\}$ contains an induced $W_5$.
Therefore, three cases remains. If $p\in S_2(v_0,v_1)$ and $q\in S_2(v_0,v_4)$, then $x$ is complete to
$\{p,q\}$ by Observation \ref{o5} (2). By definition of $p$, we have $pw\in E$ and thus $pxv_4w=C_4$.
If $p\in S_2(v_0,v_1)$ and $q\in S_2(v_2,v_3)$, then $wp\in E$. By (P9),
we have $wq\in E$. Now $zv_2qwv_0x=P_6$ or $z'v_2qwv_0x=P_6$.
Finally, $p\in S_2(v_0,v_4)$ and $q\in S_2(v_2,v_3)$.
By definition of $p$, we have $py\in E$ and hence $qy\in E$ by (P9).
We may assume that $qz\notin E$. Then $zy\notin E$ or $v_3qyz=C_4$.
Thus the fact that  $zv_3qyv_0x\neq P_6$ implies that $xy\in E$ and so $xp\in E$ or $v_4xyp=C_4$.
Now $qv_3wpxv_1=P_6$.

{\bf $C$ is of type (1,1,1,1,1)}.
Let $S_3(v_i)=\{u_i\}$ for each $i$. Note that there are at most 3 edges
within $S_3$ or $G=G_{P_4}$. We consider the following three cases.

{\bf Case a.} $S_3$ has at most two edges and does not induce a $P_3$.  Without loss of generality, we may assume
that $u_0u_1,u_1u_2,u_3u_4\notin E$. Let $C'=C\setminus \{v_1,v_4\}\cup \{u_1,u_4\}$.
Note that $u_0,u_2,u_3\notin S'_3$ and hence $|S'_3\cap (S_1\cup S_2)|\ge 3$ by the minimality of $C$.
Let $p\in S'_3\cap (S_1\cup S_2)$. If $p\in S_1(v_1)$, then $pu_0\in E$ by properties (P11) and (P12).
and thus $pu_0v_1u_1=C_4$. Hence, $S'_3\cap S_1=\emptyset$.
If $|S'_3\cap S_2(v_0,v_1)|\ge 2$ or  $|S'_3\cap S_2(v_0,v_1)|\ge 2$, then $K_5$ would occur.
Now suppose that $p,q,r\in S'_3\cap S_2(v_2,v_3)$. If $u_1$ or $u_4$ is complete to
$\{p,q,r\}$, then $K_5$ would occur. So, we may assume that $pu_1,qu_1\in E$ and $ru_4\in E$.
Since $rv_3pu_1v_0u_0\neq P_6$,  we have $rp\in E$. Replacing $q$ with $p$ we have $rq\in E$ and so
$\{v_2,v_3,p,q,r\}$ induces a $K_5$.
By (P8), we have that $|S'_3\cap S_2(v_2,v_3)|=2$
and $S'_3\cap (S_2(v_0,v_1)\cup S_2(v_0,v_4))\neq \emptyset$.
Suppose that $p\in S_2(v_0,v_1)$. We repeat the argument for $C''=C\setminus \{v_1,v_3\}\cup \{u_1,u_3\}$
and obtain that $S_2(v_0,v_4)\neq \emptyset$.  This contradicts (P8).
Hence, let $p\in S_2(v_0,v_4)$ and $q,r\in S_2(v_2,v_3)$. Note that $pu_1\in E$ by definition of $p$
and hence $u_1$ is complete to $\{p,q,r\}$. So, $S_2(v_2,v_3)=\{q,r\}$ and $S_2(v_0,v_4)=\{r\}$.
But this contradicts the fact that $|S''_3\cap S_2(v_0,v_4)|\ge 2$.

{\bf Case b.} $S_3$ does induce a $P_3$. Without loss of generality,
we assume that $u_4u_0,u_0u_1\in E$.
Let $C_1=C\setminus \{v_0,v_2\}\cup \{u_0,u_2\}$. Note that $S^1_3\cap S_1=\emptyset$.
Since $u_1,u_3\notin S^1_3$, we have $|S^1_3\cap S_2|\ge 2$ by the minimality of $C$.
If $|S^1_3\cap S_2(v_0,v_1)|\ge 2$ or $|S^1_3\cap S_2(v_1,v_2)|\ge 2$, then $K_5$ would arise.
If $p\in S^1_3\cap S_2(v_0,v_1)$ and $q\in S^1_3\cap S_2(v_1,v_2)$, then $u_1$
is complete to $\{p,q\}$ by Observation \ref{o5} (2). Also, $pu_0\in E$
by definition of $p$ and thus $\{u_0,u_1,v_0,v_1,p\}$ induces a $K_5$.
Therefore, $S_2(v_3,v_4)\neq \emptyset$.
Now we repeat the argument for $C_4=C\setminus \{v_0,v_3\}\cup \{u_0,u_3\}$
and obtain that $S_2(v_1,v_2)\neq \emptyset$. So, $S_2(v_4,v_0)=S_2(v_0,v_1)=\emptyset$.
Let $p\in S^1_3\cap S_2(v_3,v_4)$ and $q\in S^4_3\cap S_2(v_1,v_2)$.
Let $C_2=C\setminus \{v_1,v_3\}\cup \{u_1,u_3\}$ and $C_3=C\setminus \{v_2,v_4\}\cup \{u_2,u_4\}$.
Note that $|S^2_3\cap S_2|\ge 2$ and $|S^2_3\cap S_2|\ge 2$. Since $S_2(v_0,v_4)=\emptyset$
and $|S^2_3\cap S_2(v_2,v_1)|\le 1$, $S^2_3\cap S_2(v_2,v_3)\neq \emptyset$.
Let $r\in S^2_3\cap S_2(v_2,v_3)$. By definition of $r$, we have $r$ is complete to $\{u_1,u_3\}$.
Similarly, $S^3_3\cap S_2(v_2,v_3)\neq \emptyset$. If $r\in S^3_3\cap S_2(v_2,v_3)$,
then $r$ is complete to $\{u_2,u_4\}$. So, $u_0u_1ru_4=C_4$. Hence, there exists $r'\neq r$
such that $r'\in S^3_3\cap S_2(v_2,v_3)$. Thus, $S_2(v_3,v_4)=\{p\}$, $S_2(v_3,v_2)=\{r,r'\}$,
and $S_2(v_2,v_1)=\{q\}$. Now $p\in S^3_3$ and $q\in S^2_3$, i.e., $p$ (respectively $q$)
is complete to $\{u_2,u_4\}$ (respectively $\{u_1,u_3\}$).
By the fact that $ru_3\in E$ and Observation \ref{o5} (2),
we have $u_3$ is complete to $\{p,r,r'\}$ and thus $\{u_3,v_3,p,r,r'\}$ induces a $K_5$.

{\bf Case c.} $S_3$ is isomorphic to $P_3+P_2$.
Without loss of generality, we assume that $u_0u_1,u_1u_2,u_3u_4\in E$.
Let $C_i=C\setminus \{v_i\}\cup \{u_i\}$ for each $i$.
By the minimality of $C$, we have $S^i_3\cap S_2\neq \emptyset$ for each $i\neq 1$.
Let $r\in S^3_3$ and $s\in S^4_3$. If $r\in S_2(v_1,v_2)$ and $s\in S_2(v_0,v_1)$,
then $u_4sru_3=C_4$. If $r\in S_2(v_0,v_4)$ and $s\in S_2(v_2,v_3)$, let
$t\in S^2_3\cap S_2$. Note that $t\in S_2(v_3,v_4)$. By Observation \ref{o5} (2),
we have $t$ is complete to $\{u_3,u_4\}$ and so $\{u_3,u_4,v_3,v_4,t\}=K_5$.
The remaining two cases are symmetric and we may assume that
$r\in S_2(v_0,v_4)$ and $s\in S_2(v_0,v_1)$. Let $t\in S^0_3\cap S_2$.
If $t\in S_2(v_1,v_2)$, then $s$ is complete to $\{u_0,u_1\}$ by Observation \ref{o5} (2).
Hence, $\{u_0,u_1,v_0,v_1,s\}=K_5$. So, $t\in S_2(v_3,v_4)$. Then $u_4$ is complete to $\{r,t\}$.
Since $G$ is $K_5$-free, $tu_3\in E$ and thus $tru_3v_3=C_4$.

{\bf Case 6.} $|S_3|=6$. There are three possible configurations for $S_3$.

{\bf $C$ is of type (2,1,1,1,1)}. Let $S_3(v_0)=\{x,x'\}$, $S_3(v_1)=\{y\}$, $S_3(v_2)=\{r\}$,
$S_3(v_3)=\{t\}$, $S_3(v_4)=\{z\}$. We may assume that $xy\notin E$.
We also assume that $rt\notin E$ or $G=G_{2,2}$.
Let $C'=C\setminus \{v_1,v_4\}\cup \{y,z\}$ be an induced $C_5$.
As $xy\notin E$, we have $S'_3\cap (S_1\cup S_2)\neq \emptyset$ by the minimality of $C$.
Let $p\in S'_3$. Then $p$ is complete to $\{y,z\}$.
It is easy to check that $p\in (N(v_0)\cap (S_1\cup S_2))\cup S_2(v_2,v_3)$.
If $p\in S_1(v_0)$, then $px\in E$ and so $pxv_1y=C_4$.
If $p\in S_1(v_0,v_1)$, then the fact that $pyv_2v_3v_4x\neq P_6$ implies that $px\in E$.
Thus $xz\in E$ or $pxv_4z=C_4$. Hence, $x'z\notin E$ and thus $x'p\notin E$.
So, $x'\notin S'_3$. By symmetry, if $p\in S_2(v_0,v_4)$, then $x'\notin S'_3$.
If $p\in S_2(v_2,v_3)$, then by symmetry we assume that $py\in E$.
Since $tv_3pyv_0x$ does not induce a $P_6$, we have $tp\in E$. Therefore $p$ is the only vertex
in $S_2(v_2,v_3)$ that is adjacent to $y$ otherwise $K_5$ would occur.
Thus there is also at most one vertex in $S_2(v_2,v_3)$ that is adjacent to $z$.
Also, $ry\notin E$ otherwise $tv_3ryv_0x=P_6$. By symmetry, $zt\notin E$.
Hence, $|S'_3\cap (S_1\cup S_2)|\ge 3$ and $|S'_3\cap (S_1\cup S_2)|\ge 4$
if $S'_3\cap (S_2(v_0,v_4)\cup S_2(v_0,v_1))\neq \emptyset$ by the minimality of $C$.
But now we either have a $K_5$ or contradicts (P9).

{\bf $C$ is of type (2,2,0,1,1)}. Let $S_3(v_0)=\{x,x'\}$, $S_3(v_1)=\{y,y'\}$,
$S_3(v_3)=\{t\}$, $S_3(v_4)=\{w\}$. Note that $wt\notin E$ or $G=G_{2,2}$.
We may assume that $y$ is anti-complete to $\{x,x'\}$.
Let $C'=C\setminus \{v_1,v_4\}\cup \{w,y\}$.
Thus by the minimality of $C$ we have $S'_3\cap (S_1\cup S_2)\neq \emptyset$.
Let $p\in S'_3$ and it is easy to check that $p\in S_2(v_0,v_1)\cup S_2(v_0,v_4)\cup S_2(v_2,v_3)$.
Suppose first that $p\in S_2(v_0,v_1)$. Then $p$ is complete to $\{y,w\}$. Note that $tp\notin E$ or
$tpv_0v_4=C_4$. Since $tv_4wpv_1y'\neq P_6$, we have $py'\in E$ and thus $\{v_0,v_1,y,y',p\}$ induces a $K_5$.
Suppose now that $p\in S_2(v_0,v_4)$. Again, $p$ is complete to $\{y,w\}$. Note that $tp\notin E$ or $tpyv_2=C_4$.
Then the fact that $tv_3wpyy'\neq P_6$ implies that $py'\in E$.
Now $\{x,x',y,y,v_0,v_1,v_4,p\}$ induces a Hajos graph with one additional dominating vertex.
Finally, assume that $p\in S_2(v_2,v_3)$. Then $p$ is adjacent to exactly one of $\{y,w\}$.
Suppose that $pw\in E$. Then $tp\notin E$ or $v_4wpt=C_4$. By $G$ is $K_5$-free, $w$ is not complete to $\{x,x'\}$,
say $wx\notin E$, and hence $xp\notin E$ or $xv_4wp=C_4$. Now $tv_2pwv_0x=P_6$.
Therefore, $py\in E$ and $pw\notin E$. We may assume that $xw\notin E$, and now $pv_2v_1xv_4w=P_6$.

{\bf $C$ is of type (2,2,,1,0,1)}. Let $S_3(v_0)=\{x,x'\}$, $S_3(v_1)=\{y,y'\}$,
$S_3(v_2)=\{t\}$, $S_3(v_4)=\{w\}$. We may assume that $y$ is anti-complete to $\{x,x'\}$
and $x$ is anti-complete to $\{y,y'\}$.
Let $C'=C\setminus \{v_1,v_4\}\cup \{y,w\}$.
By the minimality of $C$, we have $S'_3\cap (S_1\cup S_2)\neq \emptyset$.
Let $p\in S'_3$ and it is easy to check that $p\in S_2(v_0,v_1)\cup S_2(v_0,v_4)\cup S_2(v_2,v_3)$.
Suppose first that $p\in S_2(v_0,v_1)$. Then $p$ is complete to $\{y,w\}$.
Now Since $pyv_2v_3v_4x(x')\neq P_6$, we have $p$ is complete to $\{x,x'\}$ and thus $\{v_0,v_1,p,x,x'\}$
induces a $K_5$. Now suppose that $p\in S_2(v_0,v_4)$. Again, $p$ is complete to $\{y,w\}$.
Note that $tp\notin E$ or $pv_0v_1t=C_4$. If $py'\in E$, then $xp\in E$ or $xv_1y'pv_4$ and $v_0$
would induce a $W_5$. But now $xpy'v_1=C_4$. Hence, $py'\notin E$. Now the fact that $tv_3v_4pyy'\neq P_6$
implies that $ty\in E$ or $ty'\in E$. If $ty\in E$, then $y'ytv_3v_4x=P_6$. Otherwise,
$ty'\in E$. Then $px\notin E$ or $pxv_1y=C_4$.  Now $ty'ypv_4x=P_6$.
Thus there exist vertices $p,p'\in S'_3\cap S_2(v_2,v_3)$. Suppose that $py\in E$ and so $pw\in E$.
Since $xv_4v_3pyy'\neq P_6$, we have $py'\in E$. Note that $t$ is not complete to $\{y,y'\}$.
Thus, $tp\in E$ as $tv_3py(y')v_0x\neq P_6$ . Now $py(y')v_1t=C_4$.
Hence, $w$ is complete to $\{p,p'\}$. Now $\{y,y',v_0,v_1,v_2,p,p'v_3,w\}$
induces a $G_{3,1}$.

{\bf Case 7.} $|S_3|=7$. Suppose that
$S_3(v_0)=\{x\}$, $S_3(v_1)=\{y\}$ , $S_3(v_4)=\{z\}$,
$S_3(v_2)=\{r,r'\}$ and $S_3(v_3)=\{t,t'\}$.
We may assume that $r$ is anti-complete to $\{t,t'\}$
and $t$ is anti-complete to $\{r,r'\}$ or $K_5$ would occur.
Let $C_r=C\setminus \{v_2\}\cup \{r\}$.
Since $t,t'\notin S^r_3$ we have $|S^r_3\cap (S_1\cup S_2)|\ge 2$ by minimality of $C$.
Let $p$ and $p'$ be two vertices in the $S^r_3\cap (S_1\cup S_2)$.
Then $r$ is complete to $\{p,p'\}$. It is routine to check that $p$ and $p'$
belong to $S_2(v_0,v_1)\cup S_2(v_3,v_4)$. First suppose that $\{p,p'\}\subseteq S_2(v_0,v_1)$.
Let $C_t=C\setminus \{v_3\}\cup \{t\}$. Then there exist $q$ and $q'$ such that $q$ and $q'$ belong to
$S_2(v_4,v_0)$ or $S_2(v_1,v_2)$. If $\{q,q'\}\subseteq S_2(v_4,v_0)$ or $\{q,q'\}\subseteq S_2(v_1,v_2)$
then $\{p,p'q,q',v_1\}$ would induce a $K_5$.
Hence there must be the case that $q\in S_2(v_4,v_0)$ and $q'\in S_2(v_1,v_2)$.
By definition of $q$ and $q'$, $t$ is complete to $\{q,q'\}$, which contradicts (P10).
Therefore, $S_2(v_3,v_4)\neq \emptyset$. Repeating the argument for $C_t$ we have $S_2(v_1,v_2)\neq \emptyset$.
So, $S_2(v_0,v_4)=S_2(v_0,v_1)=\emptyset$ and thus $p,p'\in S_2(v_3,v_4)$ and $q,q'\in S_2(v_1,v_2)$.
Now Let $C_y=C\setminus \{v_1\}\cup \{y\}$ and $C_z=C\setminus \{v_4\}\cup \{z\}$. The same argument shows that
$S^y_3\cap S_2(v_2,v_3)\neq \emptyset$ and $S^z_3\cap S_2(v_2,v_3)\neq \emptyset$. As $|S_2(v_1,v_2)|\ge 2$,
we obtain that $S_2(v_2,v_3)$ contains only one vertex $u$. Thus $uy\in E$ and $uz\in E$. But now $uyv_0z=C_4$.

This completes the proof. \qed


\begin{thebibliography}{99}


%

\bibitem{Bondy} Bondy, J.A., Murty, U.S.R.: Graph Theory. In: Springer Graduate Texts in Mathematics,
vol. 244 (2008).

\bibitem{Brandstadt} Brandst\"{a}dt, A.,  Ho\`{a}ng, C. T.:
On clique separators, nearly chordal graphs, and the Maximum Weight Stable Set Problem.
Theoretical Computer Science 389, 295--306 (2007).

\bibitem{Fomin} Broersma, H.J., Fomin, F.V., Golovach, P.A., Paulusma, D.:
Three complexity results on coloring $P_k$-free graphs.
European Journal of Combinatorics, 2012 (in press).


\bibitem{Broersma} Broersma, H.J., Golovach, P.A., Paulusma, D., Song, J.: Updating the complexity
status of coloring graphs without a fixed induced learn forest. Theoret. Comput. Sci. 414,
9--19 (2012).

\bibitem{Bruce}  Bruce, D., Ho\`{a}ng, C. T., Sawada, J.:
A certifying algorithm for 3-colorability of $P_5$-free graphs.
ISAAC 2009, LNCS 5878, pp. 594--604, 2009.

\bibitem{P_7 1} Chudnovsky, M., Maceli, P., Zhong, M.:
Three-coloring graphs with no induced six-edge path I: the triangle-free case. In preparation.

\bibitem{P_7 2} Chudnovsky, M., Maceli, P., Zhong, M.:
Three-coloring graphs with no induced six-edge path II: using a triangle. In preparation.


\bibitem{SPGT} Chudnovsky, M., Robertson, N., Seymour, P., Thomas, R.:
The strong perfect graph theorem. Annals of Mathematics 64, 51--229 (2006).



\bibitem{Dabrow} Dabrowski, K., Golovach, P., Paulusma, D.:
Colouring of graphs with Ramsey-type forbidden subgraphs, submitted.

\bibitem{erdos} Erd\H{o}s, P.:
Graph theory and probability II,
Canad. J. Math. 13, 346--352 (1961).


\bibitem{Garey} Garey, M.R., Johnson, D.S.: Computers and Intractability:
A Guide to the Theory of NP-Completeness. Freeman San Faranciso, (1979).



\bibitem{Short Cycle} Golovach, P.A., Paulusma, D., Song, J.:
Coloring graphs without short cycles and long induced paths.
http://www.dur.ac.uk/daniel.paulusma/Papers/Submitted/girth.pdf, 2013.


\bibitem{Golumbic} Golumbic, M. C., Algorithmic graph theory and perfect graphs. San Diego, 1980.

\bibitem{Lovasz} Gr\"{o}tschel, M., Lov\'{a}sz, L., Schrijver, A.: Polynomial algorithms for perfect
graphs. Ann. Discrete Math. 21, 325--356 (1984). Topics on Perfect Graphs.

\bibitem{Hoang} Ho\`{a}ng, C.T., Kami\'{n}ski, M., Lozin, V.V., Sawada, J., Shu, X.:
 Deciding $k$-colorability of $P_5$-free graphs in polynomial time. Algorithmica 57, 74--81 (2010).

\bibitem{Hoang1} Ho\`{a}ng, C.T., Moore, B., Recoskiez, D.,
Sawada, J., Vatshelle, M.: Constructions of $k$-critical $P_5$-free graphs, preprint, 2013.

\bibitem{Holyer} Holyer, I.: The NP-completeness of edge coloring. SIAM J. Comput. 10, 718--720 (1981).


\bibitem{Huang MFCS} Huang, S. W.: Improved complexity results on $k$-coloring $P_t$-free graphs.
In: Proceedings of MFCS 2013, in: LNCS, vol. 8087, 2013, pp. 551-558.


\bibitem{Lozin} Kami\'{n}ski, M., Lozin, V.V.: Coloring edges and vertices of graphs without short or long cycles.
Contrib. Discrete. Mah. 2, 61--66 (2007).

\bibitem{Kral} Kr\'{a}l, D.,  Kratochv\'{i}l, J.,  Tuza, Zs., Woeginger, G.J.:
Complexity of coloring graphs without forbidden induced subgraphs.
In: Proceedings of WG 2001, in: LNCS, vol. 2204, 2001, pp. 254-262.


\bibitem{Randerath 2007}  Le, V.B., Randerath, B., Schiermeyer, I.:
On the complexity of 4-coloring graphs without long induced paths.
Theoret. Comput. Sci. 389, 330--335 (2007).

\bibitem{Leven} Leven, D., Galil, Z.: NP-completeness of finding the chromatic index of regular graphs.
J. Algorithm 4, 35--44 (1983).

\bibitem{maffray}  Maffray, F.,  Morel, G.: On $3$-Colorable $P_5$-Free Graphs.
SIAM J. Discrete Math., 26(4)  1682--1708 (2012).

\bibitem{Schiermeyer} Randerath, B., Schiermeyer, I.: 3-Colorability $\in$ P for $P_6$-free graphs.
Discrete Appl. Math. 136, 299--313 (2004).

\bibitem{Randerath} Randerath, B., Schiermeyer, I.: Vertex colouring and fibidden subgraphs-a survey.
Graphs Combin. 20, 1--40 (2004).

\bibitem{NAE} Schaefer, T. J.: The complexity of satisfiability problems. Proc. STOC 1978,
216--226 (1978).

\bibitem{Tarjan} Tarjan, R. E.: Decomposition by clique separators. Discrete Mathematics 55, 221-232 (1985).

\bibitem{Tuza} Tuza, Zs.: Graph colorings with local restrictions-a survey. Discuss. Math. Graph Theory 17,
161--228 (1997).

\bibitem{Woe} Woeginger, G.J., Sgall, J.: The complexity of coloring graphs without long induced paths.
Acta Cybernet. 15, 107--117 (2001).

\end{thebibliography}
\end{document}